\documentclass{mystyle}
\usepackage{graphicx}
\usepackage{amssymb}
\usepackage{latexsym}
\def\PsfigVersion{1.9}
\ifx\undefined\psfig\else \fi

\let\LaTeXAtSign=\@
\let\@=\relax
\edef\psfigRestoreAt{\catcode`\@=\number\catcode`@\relax}
\catcode`\@=11\relax
\newwrite\@unused
\def\ps@typeout#1{{\let\protect\string\immediate\write\@unused{#1}}}
\ps@typeout{psfig/tex \PsfigVersion}

\def\figurepath{./}

\def\@nnil{\@nil}
\def\@empty{}
\def\@psdonoop#1\@@#2#3{}
\def\@psdo#1:=#2\do#3{\edef\@psdotmp{#2}\ifx\@psdotmp\@empty \else
    \expandafter\@psdoloop#2,\@nil,\@nil\@@#1{#3}\fi}
\def\@psdoloop#1,#2,#3\@@#4#5{\def#4{#1}\ifx #4\@nnil \else
       #5\def#4{#2}\ifx #4\@nnil \else#5\@ipsdoloop #3\@@#4{#5}\fi\fi}
\def\@ipsdoloop#1,#2\@@#3#4{\def#3{#1}\ifx #3\@nnil 
       \let\@nextwhile=\@psdonoop \else
      #4\relax\let\@nextwhile=\@ipsdoloop\fi\@nextwhile#2\@@#3{#4}}
\def\@tpsdo#1:=#2\do#3{\xdef\@psdotmp{#2}\ifx\@psdotmp\@empty \else
    \@tpsdoloop#2\@nil\@nil\@@#1{#3}\fi}
\def\@tpsdoloop#1#2\@@#3#4{\def#3{#1}\ifx #3\@nnil 
       \let\@nextwhile=\@psdonoop \else
      #4\relax\let\@nextwhile=\@tpsdoloop\fi\@nextwhile#2\@@#3{#4}}
\ifx\undefined\fbox
\newdimen\fboxrule
\newdimen\fboxsep
\newdimen\ps@tempdima
\newbox\ps@tempboxa
\long\def\fbox#1{\leavevmode\setbox\ps@tempboxa\hbox{#1}\ps@tempdima\fboxrule
    \advance\ps@tempdima \fboxsep \advance\ps@tempdima \dp\ps@tempboxa
   \hbox{\lower \ps@tempdima\hbox
  {\vbox{\hrule height \fboxrule
          \hbox{\vrule width \fboxrule \hskip\fboxsep
          \vbox{\vskip\fboxsep \box\ps@tempboxa\vskip\fboxsep}\hskip 
                 \fboxsep\vrule width \fboxrule}
                 \hrule height \fboxrule}}}}
\fi
\newread\ps@stream
\newif\ifnot@eof       
\newif\if@noisy        
\newif\if@atend        
\newif\if@psfile       
{\catcode`\%=12\global\gdef\epsf@start{
\def\epsf@PS{PS}
\def\epsf@getbb#1{%
\openin\ps@stream=#1
\ifeof\ps@stream\ps@typeout{Error, File #1 not found}\else
   {\not@eoftrue \chardef\other=12
    \def\do##1{\catcode`##1=\other}\dospecials \catcode`\ =10
    \loop
       \if@psfile
	  \read\ps@stream to \epsf@fileline
       \else{
	  \obeyspaces
          \read\ps@stream to \epsf@tmp\global\let\epsf@fileline\epsf@tmp}
       \fi
       \ifeof\ps@stream\not@eoffalse\else
       \if@psfile\else
       \expandafter\epsf@test\epsf@fileline:. \\%
       \fi
          \expandafter\epsf@aux\epsf@fileline:. \\%
       \fi
   \ifnot@eof\repeat
   }\closein\ps@stream\fi}%
\long\def\epsf@test#1#2#3:#4\\{\def\epsf@testit{#1#2}
			\ifx\epsf@testit\epsf@start\else
\ps@typeout{Warning! File does not start with `\epsf@start'.  It may not be a PostScript file.}
			\fi
			\@psfiletrue} 
{\catcode`\%=12\global\let\epsf@percent=
\long\def\epsf@aux#1#2:#3\\{\ifx#1\epsf@percent
   \def\epsf@testit{#2}\ifx\epsf@testit\epsf@bblit
	\@atendfalse
        \epsf@atend #3 . \\%
	\if@atend	
	   \if@verbose{
		\ps@typeout{psfig: found `(atend)'; continuing search}
	   }\fi
        \else
        \epsf@grab #3 . . . \\%
        \not@eoffalse
        \global\no@bbfalse
        \fi
   \fi\fi}%
\def\epsf@grab #1 #2 #3 #4 #5\\{%
   \global\def\epsf@llx{#1}\ifx\epsf@llx\empty
      \epsf@grab #2 #3 #4 #5 .\\\else
   \global\def\epsf@lly{#2}%
   \global\def\epsf@urx{#3}\global\def\epsf@ury{#4}\fi}%
\def\epsf@atendlit{(atend)} 
\def\epsf@atend #1 #2 #3\\{%
   \def\epsf@tmp{#1}\ifx\epsf@tmp\empty
      \epsf@atend #2 #3 .\\\else
   \ifx\epsf@tmp\epsf@atendlit\@atendtrue\fi\fi}

\chardef\psletter = 11 
\chardef\other = 12

\newif \ifdebug 
\newif\ifc@mpute 
\c@mputetrue 

\let\then = \relax
\def\r@dian{pt }
\let\r@dians = \r@dian
\let\dimensionless@nit = \r@dian
\let\dimensionless@nits = \dimensionless@nit
\def\internal@nit{sp }
\let\internal@nits = \internal@nit
\newif\ifstillc@nverging
\def \Mess@ge #1{\ifdebug \then \message {#1} \fi}

{ 
	\catcode `\@ = \psletter
	\gdef \nodimen {\expandafter \n@dimen \the \dimen}
	\gdef \term #1 #2 #3%
	       {\edef \t@ {\the #1}
		\edef \t@@ {\expandafter \n@dimen \the #2\r@dian}%
		\t@rm {\t@} {\t@@} {#3}%
	       }
	\gdef \t@rm #1 #2 #3%
	       {{%
		\count 0 = 0
		\dimen 0 = 1 \dimensionless@nit
		\dimen 2 = #2\relax
		\Mess@ge {Calculating term #1 of \nodimen 2}%
		\loop
		\ifnum	\count 0 < #1
		\then	\advance \count 0 by 1
			\Mess@ge {Iteration \the \count 0 \space}%
			\Multiply \dimen 0 by {\dimen 2}%
			\Mess@ge {After multiplication, term = \nodimen 0}%
			\Divide \dimen 0 by {\count 0}%
			\Mess@ge {After division, term = \nodimen 0}%
		\repeat
		\Mess@ge {Final value for term #1 of 
				\nodimen 2 \space is \nodimen 0}%
		\xdef \Term {#3 = \nodimen 0 \r@dians}%
		\aftergroup \Term
	       }}
	\catcode `\p = \other
	\catcode `\t = \other
	\gdef \n@dimen #1pt{#1} 
}

\def \Divide #1by #2{\divide #1 by #2} 

\def \Multiply #1by #2
       {{
	\count 0 = #1\relax
	\count 2 = #2\relax
	\count 4 = 65536
	\Mess@ge {Before scaling, count 0 = \the \count 0 \space and
			count 2 = \the \count 2}%
	\ifnum	\count 0 > 32767 
	\then	\divide \count 0 by 4
		\divide \count 4 by 4
	\else	\ifnum	\count 0 < -32767
		\then	\divide \count 0 by 4
			\divide \count 4 by 4
		\else
		\fi
	\fi
	\ifnum	\count 2 > 32767 
	\then	\divide \count 2 by 4
		\divide \count 4 by 4
	\else	\ifnum	\count 2 < -32767
		\then	\divide \count 2 by 4
			\divide \count 4 by 4
		\else
		\fi
	\fi
	\multiply \count 0 by \count 2
	\divide \count 0 by \count 4
	\xdef \product {#1 = \the \count 0 \internal@nits}%
	\aftergroup \product
       }}

\def\r@duce{\ifdim\dimen0 > 90\r@dian \then   
		\multiply\dimen0 by -1
		\advance\dimen0 by 180\r@dian
		\r@duce
	    \else \ifdim\dimen0 < -90\r@dian \then  
		\advance\dimen0 by 360\r@dian
		\r@duce
		\fi
	    \fi}

\def\Sine#1%
       {{%
	\dimen 0 = #1 \r@dian
	\r@duce
	\ifdim\dimen0 = -90\r@dian \then
	   \dimen4 = -1\r@dian
	   \c@mputefalse
	\fi
	\ifdim\dimen0 = 90\r@dian \then
	   \dimen4 = 1\r@dian
	   \c@mputefalse
	\fi
	\ifdim\dimen0 = 0\r@dian \then
	   \dimen4 = 0\r@dian
	   \c@mputefalse
	\fi
	\ifc@mpute \then
		\divide\dimen0 by 180
		\dimen0=3.141592654\dimen0
		\dimen 2 = 3.1415926535897963\r@dian 
		\divide\dimen 2 by 2 
		\Mess@ge {Sin: calculating Sin of \nodimen 0}%
		\count 0 = 1 
		\dimen 2 = 1 \r@dian 
		\dimen 4 = 0 \r@dian 
		\loop
			\ifnum	\dimen 2 = 0 
			\then	\stillc@nvergingfalse 
			\else	\stillc@nvergingtrue
			\fi
			\ifstillc@nverging 
			\then	\term {\count 0} {\dimen 0} {\dimen 2}%
				\advance \count 0 by 2
				\count 2 = \count 0
				\divide \count 2 by 2
				\ifodd	\count 2 
				\then	\advance \dimen 4 by \dimen 2
				\else	\advance \dimen 4 by -\dimen 2
				\fi
		\repeat
	\fi		
			\xdef \sine {\nodimen 4}%
       }}

\def\Cosine#1{\ifx\sine\UnDefined\edef\Savesine{\relax}\else
		             \edef\Savesine{\sine}\fi
	{\dimen0=#1\r@dian\advance\dimen0 by 90\r@dian
	 \Sine{\nodimen 0}
	 \xdef\cosine{\sine}
	 \xdef\sine{\Savesine}}}	      

\def\psdraft{
	\def\@psdraft{0}
}
\def\psfull{
	\def\@psdraft{100}
}

\psfull

\newif\if@scalefirst
\def\psscalefirst{\@scalefirsttrue}
\def\psrotatefirst{\@scalefirstfalse}
\psrotatefirst

\newif\if@draftbox
\def\psnodraftbox{
	\@draftboxfalse
}
\def\psdraftbox{
	\@draftboxtrue
}
\@draftboxtrue

\newif\if@prologfile
\newif\if@postlogfile
\def\pssilent{
	\@noisyfalse
}
\def\psnoisy{
	\@noisytrue
}
\psnoisy
\newif\if@bbllx
\newif\if@bblly
\newif\if@bburx
\newif\if@bbury
\newif\if@height
\newif\if@width
\newif\if@rheight
\newif\if@rwidth
\newif\if@angle
\newif\if@clip
\newif\if@verbose
\def\@p@@sclip#1{\@cliptrue}

\newif\if@decmpr

\def\@p@@sfigure#1{\def\@p@sfile{null}\def\@p@sbbfile{null}
	        \openin1=#1.bb
		\ifeof1\closein1
	        	\openin1=\figurepath#1.bb
			\ifeof1\closein1
			        \openin1=#1
				\ifeof1\closein1%
				       \openin1=\figurepath#1
					\ifeof1
					   \ps@typeout{Error, File #1 not found}
						\if@bbllx\if@bblly
				   		\if@bburx\if@bbury
			      				\def\@p@sfile{#1}%
			      				\def\@p@sbbfile{#1}%
							\@decmprfalse
				  	   	\fi\fi\fi\fi
					\else\closein1
				    		\def\@p@sfile{\figurepath#1}%
				    		\def\@p@sbbfile{\figurepath#1}%
						\@decmprfalse
	                       		\fi%
			 	\else\closein1%
					\def\@p@sfile{#1}
					\def\@p@sbbfile{#1}
					\@decmprfalse
			 	\fi
			\else
				\def\@p@sfile{\figurepath#1}
				\def\@p@sbbfile{\figurepath#1.bb}
				\@decmprtrue
			\fi
		\else
			\def\@p@sfile{#1}
			\def\@p@sbbfile{#1.bb}
			\@decmprtrue
		\fi}

\def\@p@@sfile#1{\@p@@sfigure{#1}}

\def\@p@@sbbllx#1{
		\@bbllxtrue
		\dimen100=#1
		\edef\@p@sbbllx{\number\dimen100}
}
\def\@p@@sbblly#1{
		\@bbllytrue
		\dimen100=#1
		\edef\@p@sbblly{\number\dimen100}
}
\def\@p@@sbburx#1{
		\@bburxtrue
		\dimen100=#1
		\edef\@p@sbburx{\number\dimen100}
}
\def\@p@@sbbury#1{
		\@bburytrue
		\dimen100=#1
		\edef\@p@sbbury{\number\dimen100}
}
\def\@p@@sheight#1{
		\@heighttrue
		\dimen100=#1
   		\edef\@p@sheight{\number\dimen100}
}
\def\@p@@swidth#1{
		\@widthtrue
		\dimen100=#1
		\edef\@p@swidth{\number\dimen100}
}
\def\@p@@srheight#1{
		\@rheighttrue
		\dimen100=#1
		\edef\@p@srheight{\number\dimen100}
}
\def\@p@@srwidth#1{
		\@rwidthtrue
		\dimen100=#1
		\edef\@p@srwidth{\number\dimen100}
}
\def\@p@@sangle#1{
		\@angletrue
		\edef\@p@sangle{#1} 
}
\def\@p@@ssilent#1{ 
		\@verbosefalse
}
\def\@p@@sprolog#1{\@prologfiletrue\def\@prologfileval{#1}}
\def\@p@@spostlog#1{\@postlogfiletrue\def\@postlogfileval{#1}}
\def\@cs@name#1{\csname #1\endcsname}
\def\@setparms#1=#2,{\@cs@name{@p@@s#1}{#2}}
\def\ps@init@parms{
		\@bbllxfalse \@bbllyfalse
		\@bburxfalse \@bburyfalse
		\@heightfalse \@widthfalse
		\@rheightfalse \@rwidthfalse
		\def\@p@sbbllx{}\def\@p@sbblly{}
		\def\@p@sbburx{}\def\@p@sbbury{}
		\def\@p@sheight{}\def\@p@swidth{}
		\def\@p@srheight{}\def\@p@srwidth{}
		\def\@p@sangle{0}
		\def\@p@sfile{} \def\@p@sbbfile{}
		\def\@p@scost{10}
		\def\@sc{}
		\@prologfilefalse
		\@postlogfilefalse
		\@clipfalse
		\if@noisy
			\@verbosetrue
		\else
			\@verbosefalse
		\fi
}
\def\parse@ps@parms#1{
	 	\@psdo\@psfiga:=#1\do
		   {\expandafter\@setparms\@psfiga,}}
\newif\ifno@bb
\def\bb@missing{
	\if@verbose{
		\ps@typeout{psfig: searching \@p@sbbfile \space  for bounding box}
	}\fi
	\no@bbtrue
	\epsf@getbb{\@p@sbbfile}
        \ifno@bb \else \bb@cull\epsf@llx\epsf@lly\epsf@urx\epsf@ury\fi
}	
\def\bb@cull#1#2#3#4{
	\dimen100=#1 bp\edef\@p@sbbllx{\number\dimen100}
	\dimen100=#2 bp\edef\@p@sbblly{\number\dimen100}
	\dimen100=#3 bp\edef\@p@sbburx{\number\dimen100}
	\dimen100=#4 bp\edef\@p@sbbury{\number\dimen100}
	\no@bbfalse
}
\newdimen\p@intvaluex
\newdimen\p@intvaluey
\def\rotate@#1#2{{\dimen0=#1 sp\dimen1=#2 sp
		  \global\p@intvaluex=\cosine\dimen0
		  \dimen3=\sine\dimen1
		  \global\advance\p@intvaluex by -\dimen3
		  \global\p@intvaluey=\sine\dimen0
		  \dimen3=\cosine\dimen1
		  \global\advance\p@intvaluey by \dimen3
		  }}
\def\compute@bb{
		\no@bbfalse
		\if@bbllx \else \no@bbtrue \fi
		\if@bblly \else \no@bbtrue \fi
		\if@bburx \else \no@bbtrue \fi
		\if@bbury \else \no@bbtrue \fi
		\ifno@bb \bb@missing \fi
		\ifno@bb \ps@typeout{FATAL ERROR: no bb supplied or found}
			\no-bb-error
		\fi
		\count203=\@p@sbburx
		\count204=\@p@sbbury
		\advance\count203 by -\@p@sbbllx
		\advance\count204 by -\@p@sbblly
		\edef\ps@bbw{\number\count203}
		\edef\ps@bbh{\number\count204}
		\if@angle 
			\Sine{\@p@sangle}\Cosine{\@p@sangle}
	        	{\dimen100=\maxdimen\xdef\r@p@sbbllx{\number\dimen100}
					    \xdef\r@p@sbblly{\number\dimen100}
			                    \xdef\r@p@sbburx{-\number\dimen100}
					    \xdef\r@p@sbbury{-\number\dimen100}}
                        \def\minmaxtest{
			   \ifnum\number\p@intvaluex<\r@p@sbbllx
			      \xdef\r@p@sbbllx{\number\p@intvaluex}\fi
			   \ifnum\number\p@intvaluex>\r@p@sbburx
			      \xdef\r@p@sbburx{\number\p@intvaluex}\fi
			   \ifnum\number\p@intvaluey<\r@p@sbblly
			      \xdef\r@p@sbblly{\number\p@intvaluey}\fi
			   \ifnum\number\p@intvaluey>\r@p@sbbury
			      \xdef\r@p@sbbury{\number\p@intvaluey}\fi
			   }
			\rotate@{\@p@sbbllx}{\@p@sbblly}
			\minmaxtest
			\rotate@{\@p@sbbllx}{\@p@sbbury}
			\minmaxtest
			\rotate@{\@p@sbburx}{\@p@sbblly}
			\minmaxtest
			\rotate@{\@p@sbburx}{\@p@sbbury}
			\minmaxtest
			\edef\@p@sbbllx{\r@p@sbbllx}\edef\@p@sbblly{\r@p@sbblly}
			\edef\@p@sbburx{\r@p@sbburx}\edef\@p@sbbury{\r@p@sbbury}
		\fi
		\count203=\@p@sbburx
		\count204=\@p@sbbury
		\advance\count203 by -\@p@sbbllx
		\advance\count204 by -\@p@sbblly
		\edef\@bbw{\number\count203}
		\edef\@bbh{\number\count204}
}
\def\in@hundreds#1#2#3{\count240=#2 \count241=#3
		     \count100=\count240	
		     \divide\count100 by \count241
		     \count101=\count100
		     \multiply\count101 by \count241
		     \advance\count240 by -\count101
		     \multiply\count240 by 10
		     \count101=\count240	
		     \divide\count101 by \count241
		     \count102=\count101
		     \multiply\count102 by \count241
		     \advance\count240 by -\count102
		     \multiply\count240 by 10
		     \count102=\count240	
		     \divide\count102 by \count241
		     \count200=#1\count205=0
		     \count201=\count200
			\multiply\count201 by \count100
		 	\advance\count205 by \count201
		     \count201=\count200
			\divide\count201 by 10
			\multiply\count201 by \count101
			\advance\count205 by \count201
		     \count201=\count200
			\divide\count201 by 100
			\multiply\count201 by \count102
			\advance\count205 by \count201
		     \edef\@result{\number\count205}
}
\def\compute@wfromh{
		\in@hundreds{\@p@sheight}{\@bbw}{\@bbh}
		\edef\@p@swidth{\@result}
}
\def\compute@hfromw{
	        \in@hundreds{\@p@swidth}{\@bbh}{\@bbw}
		\edef\@p@sheight{\@result}
}
\def\compute@handw{
		\if@height 
			\if@width
			\else
				\compute@wfromh
			\fi
		\else 
			\if@width
				\compute@hfromw
			\else
				\edef\@p@sheight{\@bbh}
				\edef\@p@swidth{\@bbw}
			\fi
		\fi
}
\def\compute@resv{
		\if@rheight \else \edef\@p@srheight{\@p@sheight} \fi
		\if@rwidth \else \edef\@p@srwidth{\@p@swidth} \fi
}
\def\compute@sizes{
	\compute@bb
	\if@scalefirst\if@angle
	\if@width
	   \in@hundreds{\@p@swidth}{\@bbw}{\ps@bbw}
	   \edef\@p@swidth{\@result}
	\fi
	\if@height
	   \in@hundreds{\@p@sheight}{\@bbh}{\ps@bbh}
	   \edef\@p@sheight{\@result}
	\fi
	\fi\fi
	\compute@handw
	\compute@resv}

\def\psfig#1{\vbox {
	\ps@init@parms
	\parse@ps@parms{#1}
	\compute@sizes
	\ifnum\@p@scost<\@psdraft{
		\special{ps::[begin] 	\@p@swidth \space \@p@sheight \space
				\@p@sbbllx \space \@p@sbblly \space
				\@p@sbburx \space \@p@sbbury \space
				startTexFig \space }
		\if@angle
			\special {ps:: \@p@sangle \space rotate \space} 
		\fi
		\if@clip{
			\if@verbose{
				\ps@typeout{(clip)}
			}\fi
			\special{ps:: doclip \space }
		}\fi
		\if@prologfile
		    \special{ps: plotfile \@prologfileval \space } \fi
		\if@decmpr{
			\if@verbose{
				\ps@typeout{psfig: including \@p@sfile.Z \space }
			}\fi
			\special{ps: plotfile "`zcat \@p@sfile.Z" \space }
		}\else{
			\if@verbose{
				\ps@typeout{psfig: including \@p@sfile \space }
			}\fi
			\special{ps: plotfile \@p@sfile \space }
		}\fi
		\if@postlogfile
		    \special{ps: plotfile \@postlogfileval \space } \fi
		\special{ps::[end] endTexFig \space }
		\vbox to \@p@srheight sp{
			\hbox to \@p@srwidth sp{
				\hss
			}
		\vss
		}
	}\else{
		\if@draftbox{		
			\hbox{\frame{\vbox to \@p@srheight sp{
			\vss
			\hbox to \@p@srwidth sp{ \hss \@p@sfile \hss }
			\vss
			}}}
		}\else{
			\vbox to \@p@srheight sp{
			\vss
			\hbox to \@p@srwidth sp{\hss}
			\vss
			}
		}\fi

	}\fi
}}
\psfigRestoreAt
\let\@=\LaTeXAtSign

\newcommand{\nop}[1]{}

\def\bproof{\noindent{\bf Proof}.\ }
\def\punto{\hspace*{\fill}$\Box$}
\def\eproof{\punto\vspace{0.4cm}}
\def\i{{\bf{i}}}
\def\j{{\bf{j}}}
\def\g{{\bf{g}}}
\def\h{{\bf{h}}}
\def\p{{\bf{p}}}
\def\q{{\bf{q}}}
\def\l{{\bf{l}}}
\def\u{{\bf{u}}}
\def\n{{\bf{n}}}
\def\m{{\bf{m}}}
\def\compij{{\i\widetilde{..}\j}}

\def\MF{{\tilde{M}}}
\def\xd{{\overline{t}_u}}
\def\xt{{\overline{t}}}
\def\xs{{\overline{s}}}
\def\xdd{{\overline{t}_l}}
\def\Qr{{\overline{Q}}}
\def\sumr{{{sum}}}
\def\countr{{{count}}}
\def\queryr{{{query}}}
\def\k{{\bf{k}}}
\def\1{{\bf{1}}}
\def\tmax{t^{U}}
\def\tmin{t^{L}}
\def\"{``}

\def\<{\mbox{$<$}}
\def\>{\mbox{$>$}}

\newtheorem{definition}{Definition}
\newtheorem{theorem}{Theorem}
\newtheorem{proposition}{Proposition}

\newtheorem{corollary}{Corollary}

\begin{document}
\begin{frontmatter}
\title{Estimating Range Queries using\\ Aggregate Data with Integrity
Constraints:\\ a Probabilistic Approach}
\thanks{An abridged version of this paper appeared in
Proceedings of 8th International Conference on Database Theory
(ICDT 2001),
Springer, Lecture Notes in Computer Science, Vol. 1973,
year 2001, ISBN 3-540-41456-8}

\author[label1]{Francesco Buccafurri}
\ead{bucca@unirc.it}
\author[label2]{Filippo Furfaro}
\ead{furfaro@si.deis.unical.it}
\author[label2,label3]{Domenico Sacc\`{a}}
\ead{sacca@unical.it}
\address[label1]{DIMET, University Mediterranea, 89100 Reggio Calabria, Italy}
\address[label2]{DEIS, University of Calabria, 87036 Rende, Italy}
\address[label3]{ICAR-CNR, 87036 Rende, Italy}

\begin{abstract}
The problem of recovering (count and sum) range queries over
multidimensional data only on the basis of aggregate information
on such data is addressed.
This problem can be formalized as follows.
Suppose that a transformation $\tau$ producing a summary from a
multidimensional data set is used.
Now, given a data set $D$, a summary $S=\tau(D)$ and a range
query $r$ on $D$, the problem consists of studying $r$ by modelling it
as a random variable defined over the sample space of all the data sets
$D'$ such that $\tau(D') = S$.
The study of such a random variable, done by the definition of its probability
distribution and the computation of its mean value and variance, represents
a well-founded, theoretical probabilistic approach for estimating the
query only on the basis of the available information (that is the summary
$S$) without assumptions on original data.
\end{abstract}
\end{frontmatter}

\section{Introduction}
In many application contexts, such as statistical databases,
transaction recording systems, scientific databases, query
optimizers, OLAP (On-line Analytical Processing), and many others,
a multidimensional view of data is often adopted:
Data are stored in multidimensional arrays, called {\em datacubes}
\cite{GraBos96,HarRaj96}, where every range query (computing
aggregate values such as the sum of the values contained inside a
range, or the number of occurrences of distinct values) can be
answered by visiting sequentially a sub-array covering the range.
In demanding applications, in order to both save storage space and
support fast access, datacubes are summarized into lossy synopses
of aggregate values, and range queries are executed over aggregate
data rather than over raw ones, thus returning approximate
answers.
Approximate query answering is very useful when the user wants to have
fast answers, thus avoiding waiting a long time to get a precision which is
often not necessary.\\
Data aggregation and approximate answering have been first introduced
many years ago for \emph{histograms} \cite{Koo80} in the context of
selectivity estimation (i.e. estimation of query result sizes) for
query optimization in relational databases \cite{Cri81,Cia93,MurDew88,Poo97}.
In this scenario, histograms are built on the frequency distribution of
attribute values occurring in a relation, and are constructed by partitioning
this distribution into a number of non-overlapping blocks (called \emph{buckets}).
For each of these blocks, a number of aggregate data are stored, instead of
the detailed frequency distribution.
The selectivity a query is estimated by interpolating the aggregate information
stored in the histogram.\\
Later on, several techniques for compressing datacubes and allowing fast approximate
answering have been proposed in the literature in the context of OLAP applications,
where data to be summarized are called \emph{measure values} (e.g., daily income of a shop,
number of users accessing a service, etc.).
Some of these approaches use either sampling \cite{GibMat98,GMP97a,Haa97,HelHaa97}
or complex mathematical transformations (such as wavelets) to compress data
\cite{Gar*02,StoDer96,VitWan98,VitWan99}.
Indeed, the approach which turned out to be the most effective one (in terms of accuracy
of the estimates) is the histogram-based one.
In fact, both frequency distributions occurring in selectivity estimation and
measure values in OLAP datacubes are multi-dimensional data distributions, which
can be partitioned and aggregated adopting the same technique.
Therefore, due to the increasing popularity of OLAP applications (which turned out to be
particularly useful for the decision making process \cite{ChaDay97}), a renewed interest
has been devoted to histogram-based compression techniques.
Most of works on this topic mainly deal with either improving partitioning techniques
in terms of efficiency and effectiveness \cite{BuFuSaSi03,Gil*01,Gun*00,Ja*01}, or
maintaining the summary data up-to-date when the collected information changes
continuously \cite{DonIoa99,Guh*01,Guh*02,Tha*02}.

In this paper we address a different problem, which has been rather disregarded
by previous research, and which is very relevant for an effective applicability
of summarization techniques:
We focus on the analysis of estimation errors which occur when evaluating range
queries directly on summary data, without accessing original ones.
Indeed, in all previous works dealing with histogram-based summarization techniques,
either the estimation error is not studied at all, or only a rough evaluation of
upper bounds of this error is given \cite{Jag*98}.
The lack of information on the estimation error reduces the scope of applicability
of approximate query answering: approximate results are really usable only if they
are returned together with a detailed analysis of the possible error so that, if
the user is not satisfied with the obtained precision, s/he may eventually decide
to submit the query on the actual datacube.

In more detail, we study the problem of estimating count and sum range queries
issued on a compressed datacube in a rather general framework: we assume that
compression has been performed by first partitioning a given datacube into a
number of blocks using any of the various proposed techniques, and then storing
aggregate information for each block.
This aggregate information mainly consists in the sum and the number of the elements
belonging to each block.
Moreover, we assume that some integrity constraints, which are expressible in a
succinct format, are stored.
Our approach is independent on the technique used to partition the
datacube into blocks: its concern is estimating values and accuracy
of range queries using aggregate data - no matter how they have been
obtained - using just interpolation with no assumptions on the actual distribution
of data into aggregation blocks.

The evaluation of the accuracy of estimates is based on a probabilistic framework
where queries are represented by means of random variables, and is performed as
follows.
Given a datacube $D$ and the compressed datacube $S$ obtained from
$D$ by applying the histogram-based compression strategy introduced above,
we denote the transformation from $D$ to $S$ by $\tau$, thus $S =\tau(D)$.
Let now ${\mathcal D_S}$ denote  the set of all the datacubes $\bar D$
such that $\tau(\bar D) = S$.
Observe that any datacube $\bar D \in {\mathcal D_S}$ is a possible
guess of the original datacube $D$, done only on the basis of the
knowledge of $S$.
So, if we are given a range query $r$ on $D$, estimating $r$ from $S$
can be thought of as guessing the response to $r$ on $D$ by applying
the range query $r$ to any datacube $\bar D$ of $\mathcal D_S$.
According to this observation, we model the estimation of the range
query $r$ from $S$ as the mean value of the random variable defined
by $r$ on the sample set ${\mathcal D_S}$, representing {\em all}
possible guesses of $r$ compatible with the summary $S$.
In order to analyze the estimation error, we thus study this random
variable by determining its probability distribution and its variance.
Actually, our analysis considers a {\em family} of transformations
$\tau$ based on the partition of the datacube into blocks, where
each transformation stores different aggregate information for
each block and a number of integrity constraints.
The introduction of integrity constraints allows us to take into
account more detailed information than sum and count on a block,
whose exploitation may bias significantly the estimation toward the
actual value.
Indeed, integrity constraints produce a restriction of the sample space
and a reduction of the variance of the estimation w.r.t. to the case of
absence of integrity constraints.
The integrity constraints which have been considered in this work concern
the minimum number of null or non-null tuples occurring in ranges of
datacubes.
Although more complex constraints could be considered, we have restricted
our attention to this kind of constraint since they often arise in practice.
For instance, given a datacube whose dimensions are the time (in terms of days)
and the products, and the measure is the amount of daily product
sales, realistic integrity constraints are that the sales are null
during the week-end, while at least 4 times a week the sales are
not null.

\noindent \textbf{Plan of the paper.}
The paper is organized as follows.
In Section \ref{sec:comprdata}, a simple compression technique which
will be used for explaining and applying our estimation paradigm is
introduced, and integrity constraints about the number of null or
non-null tuples in the datacube ranges are formally defined.
In Section \ref{sec:probfram} the probabilistic framework for
estimating $count$ and $sum$ range queries is formalized.
In Sections \ref{sec:case1}, \ref{sec:case2} and \ref{sec:case3}
three different estimation paradigms (exploiting different classes
of aggregate information) are introduced: in Section \ref{sec:case1},
sum queries are estimated by using only the information about the
sum of the values contained in each bucket, whereas count queries
are evaluated by exploiting only the information on the number of
non null elements in each bucket.
Section \ref{sec:case2} shows how the information on the number
and the sum of the non null values contained in each bucket can be
used jointly to estimate sum and count queries.
In Section \ref{sec:case3} the estimation using integrity constraints
is formalized and in the subsequent section we elaborate on the
``positive" influence of integrity constraints on the accuracy of
query estimations, and substantiate our claim with some experimental
results obtained applying our estimation techniques to real-life data
distributions.
Some interesting applications of our theoretical framework for the
estimation of frequency distributions inside a histogram are presented
in Section \ref{sec:histo}.
Finally, in Section \ref{sec:conclusioni} conclusions and future research
lines are discussed.
In particular, we stress that our work is not certainly conclusive,
since a larger family of transformations can be considered, by
taking into account different aggregates and other integrity
constraints.
Indeed the main contribution of the work is the definition of a novel
approach for modelling and studying the issue of approximate queries
from a theoretic point of view.

\section{Datacubes, their Compressed Representation and Integrity Constraints}
\label{sec:comprdata}

\subsection{Preliminary Definitions}\label{sec:preliminaries}
\vspace*{-3mm}
In this section we give some preliminary definitions and
notations.
Let $\i = \<i_1,\dots,i_r\>$ and $\j = \<j_1,\dots,j_r\>$ be two
$r$-tuples of cardinals, with $r>0$.
We extend common operators for cardinals to tuples in the obvious
way: $\i \leq \j$ means that $i_1 \leq j_1,\dots$ $i_r \leq j_r$;
$\i + \j$ denotes the tuple $\<i_1+j_1,\dots,$ $i_r + j_r\>$ and
so on.
Given a constant $p \geq 0$, $\p^{\,r}$ (or simply $\p$, if $r$
is understood) denotes the $r$-tuple of all $p$;
for instance, if $p=1$ and $r=5$, the term \textbf{1} denotes the tuple
$\<1,1,1,1,1\>$.
Finally, $[\i..\j]=$ $[i_1..j_1,\dots,i_r..j_r]$ denotes the range
of all $r$-tuples from $\i$ to $\j$, that is
$\{\q|$ $\i \leq \q \leq \j\}$.

\begin{definition}\label{def-multidimensional}
{\em A {\em multidimensional relation} $R$  is a relation whose
scheme consists of $r>0$ {\em dimensions} (also called {\em
functional attributes}) and $s>0$ {\em measure attributes}. The
dimensions are a key for the relation so that there are no two
tuples with the same dimension value.}
\end{definition}

From now on consider given a multidimensional relation $R$.
For the sake of presentation but without loss of generality, we
assume that:
\begin{itemize}
\item
$s=1$, and the domain of the unique measure attribute is the
set of cardinals,
\item
$r \geq 1$, and the domain of each dimension $q$, with $1 \leq q
\leq r$, is the range $[1..n_q]$, where $n_q > 2$ (i.e., the
projection of $R$ on the dimensions is a subset of $[\1..\n]$,
where $\n = \<n_1,\dots,n_r\>$).
\end{itemize}

Given any range $[\i..\j]$, $\1 \leq \i \leq \j
\leq \n$, we consider the following {\em range queries} on $R$:
\vspace*{-4mm}
\begin{itemize}
\item
{\em count query}:
$count^{[\i..\j]}(R)$ denotes the number of tuples of $R$ whose dimension
values are in $[\i..\j]$;
\item
{\em sum query}: $sum^{[\i..\j]}(R)$ denotes the sum of all measure values
for those tuples of $R$ whose dimension values are in $[\i..\j]$.
\end{itemize}
Since the dimension attributes are a key, the relation $R$ can be
naturally viewed as a $[\1..\n]$ matrix $M$ of elements with values
in $\mathcal{N}$.
In the rest of the paper this matrix will be called \emph{datacube}.

\begin{definition}\label{def-datacube}
{\em The datacube $M$ corresponding to the multidimensional relation $R$
is the $[\1..\n]$ matrix of cardinals such that, for each $\i \in [\1..\n]$,
$M[\i] = v$ if the tuple $\<\i,v\>$ is in $R$ or $M[\i] = 0$, otherwise.}
\end{definition}

As a consequence of the above definition, $\i$ is a {\em null element} if
either $\<\i,0\>$ is in $R$ or no tuple with dimension value $\i$ is present
in $R$.\\
The above range queries can be now re-formulated in terms of array
operations as follows:

\begin{itemize}
\item
$count^{[\i..\j]}(R)=count(M[\i..\j])= |\{\q| \ \q \in [\i..\j] \mbox{ and } M[\q] >0\}|$;
\item
$sum^{[\i..\j]}(R)=sum(M[\i..\j]) = \sum_{\q\in [\i..\j]} M[\q]$,
\end{itemize}
where $[\i..\j]$ is any range such that $\1 \leq \i \leq \j \leq \n$.

\subsection{Compressed Datacubes}\label{sec:compressed}
\vspace*{-4mm}

We next introduce  a \emph{compressed representation} of the relation
$R$ by dividing the corresponding datacube $M$ into a number of blocks
and by storing a number of aggregate data for each of them.\\
First we need to formalize the notion of \emph{compression factor}:

\begin{definition}\label{def-compressionFactor}
{\em Given $\m=$ $\<m_1,\dots,$ $m_r\>$,  $\1 \leq \m \leq \n$, an
{\em $\m$-compression factor} for $M$ is any tuple $F=$
$\<f_1,\dots,$ $f_r\>$, such that for each dimension $q$, $1 \leq
q \leq r$, $f_q$ is a $[0..m_q]$ array for which $0=f_q[0] <
f_q[1] < \cdots$ $< f_q[m_q]=n_q$, i.e., $f_q$ divides the
dimension $q$ into $m_q$ parts.}
\end{definition}

Observe that $F$ partitions the range $[\1..\n]$ into
$m_1 \times \cdots$ $\times m_r$ blocks.
Each of these blocks, denoted as $B_\k$, corresponds to a tuple
$\k=$ $\<k_1,\dots,$ $k_r\>$ in $[\1..\m]$.
Each block $B_\k$ has range $[F^-(\k)..F^+(\k)]$, where $F^+(\k)$ and $F^{-}(\k)$
denote the tuples $\<f_1[k_1], \dots, $ $f_r[k_r]\>$ and
$\<f_1[k_1-1]+1, \dots,$ $f_r[k_r-1]+1\>$, respectively.
The size of $B_\k$ (i.e. the number of cells inside the range of $B_\k$) is
$(f_1[k_1]-f_1[k_1-1])\times \cdots$ $\times (f_r[k_r]-f_r[k_r-1])$.

\begin{figure}
\centerline{ \psfig{figure=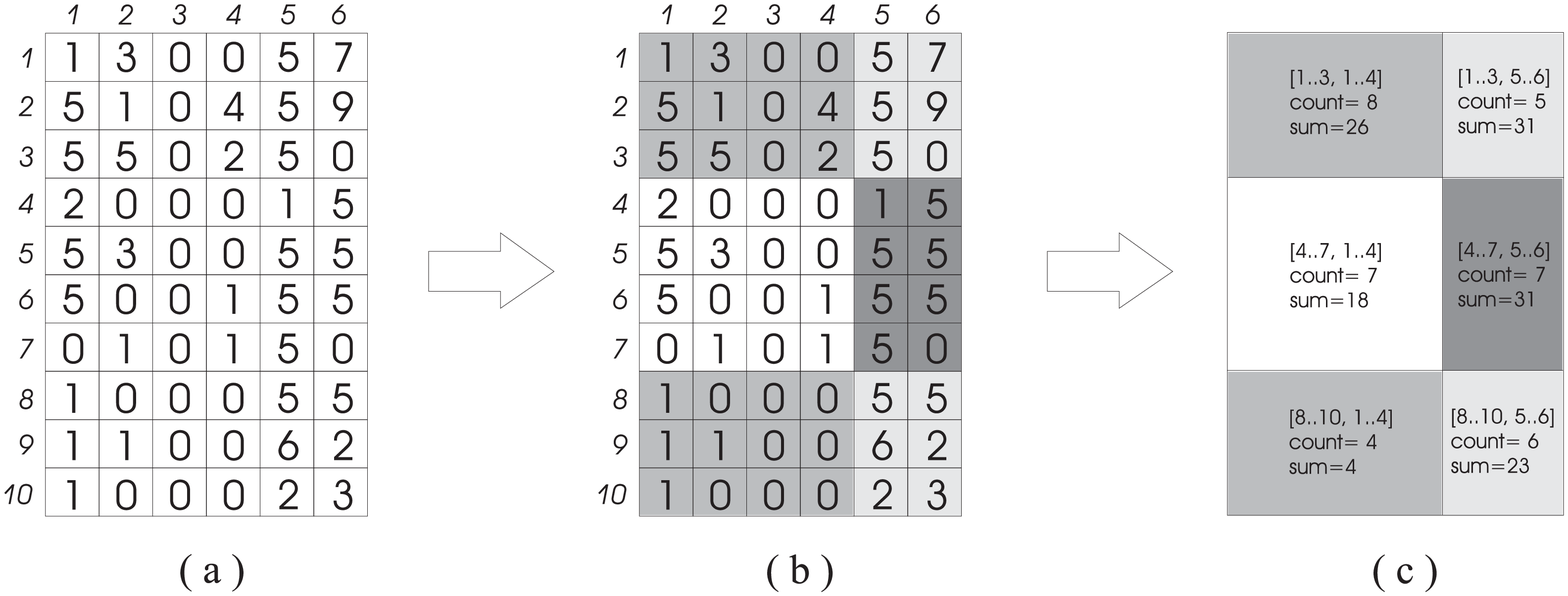,width=\textwidth} }
\caption{A two-dimensional datacube and its compressed representation}
\label{fig:new}
\end{figure}

As an example, consider the $[1..10,1..6]$ datacube $M$ in Figure
\ref{fig:new}(a), which is partitioned into 6 blocks as shown in
Figure \ref{fig:new}(b).
We have that $\m =$ $\<3,2\>$, $f_1[0]=0$, $f_1[1]=3$, $f_1[2]=7$, $f_1[3]=10$,
and $f_2[0]=0$, $f_2[1]=4$, $f_2[2]=6$.
The block $B_{\<1,1\>}$ has size $3\times 2$ and range $[1..3,1..4]$;
the block $B_{\<1,2\>}$ has size $3\times 2$ and range $[1..3,5..6]$, and so on.

\begin{definition}\label{def-compressed}
{\em Given an $\m$-compression factor $F$, a {\em ($F$-)compressed
representation} of the datacube $M$ is the pair of $[\1..\m]$
matrices $M_{count,F}$ and $M_{sum,F}$ such that for each $\k \in
[\1..\m]$, $ M_{count,F}[\k] = count (M[F^-(\k) .. F^+(\k)])$ and
$M_{sum,F}[\k] = sum (M[F^-(\k) .. F^+(\k)]).$ }
\end{definition}

The compressed representation of the datacube $M$ in Figure
\ref{fig:new}(a) is represented in Figure \ref{fig:new}(c), where
each block is associated to a triplet of values.
These values indicate, respectively, the range, the number of non-null elements
and the sum of the elements in the corresponding block.
For instance, the block $B_{\<1,1\>}$ has range $[1..3,1..4]$ and
contains 8 non-null elements with sum 26; the block $B_{\<1,2\>}$ has range
$[1..3,5..6]$ and contains 5 non-null elements with sum 29, and so on.

From now on, consider given an $\m$-compression factor $F$ and the
corresponding $F$-compressed representation of the datacube $M$.

\subsection{Integrity Constraints}
\vspace*{-4mm}
The aim of compressing a datacube is to reduce the storage space consumption of
its representation, in order to make answering range queries more efficient to
perform.
In fact queries can be evaluated on the basis of the aggregate data stored in the
compressed datacube without accessing the original one, and the amount of data that
must be extracted from the compressed datacube to answer a query is generally
smaller than the number of data that should be extracted from the original datacube.
This approach introduces some approximation, which is tolerated in all those
scenarios (such as selectivity estimation and OLAP services) where the efficiency of
query answering is mandatory and the accuracy of the answers is not so relevant.\\
The estimation of queries could be improved (in terms of accuracy) if further information
on the original data distribution inside the datacube is available.
Obviously, this additional information should be easy to be exploited so that the
efficiency of the estimation is not compromised.\\
In this section we introduce a class of integrity constraints which match these properties:
they can be stored in a succinct form (thus they can be accessed efficiently), and provide
some additional information (other than the aggregate data stored in the compressed datacube)
which can be used in query answering, as will be explained in the following sections.
%
%

Let $2^{[\1..\n]}$ be the family of all subsets of indices in $[\1..\n]$.
We analyze two types of integrity constraint:
\vspace*{-5mm}
\begin{itemize}
\item
{\em number of elements that are known to be null:} we are given a function
$LB_{=0}:  2^{[\1..\n]} \rightarrow \mathcal{N}$ returning, for any
$D$ in $2^{[\1..\n]}$, a lower bound to the number of null
elements occurring in $D$; the datacube $M$
satisfies $LB_{=0}$ if, for each $D$ in $2^{[\1..\n]}$,
$\sum_{\i \in D}count(M[\i]) \leq |D|-LB_{=0}(D)$, where $|D|$
is the number of elements of $M$ in $D$;

\item
{\em number of elements that are known to be non-null:} we are given a
function $LB_{>0}:  2^{[\1..\n]} \rightarrow \mathcal{N}$ returning,
for any $D$ in $2^{[\1..\n]}$, a lower bound for the
number of non-null elements occurring in $D$; the
datacube $M$ satisfies $LB_{>0}$ if, for each $D$ in
$2^{[\1..\n]}$, $\sum_{\i \in D}count(M[\i]) \geq LB_{>0}(D)$.
\end{itemize}

The two functions $LB_{=0}$ and $LB_{>0}$ are monotonic: for each $D', D''$ in
$2^{[\1..\n]}$, if $D' \subset D''$ then both $LB_{=0}(D') \leq LB_{=0}(D'')$
and $LB_{>0}(D') \leq LB_{>0}(D'')$ hold.
From now on, consider given the above two functions together with the compressed
representation of $M$.

We point out that the integrity constraints expressed by $LB_{=0}$ and $LB_{>0}$
often occur in practice.
For instance, consider the case of a temporal dimension with granularity {\em day} and a
measure attribute storing the amount of sales for every day.
Given any temporal range, we can easily recognize a number of {\em certain} null values,
corresponding to the holidays occurring in that range.
In similar cases, the constraints provide additional information that can be efficiently
computed with no overhead in terms of storage space on the compressed representation of $M$.

As an example on how $LB_{=0}$ and $LB_{>0}$ influences the estimation of range queries,
consider the following case.
Suppose that $LB_{=0}([4..6,1..3])=3$ and $LB_{>0}([4..6,1..3])=1$ for the two-dimensional
datacube of Figure \ref{fig:new}.
From this, we can infer that the number of non-null elements in the range $[4..6,1..3]$ is
between 1 and $(6-4+1)\times (3-1+1) - 3=$ 6.
Note that the compressed representation of $M$ in Figure 1(b) only contains the information
that the block $[4..7,1..4]$ has 7 non-nulls; so, without the knowledge about the above
constraints, we could only derive that the bounds on the number of non-null elements in
$[4..6,1..3]$ are 0 and 7.

\section{The Probabilistic Framework for Range Query Estimation}
\label{sec:probfram}
\vspace{-0.25cm}
We next introduce a probabilistic framework for estimating the answers of range queries
($sum$ and $count$) by consulting aggregate data rather than the actual datacube.
To this aim, we view queries as random variables and we give their estimation in terms
of mean and variance.

A range query $Q$ on the datacube $M$ is modelled as a random variable $\Qr$ defined by
applying $Q$ on a datacube $\MF$ extracted from a datacube population {\em compatible}
with $M$, thus consisting of  datacubes whose $F$-compressed representations coincide
(at least partially) with that of $M$.

More precisely, we have different random variables modelling $Q$, depending on what
exactly we mean for `compatible', and thus on the datacube population on which the
query is applied.
In particular, we consider the following populations:

\begin{itemize}
\item
$M^{-1}_{c,F}$ is the set of all the $[\1..\n]$ matrixes $M'$ of elements in
$\mathcal{N}$ for which $M'_{count,F} = M_{count,F}$;
\item
$M^{-1}_{s,F}$ is the set of all the $[\1..\n]$ matrixes $M'$ of elements in
$\mathcal{N}$ for which $M'_{sum,F} = M_{sum,F}$;
\item
$M^{-1}_{cs,F}$ is the set of all the $[\1..\n]$ matrixes $M'$ of elements in
$\mathcal{N}$ for which $M'_{count,F} = M_{count,F}$ and $M'_{sum,F} = M_{sum,F}$;
\item
$\Pi_{LB_{=0},LB_{>0}}(M^{-1}_{cs,F})\!\!=\!\{M'|$ $M'\!\in\!M^{-1}_{cs,F}
\wedge M'\!$ \small \emph{satisfies both} $LB_{=0}$ \emph{and} $LB_{>0}$ \normalsize$\}$
is the sub-population of $M^{-1}_{cs,F}$ which also satisfy the
integrity constraints.
\end{itemize}

On the whole, given a range $[\i..\j]$, $\1 \leq \i \leq \j \leq
\n$, of size (i.e., number of elements occurring in it) $b_{\i..\j}$,
we study the following six random variables, grouped into three cases:

\begin{description}
\item [Case 1:]
For the estimation of $count(M[\i..\j])$ we consider the population of all
datacubes having the same number of non-nulls in each block as $M$, and for
that of $sum(M[\i..\j])$ the population of all datacubes whose blocks have
the same sum of the corresponding blocks in $M$.
Thus, we study the following two random variables:
\begin{itemize}
\item [-]
The random variable $C_1(b_{\i..\j})$, computing $\countr(\MF[\i..\j])$,
where $\MF$ is extracted from the population $M^{-1}_{c,F}$.
\item [-]
The random variable $S_1(b_{\i..\j})$, computing $\sumr(\MF[\i..\j])$,
where $\MF$ is extracted from the population $M^{-1}_{s,F}$.
\end{itemize}
Note that, as will be clear in the following, both the random variables
above are only function of the size $b_{\i..\j}$ of the range size, and
not of its boundaries $\i$ and $\j$.

\item [Case 2:]
We estimate the number and the sum of the non-null elements in $M[\i..\j]$
by considering the population of all the datacubes whose blocks have both
the same sum and the same number of non-nulls as the corresponding blocks
in $M$.
Then, the random variables are:
\begin{itemize}
\item [-]
The random variable $C_2(b_{\i..\j})$, computing $\countr(\MF[\i..\j])$,
where $\MF$ is extracted from the population $M^{-1}_{cs,F}$.
\item [-]
The random variable $S_2(b_{\i..\j})$, computing $\sumr(\MF[\i..\j])$,
where $\MF$ is extracted from the population $M^{-1}_{cs,F}$.
\end{itemize}
Again, $C_2$ and $S_2$ depend only on the size of the range and not
on the range itself.

\item [Case 3:]
We consider the population of all datacubes having both the same sum
and the same number of non-nulls in each block as $M$, and, besides,
satisfying the lower bound constraints on the number of null and
non-null elements occurring in each range.
Thus, we study the following two random variables:
\begin{itemize}
\item [-]
The random variable $C_3([\i..\j])$, computing $\countr(\MF[\i..\j])$,
where $\MF$ is extracted from the population
$\Pi_{LB_{=0},LB_{>0}}(M^{-1}_{cs,F})$.
\item [-]
The random variable $S_3([\i..\j])$, computing $\sumr(\MF[\i..\j])$,
where $\MF$ is extracted from the population
$\Pi_{LB_{=0},LB_{>0}}(M^{-1}_{cs,F})$.
\end{itemize}

In this case, differently from the previous ones, the examined
random variables are function of the range $[\i..\j]$ (not only of
its size) as the value returned by  $LB_{=0}$ and $LB_{>0}$ depend
on the considered range.
\end{description}

We observe that Case 2 can be derived from the more general Case 3
but, for the sake of presentation, we first present the simpler
case, and then we move to the general one. Actually the results of
Case 2 will be stated as corollaries of the corresponding ones of
Case 3 and their proofs will be postponed in the Appendix.

For each random variable above, say $cs(\MF[\i..\j])$ (where $cs$
stands for $count$ or $sum$), we have to determine its probability
distribution and then its mean and variance.
Concerning the mean $E\left(cs(\MF[\i..\j])\right)$, due to the
linearity of $E$, we have:
\noindent
 \[ E(cs(\MF[\i..\j]) =  \sum_{B_\q
\in TB_F(\i..\j)} M_{cs,F}[\q] + \sum_{B_\k \in PB_F(\i..\j)}
E(cs(\MF[\i_\k..\j_\k])) \]

\vspace*{-5mm}
where:\vspace*{-5mm}
\begin{enumerate}
\item
$TB_F(\i..\j)$ returns the set of blocks $B_\q$ that are totally
contained in the range $[\i..\j]$, i.e. every block $B_\q$ such that
both $\i \leq F^-(\q)$ and $F^+(\q) \leq \j$,
\item
$PB_F(\i..\j)$ returns the set of blocks $B_\k$ that are partially
inside the range, i.e. $B_\k \not \in TB_F(\i..\j)$ and either $\i
\leq F^-(\k) \leq \j$ or $\i \leq F^+(\k) \leq \j$, and
\item
for each $B_\k \in PB_F(\i..\j)$,  $\i_\k$ and $\j_\k$ are the
boundaries of the portion of the block $B_\k$ which overlaps the
range $[\i..\j]$, i.e., $[\i_\k..\j_\k]=$ $[\i..\j] \cap
[F^-(\k)..F^+(\k)]$.
\end{enumerate}

For instance, consider the datacube in Figure \ref{fig:new}(a) and the
range $[\i..\j]$ whose boundaries are $\i = \<4,3\>$ and $\j= \<8,6\>$.
Then the block $B_{\<2,2\>}$ is totally contained in the $[\i..\j]$, the
blocks $B_{\<2,1\>}$, $B_{\<3,1\>}$, $B_{\<3,2\>}$ are partially contained
in $[\i..\j]$, whereas the blocks $B_{\<1,1\>}$, $B_{\<1,2\>}$ are outside
$[\i..\j]$.

Concerning the variance, we assume statistical independence between
the measure values of different blocks, so that its value is
determined by summing the variances of all the partially overlapped
blocks, thus introducing no covariance:
\[\hspace*{2cm}
\sigma^2(cs(\MF[\i..\j]) = \sum_{B_\k \in PB_F(\i..\j)} \sigma^2(
cs(\MF[\i_k..\j_k])).
\]

\vspace*{-5mm}
It turns out that we only need to study the estimation of a query
inside one block, as all other cases can be easily re-composed from
this basic case:
the estimate of a query involving more than one block is the sum of the
estimates for each of the blocks involved, and the same holds for the variance.

Therefore, from now on we assume that the query range $[\i..\j]$ is
strictly inside one single block, say the block $B_\k$, i.e.
$F^-(\k) \leq \i \leq \j \leq F^+(\k)$.
We use the following notations and assumptions:
\begin{enumerate}
\item
$b$ is the size of $B_\k$, that is the total number of null and
non-null elements in $B_\k$;
\item
$b_{\i..\j}$ is the size of the query range $[\i..\j]$, that is the number of elements in the range
($1 \leq b_{\i..\j} < b$);
\item
$t= M_{count,F}[\k]$ is the number of non-null elements in $B_\k$ ($1 \leq t \leq b$);
\item
$s = M_{sum,F}[\k]$ is the sum of the elements in $B_\k$.
\end{enumerate}

\section{Case 1: using the number and the sum of non-null
elements separately}\label{sec:case1}
 \vspace{-0.25cm}

In this section we study the estimation of count and sum queries on the basis
of the sum and count information given for each block (that is, the sum $s$
of the elements occurring in each block, and the number $t$ of non null elements
in it).\\
Let us first perform the estimation of the range query $count(M[\i..\j])$.
Notice that the random variable representing the answer of a count query depends
on the size $b_{\i..\j}$ of the range $[\i..\j]$ involved in the query, rather than
on the position of the range in the block.

\begin{theorem}\label{prop-count1}
Let $C_1(b_{\i..\j}) = \countr(\MF[\i..\j])$ be the integer
random variable ranging from 0 to $t$ defined by extracting
$\MF$ from the datacube population $M^{-1}_{c,F}$.
Then:
\begin{enumerate}
\item
the probability distribution $P(C_1(b_{\i..\j}) = t_{\i..\j} )$ is:

\hspace*{-5mm}
$
P =
\left\{
    \begin{tabular}{ll}
    $\frac {
        {\scriptsize \left (
            \begin{tabular}{c} \footnotesize$b_{\i..\j}$ \\ \footnotesize$t_{\i..\j}$ \end{tabular}
        \right)}
        \cdot
        {\scriptsize \left (
            \begin{tabular}{c} \footnotesize$b-b_{\i..\j}$ \\ \footnotesize$t-t_{\i..\j}$ \end{tabular}
        \right)}
    }
    {
        {\scriptsize \left (
            \begin{tabular}{c} \footnotesize$b$\\ \footnotesize$t$ \end{tabular}
        \right)}
    }$
    & \small if $max\{0,b_{\i..\j}\!-\!(b\!-\!t)\}\!\leq\!t_{\i..\j}\!\leq\!min\{t,b_{\i..\j}\}$\\
    \small 0 & \small otherwise \\
\end{tabular}
\right.
$

\item
mean and variance are, respectively:\\

$E(C_1(b_{\i..\j}))=(b_{\i..\j}/b) \cdot  t$\\

$\sigma^2(C_1(b_{\i..\j})) = t  \cdot  (b-t)  \cdot  b_{\i..\j}  \cdot \frac{b-b_{\i..\j}}{b^2 \cdot  (b-1)}$
\end{enumerate}
\end{theorem}
{\bf Proof.}
It is easy to see that the probability that the number of non-null elements
is $t_{\i..\j}$ corresponds to the probability of extracting $t_{\i..\j}$ times
the value $1$ in $b_{\i..\j}$ trials, using a binary variable (with values
$\epsilon$, corresponding to a null value, and $1$, corresponding to non-null)
in a sample set composed of $b$ variables, with probability of finding $1$ equal
to $t/b$.
This case is known to be characterized by the above expression, that is called
\emph{hypergeometric distribution} \cite{Fel68}.
Thus, mean and variance are those of a random variable following a hypergeometric
distribution.
\eproof

\begin{figure}[h]
\centerline{ \psfig{figure=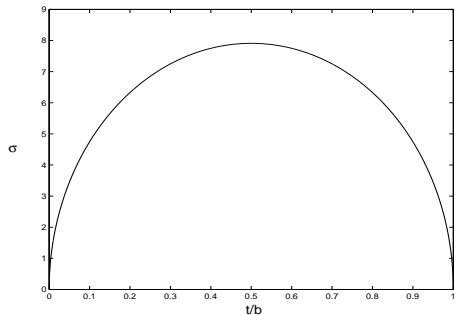,width=6cm,height=4.2cm} }
\caption{$\sigma(C_1(b_{\i..\j}))$ versus $t/b$ for a block of
size $b=1000$ and a query of size $b_{\i..\j}=b/2$}
\label{e1c}
\end{figure}

The diagram in Fig. \ref{e1c} shows how the variance of
$C_1(b_{\i..\j})$ changes when we vary $t/b$ for a query of size
$b_{\i..\j}=b/2$ and a block of size $b=1000$.\\
The estimated error is maximum for $t=b/2$ and behaves
symmetrically for $t>b/2$ and $t<b/2$. This result can be
explained by observing that, when $t=b/2$, the uncertainty in the
distribution of null elements is maximum, since the probability
that a fixed element inside the block is null is the same as it is
not null.
The variance is symmetric w.r.t. $t=b/2$ as the error which occurs when
we estimate $count([\i..\j])$ on a block of size $b$ containing $t$ non
null elements is equal to the error of the estimate of the same range
query over a block with the same size $b$, but containing $b-t$ non null
elements.

The behavior of $\sigma(C_1(b_{\i..\j}))$ w.r.t. $b_{\i..\j}/b$ is
analogous to the behavior of $\sigma(C_1(b_{\i..\j})$ w.r.t.
$t/b$: The estimated error is maximum for $b_{\i..\j}=b/2$, and is
symmetric for $b_{\i..\j}>b/2$ and $b_{\i..\j}<b/2$.
The maximum uncertainty in the estimated result is reached when the
size of the query is an half of the size of the whole block.
The estimation becomes more accurate as $b_{\i..\j}$ gets near to $b$
or to $0$: When $b_{\i..\j}=b$ the computed answer of the query is
exact and is given by $t$, whereas if $b_{\i..\j}=0$ the returned
answer is zero.

The maximum estimation error which may occur when $E(C_1(b_{\i..\j}))$
is returned as the answer of the range query $count(b_{\i..\j})$, denoted
by $err_{C_1}^{MAX}$, is quantified next.

\begin{proposition}
\label{teo:maxerrcount1} \
\begin{center}
$
err_{C_1}^{MAX} = \small
max \left\{
    \frac{b_{\i..\j}}{b}\cdot t - max\{0, t- (b-b_{\i..\j})\}, \ \
    min \{ t, b_{\i..\j}\} - \frac{b_{\i..\j}}{b}\cdot t
    \right \}
$
\end{center}
\end{proposition}
{\bf Proof.}
The maximum error can be obtained when the actual number of non null elements
inside the range of the query is either minimum (i.e.
$count(b_{\i..\j})=max\{0, t- (b-b_{\i..\j})\}$)
or maximum (i.e. $count(b_{\i..\j})=min \{ t, b_{\i..\j}\}$).
\eproof

Let us now study the random variable $sum(\MF[\i..\j])$
representing the answer of a sum query on the range $[\i..\j]$
assuming only the knowledge of the sum $s$ of the elements
occurring in the block $\k$. Once again, the estimated value
depends on the size of the range involved in the query and not on
its actual position in the block.

\begin{theorem}\label{prop-sum1}
Let $S_1(b_{\i..\j}) = \sumr(\MF[\i..\j])$ be the integer random
variable ranging from 0 to $s$ defined by extracting $\MF$ from
the datacube population $M^{-1}_{s,F}$. Then:
\begin{enumerate}
\item
the probability distribution of $S_1(b_{\i..\j})$ is:\\
$\hspace*{-3mm}
P(S_1(b_{\i..\j})\!=\!s_{\i..\j})\! =\!\!
\left\{
\begin{array}{ll}
\frac {
    {\scriptsize
        \left ( \!\!\begin{tabular}{c} \footnotesize$b_{\i..\j}\!+\!s_{\i..\j}\!-\!1$  \\ \footnotesize$s_{\i..\j}$ \end{tabular} \!\!\right)
    }
    \!\!\cdot\!
    {\scriptsize
        \left( \!\!\begin{tabular}{c} \footnotesize$b\!-\!b_{\i..\j}\!+\!s\!-\!s_{\i..\j}\!-\!1$ \\ \footnotesize$s\!-s_{\i..\j}$ \end{tabular} \!\!\right)
    }
}
{
    {\scriptsize
        \left ( \begin{tabular}{c} \footnotesize$b+s-1$  \\ \footnotesize$s$ \end{tabular} \right)
    }
}
        & \ \ \mbox{if
$0\!\leq\!s_{\i..\j}\leq\!s$}\\
0 & \ \ \mbox{otherwise} \\
\end{array}
\right.
$

\item
mean and variance are, respectively:
\[
E(S_1(b_{\i..\j}))= (b_{\i..\j}/b) \cdot  s
\]

\[
\sigma^2(S_1(b_{\i..\j}))=b_{\i..\j}  \cdot s \cdot \frac{(b-b_{\i..\j}) \cdot (b+s) }{b^2 \cdot  (b+1)} \\
\]
\end{enumerate}
\end{theorem}
\bproof (1) We can see the block $\k$ as a vector $V$ of $b$
elements which can assume values between 0 and $s$. Let
$V_{b_{\i..\j}}$ be the portion of $V$ of size $b_{\i..\j}$
containing the elements inside the range of the query, and let
$V_{b-b_{\i..\j}}$ be the remainder part of $V$. The random
variable $S_1(b_{\i..\j})$ represents the sum of the elements
belonging to $V_{b_{\i..\j}}$. The probability that
$S_1(b_{\i..\j})$ assumes the value $s_{\i..\j}$ can be obtained
by considering all possible value assignments to the elements in
$V_{b_{\i..\j}}$ so that their sum is $s_{\i..\j}$, combined with
all possible value assignments to the elements of
$V_{b-b_{\i..\j}}$ so that the total sum is $s$. The above
considered assignments represent the cases of success. The number
of possible cases can be similarly obtained by considering all
possible value assignments to the $b$ elements in $V$ so that
their sum is $s$.\\
The number of all the possible assignments from the domain of
cardinals to $y$ elements whose sum is $z$ is equal to the number
of multisets with elements taken from the set $\{ 1,...,y \}$
and having cardinality $z$:
{\tiny
$\left( \begin{tabular}{c}  \normalsize $y+z-1$ \\ \normalsize $z$ \end{tabular} \right)$}.\\
Thus, the number of possible assignments for
the elements in the portion $V_{b_{\i..\j}}$ is:
$A=$
{\tiny
$\left( \!\!\begin{tabular}{c} \normalsize $b_{\i..\j} + s_{\i..\j} - 1$ \\ \normalsize $s_{\i..\j}$ \end{tabular} \!\!\right)$},
whereas the assignments for $V_{b-b_{\i..\j}}$ are:
$B=$
{\tiny
$\left( \!\!\begin{tabular}{c} \normalsize $(b\!-\!b_{\i..\j}) + (s\!-\!s_{\i..\j}) - 1$\\ \normalsize $(s\!-\!s_{\i..\j})$ \end{tabular} \!\!\right)$}.
Analogously, there are
$C=$
{\tiny
$\left( \!\!\begin{tabular}{c} \normalsize $b+s-1$\\ \normalsize $s$ \end{tabular} \!\!\right)$}
different assignments of cardinals to the elements in the whole $V$
such that the sum is $s$.
Hence, the probability that the sum inside a range of size $b_{\i..\j}$
is $s_{\i..\j}$ is given by: $\frac {A \cdot B}{C}$

\noindent (2)  Consider the vectors $V$, $V_{b_{\i..\j}}$ and
$V_{b-b_{\i..\j}}$ defined above.
The event $(S_1([\i..\j]) = s_{\i..\j})$ is equivalent to the
following event: The sum of all the elements in $V_{b_{\i..\j}}$
is $s_{\i..\j}$.
Let $V[i]$ be a random variable corresponding to the $i$-th
element of $V$.\\
From $s = \sum_{1 \leq i \leq b} V[i]$, we derive $s = \sum_{1
\leq i \leq b} E(V[i])$ by linearity of the operator $E$. The mean
of the random variable $V[i]$ is equal to the mean of the random
variable $V[j]$, for any $i,j$, $1 \leq i,j \leq b$: For symmetry,
the probability that an element of $V$ assumes a given value is
independent on the position of this element inside the vector. Let
denote by $m$ this mean.
From the above formula for $s$ it follows that $ m \cdot b = s$,
thus $m= s/b$.
Consider now the vector $V_{b_{\i..\j}}$.
Let $S'$ be the random variable representing the sum of all the
elements of $V_{b_{\i..\j}}$.
Then $E(S')= b_{\i..\j} \cdot m$.
Hence, $E(S')= b_{\i..\j} \cdot s/b$.

The variance can be obtained using its definition. The detailed
proof is rather elaborated and, for the sake of presentation, is
included in the Appendix as Claim 1. \eproof

The maximum estimation error which may occur while returning
$E(S_1(b_{\i..\j}))$ as the answer of the range query
$sum(b_{\i..\j})$, denoted by $err_{S_1}^{MAX}$, is quantified
next.

\begin{proposition}
\label{prop:maxerrsum1} \
\begin{center}
$
err_{S_1}^{MAX} = \small
max \left\{
    \frac{b_{\i..\j}}{b}\cdot s ,
    s - \frac{b_{\i..\j}}{b}\cdot s
    \right \}
$
\end{center}
\end{proposition}
{\bf Proof.} The maximum error occur when  the elements inside the
range of the query are either all null or all non-null. \eproof

\begin{figure}[h]
\centerline{ \psfig{figure=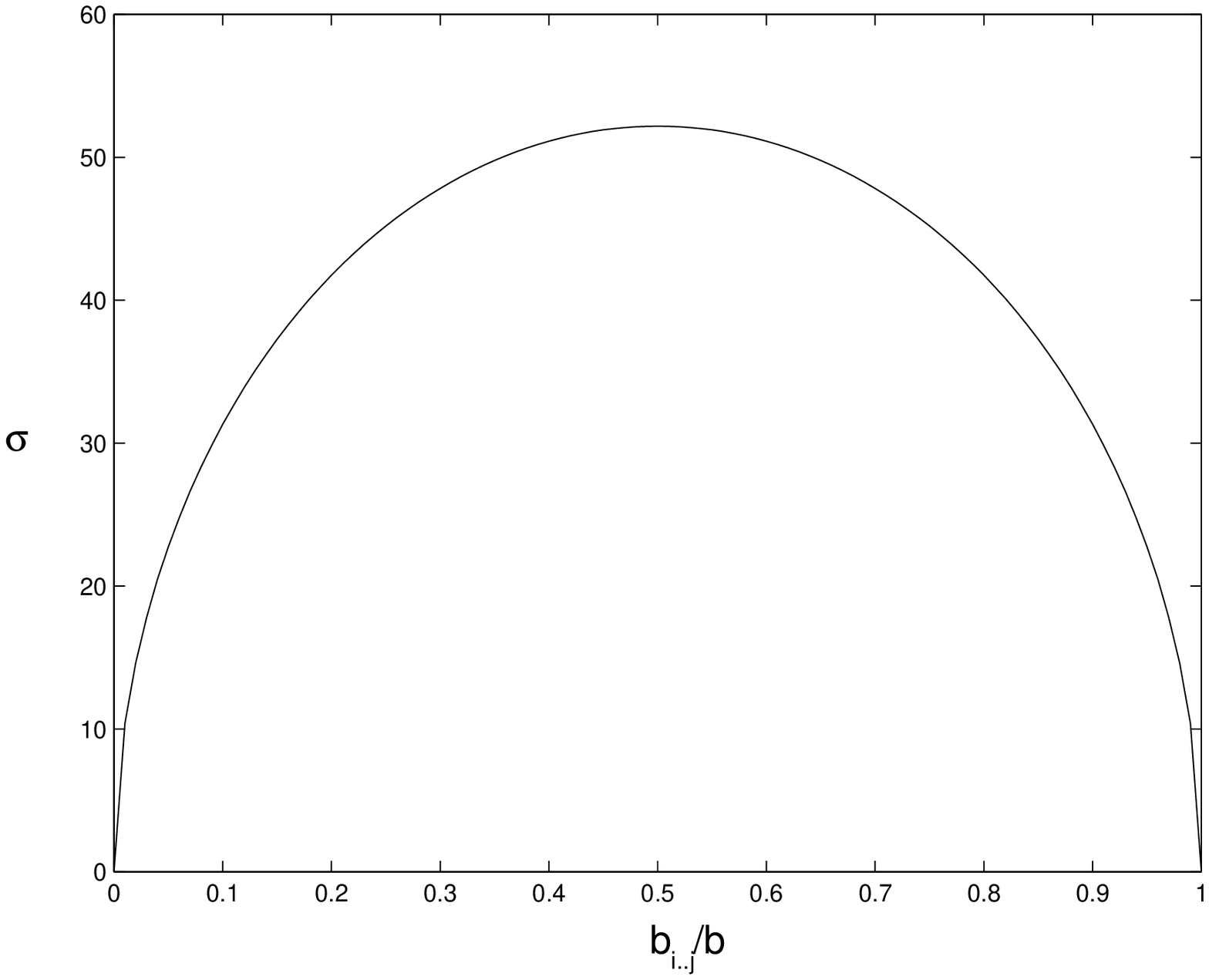,width=6cm,height=4.2cm}
             \hspace{5mm} \psfig{figure=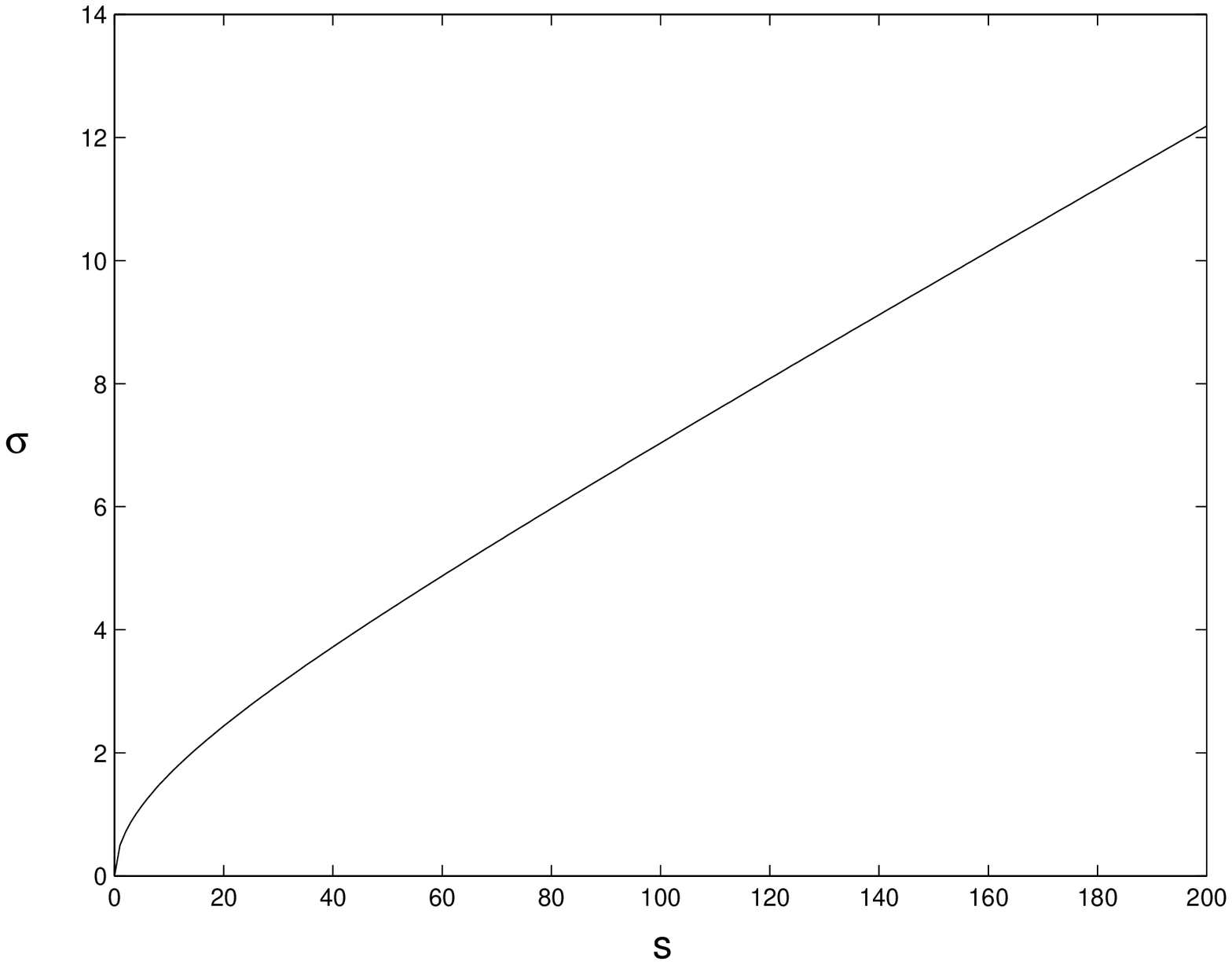,width=6cm,height=4.2cm}  }
\caption{$\sigma(S_1(b_{\i..\j}))$ versus $b_{\i..\j}/b$ and
$s$ for a block of size $b=1000$}
\label{e1s12}
\end{figure}

In the diagrams of Fig. \ref{e1s12} we show how the standard
deviation of $S_1(b_{\i..\j})$ changes, respectively, when we vary
$b_{\i..\j}/b$ (with $s=1000$), and when we vary $s$ (with
$b_{\i..\j}=b/2$) for a query over a block of size $b=100$.
The behavior of the estimated error w.r.t. $b_{\i..\j}/b$ is the same
as that of $\sigma(C_1(b_{\i..\j}))$: The standard deviation is
maximum for $b_{\i..\j}=b/2$ and is symmetric for $b_{\i..\j}>b/2$
and $b_{\i..\j}<b/2$.
As shown in the diagram on the right-hand side of Fig. \ref{e1s12},
the estimated error increases as the sum of the elements contained
in the block increases:
this result is rather expected, as the variance can be thought of as
an estimate of the absolute error.

\section{Case 2: using the number and the sum of non-null elements jointly}
\label{sec:case2} \vspace{-0.25cm}
We now perform the estimation of count and sum queries by exploiting sum
and count aggregate information simultaneously.
This issue consists in studying the conjunction of two events:
The value of the sum (in a range of size $b_{\i..\j}$) is $s_{\i..\j}$,
and the number of non nulls (in the same range) is $t_{\i..\j}$.
In this case, count and sum queries are evaluated on datacubes belonging
to $M^{-1}_{cs,F}$.
More precisely, we have to study the joint probability distribution of the
two random variables (representing the answer of the count query and the sum
query), in order to derive the two probability distributions.
As this case can be viewed as a specialization of Case 3 (where also integrity
constraints will be exploited to evaluate the estimates -- see Section \ref{sec:case3}),
results on this estimation strategy are formalized in the following corollaries,
whose proofs are reported after the proofs of the corresponding theorems of Case 3.

\begin{corollary}\label{prop-conj0}
Let $C_2(b_{\i..\j}) = \countr(\MF[\i..\j])$ and $S_2(b_{\i..\j}) = \sumr(\MF[\i..\j])$
be two integer random variables ranging, respectively, from 0 to $t$ and from $0$ to $s$,
defined by extracting $\MF$ from the datacube population $M^{-1}_{cs,F}$.
Then the joint probability distribution
$P(C_2(b_{\i..\j}) = t_{\i..\j}, S_2(b_{\i..\j}) = s_{\i..\j})$ is given by:\\

P=
$\left\{
    \begin{tabular}{ll}
        $\frac{
                {\mbox{\normalsize $Q(b_{\i..\j},t_{\i..\j},s_{\i..\j}) \cdot Q(b-b_{\i..\j},t-t_{\i..\j},s-s_{\i..\j})$}}
              }
              {
                {\mbox{\normalsize $Q(b,t,s)$}}
              }$
        &
        \small if:
        \begin{tabular}{l}
            \small $0 \leq t_{\i..\j} \leq b_{\i..\j}$,\\
            \small $t_{\i..\j} \leq s_{\i..\j} \leq s$
        \end{tabular}\\
        \small $0$
        &
        \small otherwise\\
\end{tabular}
\right.
$

\noindent
where $Q(x,y,z)$ is equal to:

$ Q(x,y,z)= \small
\left\{
   \begin{tabular}{ll}
        0 & if $\ (y\!=\!0 \wedge z\!>\!0) \ \vee \ (y\!>\!0 \wedge z\!<\!y) \ \vee \ y\!>\!x$ \\
        1 & if $y =0 \wedge z=0$ \\
        $\!\!\!{\scriptsize
            \left ( \!\begin{tabular}{c} \small$x$  \\ \small$y$ \end{tabular} \!\right)
         }
         \!\!\cdot\!\!
         {\scriptsize
            \left ( \!\begin{tabular}{c} \small$z-1$  \\ \small$z-y$ \end{tabular} \!\right)
         }$
        &
        otherwise
\end{tabular}
\right.
$

\end{corollary}


With the next corollary we formalize a first result about the
estimation of the count query using both count and sum
information: that is, the estimation of the count query cannot
exploit the aggregate information about the sum of the elements in
a block.
Therein, we derive the probability distribution of $C_2(b_{\i..\j})$,
its mean and its variance.
In particular, we obtain that the probability distribution of $C_2(b_{\i..\j})$
coincides with that of $C_1(b_{\i..\j})$, representing the answer
of the count query when only the knowledge of $t$ is given.

\begin{corollary}\label{prop-conj}
The probability distribution of the random variable $C_2(b_{\i..\j})$ defined
in Corollary \ref{prop-conj0} is:
$P(C_2(b_{\i..\j})\!=\!t_{\i..\j}) = P(C_1(b_{\i..\j})\!=\!t_{\i..\j})$,
where $C_1(b_{\i..\j})$ is the random variable defined in Theorem
\ref{prop-count1}.
\end{corollary}

From the corollary above, it follows that also mean and variance
of $C_2(b_{\i..\j})$ are the same as those of $C_1(b_{\i..\j})$,
as well as the maximum estimation error which may occur while
returning $E(C_2(b_{\i..\j}))$ as the answer of the range query
$count(M[\i..\j])$ is the same as that of  Case 1 (Proposition
\ref{teo:maxerrcount1}).

Now, we derive mean and variance of the random variable $S_2(b_{\i..\j})$,
representing the estimated answer of a sum query given the knowledge of $t$
and $s$.
Its probability distribution is given by
$P\left(S_2(b_{\i..\j})= s_{\i..\j}\right) =
\sum_{0 \leq t_{\i..\j} \leq t} P(C_2(b_{\i..\j}) = t_{\i..\j},S_2(b_{\i..\j})=s_{\i..\j} )$,
according to the definition of joint probability distribution.

\begin{corollary}\label{prop-sumj}
Mean and variance of the random variable $S_2(b_{\i..\j})$ defined in Theorem \ref{prop-conj0} are, respectively:
\[
E(S_2(b_{\i..\j})) = \frac{b_{\i..\j}}{b} \cdot s
\]

\[
\sigma^2(S_2(b_{\i..\j})) = \frac{s \cdot b_{\i..\j} \cdot
(b-b_{\i..\j})}{b^2 \cdot (b-1) \cdot (t+1)} \cdot [ b
\cdot (2 \cdot s - t + 1) - s \cdot (t + 1)].
\]
\end{corollary}

Next we derive the maximum error $err_{S_2}^{MAX}$ produced by estimating
the answer of the range query $sum(M([\i..\j]))$ by means of $E(S_2(b_{\i..\j}))$.

\begin{proposition}
\label{prop:maxerrsum2}
\ \\
\begin{center}
$
err_{S_2}^{MAX} = \small
max \left\{
    \frac{b_{\i..\j}}{b}\!\cdot\! s - max\{0, t\!-\!(b\!-\!b_{\i..\j})\}, \ \
    s-max\{0, t\!-\!b_{\i..\j}\} - \frac{b_{\i..\j}}{b}\cdot s
    \right \}
$
\end{center}
\end{proposition}
{\bf Proof.}
As non-null elements have a value equal to or greater than 1, the
minimum value of the sum inside $[\i..\j]$ is given by the minimum
number of non null elements occurring in this range, that is
$max\{0, t- (b-b_{\i..\j})\}$.
The maximum value of $sum(b_{\i..\j})$ is reached when the number
of elements outside the range of the query is minimum and all of
them have minimum value (i.e. $1$).
As the minimum of $count([\i..\j])$ is given by $max\{0, t-b_{\i..\j}\}$,
it holds that the maximum value of $sum(b_{\i..\j})$ is given by:
$s-max\{0, t-b_{\i..\j}\}$.
The formula expressing the maximum error is obtained by considering
the cases when $sum(b_{\i..\j})$ is either maximum or minimum.
\eproof

\begin{figure}[h]
\centerline{ \psfig{figure=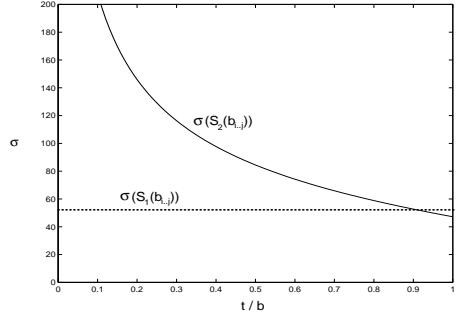,width=6cm,height=4.2cm} }
\caption{$\sigma(S_1(b_{\i..\j}))$ and $\sigma(S_2(b_{\i..\j}))$
versus $t/b$ for $b=100$ and $s=1000$} \label{e2s}
\end{figure}

The main consequence of Corollary \ref{prop-conj} is that the
knowledge of $s$ does not influence the estimation of the answer
of a count query:
The probability distribution of $C_2(b_{\i..\j})$ coincides to that
of $C_1(b_{\i..\j})$.
On the other hand, the knowledge of the number of null elements in
each block changes the estimation of the answer of a sum query:
The probability distribution of $S_2(b_{\i..\j})$ is different w.r.t.
that of $S_1(b_{\i..\j})$.
Indeed, the two random variables have the same mean but different
variances.
In Fig. \ref{e2s} we show $\sigma(S_2(b_{\i..\j}))$ (dashed line) and
$\sigma(S_1(b_{\i..\j}))$ (dotted line) versus $t/b$ for a query
of size $50$ on a block of size $100$ whose elements have sum
$1000$.

The standard deviation $\sigma(S_2(b_{\i..\j}))$ is a decreasing function of $t$:
as $t$ gets near $b$, $\sigma(S_2(b_{\i..\j}))$ decreases, and reaches a minimum
for $t=b$.
The influence of $t$ on the value of the estimated error can be strong.
For instance, if $t=b/3$ the value of $\sigma(S_1(b_{\i..\j}))$ is approximately
an half of the value of $\sigma(S_2(b_{\i..\j}))$.
The measure of the error provided by $\sigma(S_2(b_{\i..\j}))$ is generally
greater than that obtained by means of $S_1(b_{\i..\j})$, but is more
truthful.
For instance, if $t$ has a `small' value (w.r.t. $b$), we have that
$\sigma(S_2(b_{\i..\j})) \gg \sigma(S_1(b_{\i..\j}))$.
In this scenario, $\sigma(S_2(b_{\i..\j}))$ provides a better description of the
case, since when $t \ll b$ the block is very sparse, the sum is distributed among
few elements and, as there is no information about the exact position of non null
elements, there is no way to decide whether the non null elements are inside
or outside the range of the query.

Note that for $t \cong b$ , $\sigma(S_2(b_{\i..\j})) < \sigma(S_1(b_{\i..\j}))$.
Indeed, $\sigma(S_1(b_{\i..\j}))$ is an increasing function of $s$ and,
when $t=b$, evaluating $\sigma(S_2(b_{\i..\j}))$ is the same as
evaluating $\sigma(S_1(b_{\i..\j}))$ over a block of the same size
(i.e. $b$) whose elements have sum $s-b$.

\section{Case 3: using integrity constraints}
\label{sec:case3}
\vspace*{-0.5cm}

In this section we show how the knowledge of both lower bounds and
upper bounds on the number of non-null elements derived by the
functions $LB_{=0}$ and $LB_{>0}$ can be exploited in the
estimation process. We use the following additional notations:
\begin{enumerate}
\item $\tmax_{\i..\j}= b_{\i..\j} - LB_{=0}([\i..\j])$ and
$\tmin_{\i..\j}=$ $LB_{>0}([\i..\j])$ are respectively an upper
bound and  a lower bound on the number of non-null elements in the
range $[\i..\j]$; \item $\tmax_{\compij}= b_{\compij} -
LB_{=0}([\compij])$ and $\tmin_{\compij}=$ $LB_{>0}([\compij])$
are respectively an upper bound and  a lower bound on the number
of non-null elements in the block $B_k$ outside the range
$[\i..\j]$ --- $[\compij]$ denotes the set of elements that are in
$B_\k$ but not in the range $[\i..\j]$;

\item $\tmax =$ $\tmax_{\i..\j}+\tmax_{\compij}=$ $b -
LB_{=0}([\i..\j]) - LB_{=0}([\compij])$ and $\tmin =$
$\tmin_{\i..\j}+\tmin_{\compij}=$ $LB_{>0}([\i..\j]) +
LB_{>0}([\compij])$,
that is $\tmax$ and $\tmin$ are respectively an
upper bound and a lower bound on the number of non-null elements
in $B_\k$.
\end{enumerate}

We define the random variables $\countr(\MF[\i..\j])$ and
$\sumr(\MF[\i..\j])$ by extracting $\MF$ from the population
$\Pi_{LB_{=0},LB_{>0}}(M^{-1}_{cs,F})$.
We point out that, differently from the previous cases, the random variable
representing the answer of a query also depends on the position of
the range $[\i..\j]$ in the block and not only on its size.
This is because integrity constraints contain information about the
position of null elements in the block, and two distinct ranges of
the same size and belonging to the same block may have different
upper bounds and lower bounds on the number of null and non-null
elements.

\begin{theorem}\label{prop-probCase2}
Let $C_3([\i..\j]) = \countr(\MF[\i..\j])$ and
$S_3([\i..\j]) = \sumr(\MF[\i..\j])$ be two integer random variables
ranging, respectively, from $0$ to $t$ and from $0$ to $s$, and defined over
the datacube population $\Pi_{LB_{=0},LB_{>0}}(M^{-1}_{cs,F})$.
Then, for each $t_{\i..\j}$ and $s_{\i..\j}$, such that
$\tmin_{\i..\j} \leq t_{\i..\j} \leq \tmax_{\i..\j}$, and
$0 \leq s_{\i..\j} \leq s$,
the joint probability distribution
$P(C_3([\i..\j]) = t_{\i..\j}, S_3([\i..\j]) =s_{\i..\j})$ is equal to:

P=
$\left\{
    \begin{tabular}{ll}
        $\frac{
            \mbox{ \small $N(\tmax_{\i..\j},t_{\i..\j},s_{\i..\j},\tmin_{\i..\j})$}
            \ \cdot \
            \mbox{\small $N(\tmax_{\compij},t_{\compij},s_{\compij}, \tmin_{\compij})$}
         }
         {
            \mbox{ \small $N(\tmax,t,s,\tmin)$}
         }
        $
        &
        \small where:
        \begin{tabular}{l}
            \small $\tmin_{\i..\j} \leq t_{\i..\j} \leq \tmax_{\i..\j}$, \\
            \small $t_{\i..\j} \leq s_{\i..\j} \leq s$
        \end{tabular}\\
        \small $0$
        &
        \small otherwise\\
\end{tabular}
\right.
$

where $t_{\compij}= t-t_{\i..\j}$, $s_{\compij}=$ $s-s_{\i..\j}$, and
\[
\begin{array}{ll}
N(\xd,\xt,\xs, \xdd)=  & \left\{ \begin{array}{ll} 0 & \ \
\mbox{\small if \ $\xt\!>\!\xd \vee \xt\!>\!\xs  \vee (\xt\!=\!0 \wedge \xs\!>\!0)$ } \\
1 & \ \   \mbox{\small if \ $\xt=0 \, \wedge \, \xs=0$}\\
{\scriptsize
\left ( \begin{tabular}{c} \footnotesize$\xd - \xdd$ \\ \footnotesize$\xt - \xdd$ \end{tabular} \right)
}
\cdot
{\scriptsize
\left ( \begin{tabular}{c} \footnotesize$\xs-1$  \\ \footnotesize$\xs - \xt$ \end{tabular} \right)
}
& \ \ \ \ otherwise
\end{array}
\right.
\end{array}
\]
\end{theorem}

\bproof
$N(x,y,z,v)$ represents the number of configurations of a vector of
size $x$ containing $y$ non null elements with sum $z$ such that we
know the exact position of $v$ of them.
If $y=0$ and $z=0$ there is an unique configuration (all elements are
null), and so $N(x,y,z,v)=1$.
Furthermore, it is not possible that $y=0$ and $z>0$ (if the sum is
greater than $0$ there must be at least one non null element), or that
$y>0$ and $z<y$ (each non null element has at least value $1$), or that
$y>x$ (the number of non null elements cannot be greater than the size
of the vector): in such cases $N(x,y,z,v)=0$.

Otherwise, $N(x,y,z,v)$ can be obtained by disposing on $x-v$ positions
the $y-v$ non null elements of which we don't know the exact position
(that can be accomplished in
{\tiny
$\left( \!\!\begin{tabular}{c} \normalsize $x-v$ \\ \normalsize $y-v$ \end{tabular} \!\!\right)$}
different ways) and, for each of these configurations, by distributing
the sum $s$ on $y$ elements.
This can generate
{\tiny
$\left( \!\!\begin{tabular}{c} \normalsize $z - 1$ \\ \normalsize $z-y$ \end{tabular} \!\!\right)$}
different configurations.
The value of $N(x,y,z,v)$ is given by the product of these two quantities.

The probability distribution does not change if we remove
from the block $B_\k$ the elements which are certainly null,
according to the constraints expressed by $LB_{=0}$.
The block $B'_\k$ we obtain removing such elements can be seen as a vector
$V$ of size $\tmax$, and the query re-formulated over $B'_\k$ defines a
sub-vector $V_{\tmax_{\i..\j}}$ of $V$ which has size $\tmax_{\i..\j}$.

In order to evaluate the total number of ``successful" configurations for the
entire vector $V$, we have to observe that for each successful configuration
for the portion $V_{\tmax_{\i..\j}}$ we have a number of configurations for the
remainder portion of the vector, say $V_{\tmax-\tmax_{\i..\j}}$, which
is equal to the number of ways of disposing $t-t_{\i..\j}$ non null elements
on $\tmax-\tmax_{\i..\j}$ places, having that their sum is $s-s_{\i..\j}$.
Thus, the cases of success are given by
$N(\tmax_{\i..\j},t_{\i..\j},s_{\i..\j},\tmin_{\i..\j}) \cdot
N(\tmax_{\compij},t_{\compij},s_{\compij}, \tmin_{\compij})$
appearing as numerator in the expression of the statement.
The denominator $N(\tmax,t,s,\tmin)$ can be similarly obtained by
considering that the number of possible cases are all the
configurations of the vector $V$ such that the number of non null
elements is $t$, the sum is $s$, and satisfying both $LB_{=0}$ and $LB_{>0}$.\\
\eproof

Results stated in Theorem \ref{prop-probCase2} can be used to prove Corollary \ref{prop-conj0}.
We recall that Corollary \ref{prop-conj0} concerns the definition of the probability distribution
of the random variables $C_2(b_{\i..\j})$ and $S_2(b_{\i..\j})$ defined in Case 2, where no
integrity constraints were considered.

{\bf Proof of Corollary \ref{prop-conj0}}
We first observe that Case 2 corresponds to Case 3 with trivial bounds,
i.e., $LB_{=0}([\i..\j]) = LB_{>0}([\i..\j]) = 0$.
Then $t_{\i..\j}^U=b_{\i..\j}$, $t_{\i..\j}^L=0$; so the expression for
$P(C_3([\i..\j]) = t_{\i..\j}, S_3([\i..\j]) =s_{\i..\j})$ (see
Theorem \ref{prop-probCase2}) reduces to the one of
$P(C_2([\i..\j]) = t_{\i..\j}, S_2([\i..\j]) =s_{\i..\j})$.
\eproof

\begin{theorem}\label{prop-countCase2}
Let $C_3([\i..\j])$ be the random variable defined in Theorem \ref{prop-probCase2}. Then:
\begin{enumerate}
\item
the probability distribution $P(C_3([\i..\j]) = t_{\i..\j} )$ is:

\hspace*{-5mm}
$
P =
\left\{
    \begin{tabular}{ll}
    $\frac {
        {\scriptsize
            \left ( \begin{tabular}{c} \footnotesize$\tmax_{\i..\j}-\tmin_{\i..\j}$ \\
                                       \footnotesize$t_{\i..\j}-\tmin_{\i..\j}$
                    \end{tabular}
            \right)
        }
        \cdot
        {\small
            \left ( \begin{tabular}{c} \footnotesize$\tmax_{\widetilde{\i..\j}}-\tmin_{\widetilde{\i..\j}}$ \\
                                       \footnotesize$t-t_{\i..\j}-\tmin_{\widetilde{\i..\j}}$
                    \end{tabular}
            \right)
        }
    }
    {
        {\scriptsize
            \left ( \begin{tabular}{c} \footnotesize$\tmax-\tmin$ \\
                                       \footnotesize$t-\tmin$
                    \end{tabular}
            \right)
        }
    }$
    & $\small if \ \ \ \begin{array}{c} t_{\i..\j}\geq max\{\tmin_{\i..\j},t-\tmax_{\widetilde{\i..\j}}\}\\
                                 and\ \ t_{\i..\j}\!\leq\!min\{t,\tmax_{\i..\j}\}
                \end{array}$\\
    \small 0 & \small otherwise \\
\end{tabular}
\right.
$

\item
Mean and variance of the random variable $C_3([\i..\j])$ are:
\[
\begin{array}{ll}
 \hspace*{-1.5cm}E(C_3([\i..\j]))\!= \!\left \{
                    \begin{array}{ll}
                    \tmin_{\i..\j} + \frac{\tmax_{\i..\j} - \tmin_{\i..\j}}{\tmax - \tmin} \cdot ( t - \tmin) &
                    \mbox{\ \small if \ $\tmax > \tmin$}\\
                    \tmin_{\i..\j} & \mbox{\ \small if \ $\tmax = \tmin$}
                    \end{array}
                    \right .
& \ (1)\\
\ & \ \\
\hspace*{-1.5cm} \sigma^2(C_3([\i..\j]))\! =\!
                    \left \{
                    \begin{array}{ll}
                    \frac{\tmax_{\i..\j} - \tmin_{\i..\j}}{\tmax - \tmin}
                    \!\cdot\!
                    (t\! -\! \tmin)\! \cdot\!\!
                    \frac { [(\tmax \!\!- \tmin) - (\tmax_{\i..\j} -\! \tmin_{\i..\j})] \cdot (\tmax \!\!- t) }
                    {(\tmax - \tmin) \cdot (\tmax - \tmin -1)} &
                    \mbox{\ \small if \ $\tmax\!>\!\tmin\!+\!1$}\\
                    0 & \mbox{\ \small if \ $\tmin\!\leq\!\tmax\!\leq\!\tmin\!+\!1$}
                    \end{array}
                    \right .
& \ (2)
\end{array}
\]
\end{enumerate}
\end{theorem}

\bproof
(1)
The probability distribution of $C_3([\i..\j])$ can be obtained by considering that
$P \left( C_3([\i..\j] = \tmin_{\i..\j} \right ) =
\sum_{s_{\i..\j}=0}^s P(C_3([\i..\j]) = t_{\i..\j}, S_3([\i..\j]) =s_{\i..\j})$, and
applying the following equation:
\[
\mbox{\LARGE $\sum$}_{s_{\i..\j}=t_{\i..\j}}^{s}
        {\scriptsize
            \left ( \begin{tabular}{c} \footnotesize$s_{\i..\j}-1$ \\
                                       \footnotesize$s_{\i..\j}-t_{\i..\j}$
                    \end{tabular}
            \right)
        }
        \cdot
        {\scriptsize
            \left ( \begin{tabular}{c} \footnotesize$s-s_{\i..\j}-1$ \\
                                       \footnotesize$s-s_{\i..\j}-(t-t_{\i..\j})$
                    \end{tabular}
            \right)
        }
    =
       {\scriptsize
            \left ( \begin{tabular}{c} \footnotesize$s-1$ \\
                                       \footnotesize$s-t$
                    \end{tabular}
            \right)
        }
\]
\vspace*{-5mm}which holds as both its left-hand side term and right-hand side term
represent the number of sets containing $t$ cardinals (strictly greater than $0$)
with sum $s$.

(2: \emph{computation of the mean})
If $\tmax=\tmin$ it is the case that all null and non null elements
are located by integrity constraints.
Therefore, $P \left( C_3([\i..\j] = \tmin_{\i..\j} \right ) = 1$.
Otherwise, if $\tmax>\tmin$ we can reason as follows.
The block $\k$ can be viewed as a vector $V$ of $b$
elements whose values range between 0 and $s$.
Let $V_{[\i..\j]}$ be the portion of $V$ corresponding to the range
$[\i..\j]$, and let $V_{[\compij]}$ be the remainder part of $V$.
The event $(C_3([\i..\j]) = t_{\i..\j})$ is equivalent to the
following event:
The sum of all elements in $V_{[\i..\j]}$ is $s_{\i..\j}$.
Let $V[i]$ be a random variable which assumes the value $1$ if
the $i$-th element of $V$ is not null, the value $0$ otherwise.

From $t= \sum_{1 \leq l \leq b} V[i]$, we derive $t = \tmin +
\sum_{1 \leq l \leq b \wedge LB_{>0}(l)=0 \wedge LB_{=0}(l)} V[l]$
$\tmin + \sum_{1 \leq l \leq b \wedge LB_{>0}(l)=0 \wedge
LB_{=0}(l)} E(V[l])$ by linearity of the operator $E$.
The mean of the random variable $V[i]$ is equal to the mean of the
random variable $V[j]$, for any $i,j$ s.t. $1 \leq i,j \leq b$ and
$LB_{>0}(i)=LB_{>0}(j)=LB_{=0}(i)=LB_{=0}(j)=0$:
For symmetry, the positions which are not localized by the integrity
constraints have the same probability of containing null or non null elements.
Let $m$ be the mean $E(V[i])$.
From the above formula for $t$, it follows that $m \cdot (\tmax-\tmin) = t-\tmin$.

Consider now the vector $V_{[\i..\j]}$.
Since $C_3([\i..\j])$ can be seen as the random variable representing
the number of non null elements of $V_{[\i..\j]}$, we have that:
$E(C_3([\i..\j]))= \tmin_{\i..\j} + ( \tmax_{\i..\j} - \tmin_{\i..\j} ) \cdot m$.
Hence, $E(C_3([\i..\j]))= \tmin_{\i..\j}+ \frac {\tmax_{\i..\j} - \tmin_{\i..\j}}{\tmax-\tmin}
\cdot (t-\tmin)$.

\noindent (2: \emph{computation of the variance})
If $\tmax=\tmin$, as explained for part (1) of this proof,
it holds that $P \left( C_3([\i..\j] = \tmin_{\i..\j} \right ) = 1$, and
therefore $\sigma^2 \left( C_3([\i..\j] \right )$.
If $\tmax=\tmin+1$, two cases can occur: 1) $t=\tmax$, or 2)$t=\tmin$.
In the former case, $P \left( C_3([\i..\j] = \tmax_{\i..\j} \right ) = 1$
holds, whereas in the latter one
$P \left( C_3([\i..\j] = \tmin_{\i..\j} \right ) = 1$.
In both cases, we have that:
$\sigma^2 \left( C_3([\i..\j] \right ) = 0$.

The formula expressing $\sigma^2$ for $\tmax>\tmin+1$  can be
obtained using the definition of variance. The detailed proof is
rather elaborated and, for the sake of presentation, is reported
in Appendix as Claim 2. \eproof

Results stated in Theorem \ref{prop-probCase2} can be used to prove Corollary 2, as the random
variable $C_2(b_{\i..\j})$ can be seen as a special case of $C_3([\i..\j])$.

{\bf Proof of Corollary 2}
As shown in the proof of Corollary 1, $t_{\i..\j}^U=b_{\i..\j}$,
$t_{\i..\j}^L=0$.
For the same reasons, $t^U=b$ and $t^L=0$.
By performing these substitutions, the statement of Theorem \ref{prop-countCase2} reduces
to that of Corollary 2.
 \eproof

Let us now quantify the maximum estimation error $err_{C_3}^{MAX}$
which may occur while returning $E(C_3([\i..\j]))$ as the answer
of the range query $count([\i..\j])$.
\begin{proposition}
\label{prop:maxerrcount3} \

\begin{center}
$
err_{C_3}^{MAX} = \small
max \left\{
    E(C_3([\i..\j]) - max\{ \tmin_{\i..\j}, t- (\tmax-\tmax_{\i..\j}) \}, \right.$
$   \left. \hspace*{2.9cm}   min \{ \tmax_{\i..\j}, t-(\tmin-\tmin_{\i..\j}) \} - E(C_3([\i..\j])
    \right \}
$
\end{center}
\end{proposition}
{\bf Proof.}
The minimum number of non null elements which could be contained in
the range of the query is given by:
$max\{ \tmin_{\i..\j} , t-\tmax_{\compij} \}=$
$max\{\tmin_{\i..\j}, t- (\tmax-\tmax_{\i..\j}) \}$,
whereas the maximum of $count([\i..\j])$ is:
$min \{ \tmax_{\i..\j}, t-\tmin_{\compij} \}=$
$min \{ \tmax_{\i..\j}, t-(\tmin-\tmin_{\i..\j}) \}.$
The formula of the maximum error is obtained by considering the cases
where the actual number of non null elements inside the range of the query is
either minimum or maximum.
\eproof

We now focus our attention on the random variable $S_3([\i..\j])$, whose
mean and variance are computed in the following theorem.
Results stated in this theorem will be used, in the following, to prove Corollary 3.

\begin{theorem}\label{prop-sumCase2}
Mean and variance of the random variable $S_3([\i..\j])$ defined in
Theorem \ref{prop-probCase2} are:

$E(S_3([\i..\j])) = \left \{
                    \begin{array}{ll}
                    \tmin_{\i..\j} \cdot \frac{s}{t} +
                    (\tmax_{\i..\j}- \tmin_{\i..\j}) \cdot
                    \frac{s}{t} \cdot \frac{t-\tmin}{\tmax-\tmin}
                    &
                    \mbox{\ \ \ \small if \ $\tmax > \tmin$}\\
                    \tmin_{\i..\j} \cdot \frac{s}{t} & \mbox{\ \ \ \small if \ $\tmax = \tmin$}
                    \end{array}
                    \right .$

$\sigma^2(S_3([\i..\j]))\!=\!
                    \left \{
                    \begin{array}{ll}
                        \begin{array}{l}
                        \alpha\!\cdot\!(\tmax_{\i..\j}\!-\!\tmin_{\i..\j})\!\cdot\!
                        \frac {t-\tmin}{\tmax- \tmin}\!\cdot\!
                        \left[ 1\!+\!(\tmax_{\i..\j}\!-\!\tmin_{\i..\j}\!-\!1)\!\cdot\!
                        \frac{t- \tmin -1}{\tmax - \tmin -1} \right ]\!+ \\
                        (\!\beta\!+\!2\!\cdot\!\alpha\!\cdot\!\tmin_{\i..\j})\!\cdot\!
                        (\!\tmax_{\i..\j}\!-\!\tmin_{\i..\j})\!\cdot\!
                        \frac{t- \tmin}{\tmax - \tmin} +\!
                        (\alpha\!\cdot\!{\tmin_{\i..\j}}^2\!+\!\beta\!\cdot\!\tmin_{\i..\j})\!-\!\gamma^2
                        \end{array}
                        &
                        \mbox{\ \ \small if \ $\tmax\!\!>\!\!\tmin\!+\!1$}\\
                        & \\
                        \alpha \cdot (\tmax_{\i..\j}-\tmin_{\i..\j}) \cdot \frac {t- \tmin}{\tmax- \tmin}
                        &
                        \mbox{\ \ \small if \ $\tmax = \tmin$}\\
                    \end{array}
                    \right .$

where:

\[
\alpha=\frac{s \cdot (s+1)}{t \cdot (t+1)}, \beta=\frac{s \cdot
(s-t)}{t \cdot (t+1)} \mbox{, and } \ \gamma=E \left( S_3([\i..\j]) \right).
\]
\end{theorem}
{\bf Proof.}
Let us first prove the formula expressing $E \left( S_3([\i..\j]) \right)$.
We assume that $\tmax>\tmin+1$, as the proof for the case $\tmax=\tmin+1$
is trivial.

\[
\hspace*{-.9cm}
E\left(S_3([\i..\j])\right)=\sum_{s_{\i..\j}=0}^{s}
\sum_{t_{\i..\j}=0}^{t} s_{\i..\j} \cdot
P\left(C_3([\i..\j])=t_{\i..\j},S_3([\i..\j])=s_{\i..\j}\right)=
\]

\[
\hspace*{-1cm}
= \hspace*{-.4cm} \sum_{t_{\i..\j}=\tmin_{\i..\j}}^{t-(\tmin\!-\tmin_{\i..\j})}
\sum_{s_{\i..\j}=t_{\i..\j}}^{s-t+t_{\i..\j}}
\!\!\!s_{\i..\j} \!\cdot\!\!
\frac {
    {\scriptsize \left ( \!\!\begin{tabular}{c} \footnotesize$\tmax_{\i..\j}\! - \!\tmin_{\i..\j}$\\
                                                \footnotesize$t_{\i..\j}\! - \!\tmin_{\i..\j}$
                             \end{tabular} \!\!\right )
    }
    \!\!\cdot\!\!
    {\scriptsize
        \left ( \!\!\begin{tabular}{c} \footnotesize $s_{\i..\j}\!-\!1$ \\ \footnotesize $s_{\i..\j}\!-\!t_{\i..\j}$ \end{tabular} \!\! \right )
    }
    \!\!\cdot\!\!
    {\scriptsize
        \left ( \!\!\begin{tabular}{c}  \footnotesize$\tmax\!\!- \!\tmax_{\i..\j} - (\tmin\!\!-\!\tmin_{\i..\j})$ \\
                                    \footnotesize$t\!-\!t_{\i..\j} - (\tmin\!\!-\!\tmin_{\i..\j}$)
                \end{tabular} \!\!\right )
    }
    \!\!\cdot\!\!
    {\scriptsize \left ( \!\!\begin{tabular}{c} \footnotesize$s - s_{\i..\j} - 1$ \\
                                            \footnotesize$s\!\!-\!s_{\i..\j}\! - \!t \!+ \!t_{\i..\j}$ \end{tabular} \!\!\right )
    }
}
{
    {\scriptsize
        \left (\!\! \begin{tabular}{c} \footnotesize$\tmax\!-\!\tmin$ \\ \footnotesize$t\!-\!\tmin$ \end{tabular} \!\!\right )
    }
    \!\cdot\!
    {\scriptsize
        \left (\!\! \begin{tabular}{c} \footnotesize$s-1$ \\ \footnotesize$s-t$ \end{tabular} \!\!\right )
    }
}
\!\!=
 \]

\[
\hspace*{-1cm}
= \hspace*{-.4cm}
\sum_{t_{\i..\j}=\tmin_{\i..\j}}^{t-(\tmin\!-\tmin_{\i..\j})}
\!\left [
\frac {
    {\scriptsize
        \left ( \!\!\begin{tabular}{c}  \footnotesize$\tmax_{\i..\j}\!-\!\tmin_{\i..\j}$ \\
                                        \footnotesize$t_{\i..\j}\!-\!\tmin_{\i..\j}$ \end{tabular} \!\!\right )
    }
    \!\!\cdot\!\!
    {\scriptsize
        \left ( \!\!\begin{tabular}{c}  \footnotesize$\tmax\!\!-\!\tmax_{\i..\j}\!-\!\tmin\!\!+\!\tmin_{\i..\j}$ \\
                                    \footnotesize$t\!\!-\!t_{\i..\j}\!-\!\tmin\!+\!\tmin_{\i..\j}$ \end{tabular} \!\!\right )
    }
}
{
    {\scriptsize
        \left ( \!\!\begin{tabular}{c} \footnotesize$\tmax\!\!-\!\tmin$ \\ \footnotesize$t\!-\!\tmin$ \end{tabular} \!\!\right )
    }
    \!\!\cdot\!\!
    {\scriptsize
        \left ( \!\!\begin{tabular}{c} \footnotesize$s\!-\!1$ \\ \footnotesize$s\!-\!t$ \end{tabular} \!\!\right )
    }
}
\!\cdot\!\!\!
\sum_{s_{\i..\j}=t_{\i..\j}}^{s-t+t_{\i..\j}} s_{\i..\j} \!\cdot\!\!
    {\scriptsize
        \left ( \begin{tabular}{c} \footnotesize$s_{\i..\j}\!-\!1$ \\ \footnotesize$s_{\i..\j}\!-\!t_{\i..\j}$ \end{tabular} \right )
    }
    \!\!\cdot\!\!
    {\scriptsize
        \left ( \begin{tabular}{c} \footnotesize$s\!\!-\!\!s_{\i..\j}\!-\!\!1$ \\ \footnotesize$s\!\!-\!\!s_{\i..\j}\!-\!\!t\!+\!t_{\i..\j}$ \end{tabular} \right )
    }
\right ]
\]

\noindent The term:

\[
\hspace*{-.9cm}
\sum_{s_{\i..\j}=t_{\i..\j}}^{s-t+t_{\i..\j}} s_{\i..\j} \cdot\!
{\tiny
\left ( \begin{tabular}{c} \footnotesize$s_{\i..\j}-1$ \\ \footnotesize$s_{\i..\j}-t_{\i..\j}$ \end{tabular} \right )
}
\!\cdot\!
{\tiny
\left ( \begin{tabular}{c} \footnotesize$s - s_{\i..\j} - 1$ \\ \footnotesize$s-s_{\i..\j} - t + t_{\i..\j}$ \end{tabular} \right )
}
\]

can be re-written, by replacing $s_{\i..\j}$ with $S_{\i..\j}$ +
$t_{\i..\j}$, as:

\[
\hspace*{-.9cm}
\sum_{s_{\i..\j}=t_{\i..\j}}^{s-t+t_{\i..\j}} s_{\i..\j} \cdot\!
{\scriptsize
\left ( \begin{tabular}{c} \footnotesize$s_{\i..\j}-1$ \\ \footnotesize$s_{\i..\j}-t_{\i..\j}$ \end{tabular} \right )
}
\!\cdot\!
{\scriptsize
\left ( \begin{tabular}{c} \footnotesize$s - s_{\i..\j} - 1$ \\ \footnotesize$s-s_{\i..\j} - t + t_{\i..\j}$ \end{tabular} \right )
}
= \]

\[
\hspace*{-.9cm}
= \sum_{S_{\i..\j}=0}^{s-t} (S_{\i..\j}+t_{\i..\j}) \cdot
{\scriptsize
\left ( \begin{tabular}{c} \footnotesize$t_{\i..\j} + S_{\i..\j}-1$ \\ \footnotesize$S_{\i..\j}$ \end{tabular} \right )
}
\!\cdot\!
{\scriptsize
\left ( \begin{tabular}{c} \footnotesize$- t_{\i..\j} + s - S_{\i..\j} - 1$ \\ \footnotesize$s-S_{\i..\j} - t$ \end{tabular} \right )
}
=
\]

\begin{equation}\label{filuno}
\hspace*{-.9cm}
= \sum_{S_{\i..\j}=0}^{S} (S_{\i..\j}+t_{\i..\j}) \cdot
{\scriptsize
\left (
\begin{tabular}{c} \footnotesize$t_{\i..\j} + S_{\i..\j}-1$ \\ \footnotesize$S_{\i..\j}$ \end{tabular} \right )
}
\cdot
{\scriptsize
\left ( \begin{tabular}{c} \footnotesize$t - t_{\i..\j} + S - S_{\i..\j} - 1$ \\ \footnotesize$S-S_{\i..\j}$ \end{tabular} \right )
}
\end{equation}
where: $S = s-t $.

Since
${\scriptsize
\left ( \begin{tabular}{c} \footnotesize$x$ \\ \footnotesize$y$ \end{tabular} \right )
}
= \frac {\mbox{\footnotesize $x-y+1$}}{\mbox{\footnotesize $y$}} \cdot
{\scriptsize
\left ( \begin{tabular}{c} \footnotesize$x$ \\ \footnotesize$y-1$ \end{tabular} \right )
}
$
, it results that:
\[
\hspace*{-1cm}
\sum_{S_{\i..\j}=0}^{S} S_{\i..\j} \cdot
{\scriptsize
\left (
\begin{tabular}{c} \footnotesize$t_{\i..\j} + S_{\i..\j}-1$ \\ \footnotesize$S_{\i..\j}$ \end{tabular} \right )
}
\cdot
{\scriptsize
\left ( \begin{tabular}{c} \footnotesize$t - t_{\i..\j} + S - S_{\i..\j} - 1$ \\ \footnotesize$S-S_{\i..\j}$ \end{tabular} \right )
}
=
\]

\[
\hspace*{-1cm}
= \sum_{S_{\i..\j}=1}^{S} t_{\i..\j}  \cdot
{\scriptsize
\left (
\begin{tabular}{c} \footnotesize$t_{\i..\j} + S_{\i..\j}-1$ \\ \footnotesize$S_{\i..\j}-1$ \end{tabular} \right )
}
\cdot
{\scriptsize
\left ( \begin{tabular}{c} \footnotesize$t - t_{\i..\j} + S - S_{\i..\j} - 1$ \\ \footnotesize$S-S_{\i..\j}$ \end{tabular} \right )
}
= \]

\[
\hspace*{-1cm}
= \sum_{Q_{\i..\j}=0}^{S-1} t_{\i..\j} \cdot
{\scriptsize
\left ( \begin{tabular}{c} \footnotesize$(t_{\i..\j}\!+\!1)\!+\!Q_{\i..\j}\!-\!1$ \\ \footnotesize$Q_{\i..\j}$ \end{tabular} \right )
}
\cdot
{\scriptsize
\left ( \begin{tabular}{c} \footnotesize$(t\!+\!1)\!-\!(t_{\i..\j}\!+\!1)\!+\!(S\!-\!1)\!-\!Q_{\i..\j}\!-\!1$ \\ \footnotesize$(S-1) - Q_{\i..\j}$ \end{tabular} \right )
}
\]
\noindent
where: $Q_{\i..\j} = S_{\i..\j} -1$.

\noindent Now observe that the following holds:
\begin{equation}
\label{s1tre}
\sum_{k=0}^{z}
{\scriptsize
\left ( \begin{tabular}{c} \footnotesize$y+k-1$ \\ \footnotesize$k$ \end{tabular} \right )
}
\cdot
{\scriptsize
\left ( \begin{tabular}{c} \footnotesize$x-y+z-k-1$ \\ \footnotesize$z-k$ \end{tabular} \right )
}
=
{\scriptsize
\left ( \begin{tabular}{c} \footnotesize$x+z-1$ \\ \footnotesize$z$ \end{tabular} \right )
}
\end{equation}
\noindent since both the above terms represent the number of sets
containing $x$ naturals (including zero) such that their sum is
$z$.

\noindent
Then, by applying formula (\ref{s1tre}) we obtain:
\[
\hspace*{-1cm}
\sum_{S_{\i..\j}=0}^{S} S_{\i..\j} \cdot
{\scriptsize
\left (
\begin{tabular}{c} \footnotesize$t_{\i..\j} + S_{\i..\j}-1$ \\ \footnotesize$S_{\i..\j}$ \end{tabular} \right )
}
\cdot
{\scriptsize
\left ( \begin{tabular}{c} \footnotesize$t - t_{\i..\j} + S - S_{\i..\j} - 1$ \\ \footnotesize$S-S_{\i..\j}$ \end{tabular} \right )
}
=
\]

\[
\hspace*{-1cm}
= \sum_{Q_{\i..\j}=0}^{S-1} t_{\i..\j} \cdot\!
{\scriptsize
\left ( \begin{tabular}{c} \footnotesize$(t_{\i..\j}\!+\!1)\!+\!Q_{\i..\j}\!-\!1$ \\ \footnotesize$Q_{\i..\j}$ \end{tabular} \right )
}
\!\!\cdot\!\!
{\scriptsize
\left ( \begin{tabular}{c} \footnotesize$(t\!+\!1)\!-\!(t_{\i..\j}\!+\!1)\!+\!(S\!-\!1)\!-\!Q_{\i..\j}\!-\!1$\\ \footnotesize$(S-1) - Q_{\i..\j}$ \end{tabular} \right ) =
}
\]
\[
\hspace*{-1cm}
= t_{\i..\j} \cdot
{\scriptsize\left ( \begin{tabular}{c} \footnotesize$t+S-1$ \\ \footnotesize$S-1$ \end{tabular} \right )
}
=
t_{\i..\j} \cdot \frac{S}{t} \cdot
{\scriptsize\left ( \begin{tabular}{c} \footnotesize$t+S-1$ \\ \footnotesize$S$ \end{tabular} \right )
}
=
t_{\i..\j} \cdot \frac{s-t}{t} \cdot
{\scriptsize
\left ( \begin{tabular}{c} \footnotesize$s-1$ \\ \footnotesize$s-t$ \end{tabular} \right )
}
\]

\noindent and:

\[
\hspace*{-1cm}
\sum_{S_{\i..\j}=0}^{S} t_{\i..\j} \cdot
{\scriptsize
\left ( \begin{tabular}{c} \footnotesize$t_{\i..\j} + S_{\i..\j}-1$ \\ \footnotesize$S_{\i..\j}$ \end{tabular} \right )
}
\cdot
{\scriptsize
\left ( \begin{tabular}{c} \footnotesize$t - t_{\i..\j} + S - S_{\i..\j} - 1$ \\ \footnotesize$S-S_{\i..\j}$ \end{tabular} \right )
}
=
\]
\[
\hspace*{-.9cm}
= t_{\i..\j} \cdot
{\scriptsize
\left ( \begin{tabular}{c} \footnotesize$t+S-1$ \\ \footnotesize$S$ \end{tabular} \right )
}
=
t_{\i..\j} \cdot
{\scriptsize
\left ( \begin{tabular}{c} \footnotesize$s-1$ \\ \footnotesize$s-t$ \end{tabular} \right )
}
\]

\noindent By replacing these two terms in (\ref{filuno}), we
obtain:

\[
\hspace*{-1cm}
\sum_{s_{\i..\j}=t_{\i..\j}}^{s-t+t_{\i..\j}} s_{\i..\j} \cdot
{\scriptsize
\left ( \begin{tabular}{c} \footnotesize$s_{\i..\j}-1$ \\ \footnotesize$s_{\i..\j}-t_{\i..\j}$ \end{tabular} \right )
}
\cdot
{\scriptsize
\left ( \begin{tabular}{c} \footnotesize$s - s_{\i..\j} - 1$ \\ \footnotesize$s-s_{\i..\j} - t + t_{\i..\j}$ \end{tabular} \right )
}=
t_{\i..\j} \cdot \frac{s}{t} \cdot
{\scriptsize
\left ( \begin{tabular}{c} \footnotesize$s-1$ \\ \footnotesize$s-t$ \end{tabular} \right )
}
\]

\noindent so that:
\begin{equation}
\label{fildue}
\hspace*{-.9cm}
E\left(S_3([\i..\j])\right)=\!\!\!
\sum_{t_{\i..\j}=\tmin_{\i..\j}}^{t-(\tmin-\tmin_{\i..\j})} \left
[ \frac { \left ( \begin{array}{c} \tmax_{\i..\j} -\tmin_{\i..\j}
\\ t_{\i..\j} - \tmin_{\i..\j} \end{array} \right )
\cdot \left ( \begin{array}{c} \tmax - \tmax_{\i..\j} - \tmin + \tmin_{\i..\j} \\
t - t_{\i..\j} - \tmin + \tmin_{\i..\j}
\end{array} \right ) } { \left ( \begin{array}{c} \tmax -\tmin \\ t -\tmin
\end{array} \right )  }
 \cdot
t_{\i..\j} \cdot \frac{s}{t} \right ]
\end{equation}

\noindent Moreover, it holds that:

\[
\hspace*{-.9cm}
\sum_{t_{\i..\j}=\tmin_{\i..\j}}^{t-(\tmin-\tmin_{\i..\j})}
\!\!t_{\i..\j} \!\cdot\!\!
{\tiny
\left ( \begin{tabular}{c} \footnotesize$\tmax_{\i..\j} - \tmin_{\i..\j}$ \\ \footnotesize$t_{\i..\j} - \tmin_{\i..\j}$ \end{tabular} \right )
}
\!\cdot\!
{\tiny \left ( \begin{tabular}{c} \footnotesize$\tmax\!-\!\tmin\!-\!\tmax_{\i..\j}\!+\!\tmin_{\i..\j}$\\ \footnotesize$t\!-\!\tmin\!-\!t_{\i..\j}\!+\!\tmin_{\i..\j}$\end{tabular} \right )
}
=
\]
\[
\hspace*{-.9cm}
=\sum_{h_{\i..\j}=0}^{m} (h_{\i..\j}+\tmin_{\i..\j}) \!\cdot\!\!
{\tiny
\left (
\begin{tabular}{c} \footnotesize$l_{\i..\j}$ \\ \footnotesize$h_{\i..\j}$ \end{tabular} \right )
}
\!\!\cdot\!\!
{\tiny
\left ( \begin{tabular}{c} \footnotesize$n\!-\!l_{\i..\j}$\\
                           \footnotesize$m\!-\!h_{\i..\j}$
        \end{tabular}
\right )
}
\]

\noindent where: $h_{\i..\j}= t_{\i..\j}-\tmin_{\i..\j}$,
\hspace{5mm} $l_{\i..\j}= \tmax_{\i..\j}-\tmin_{\i..\j}$,
\hspace{5mm}
$m= t- \tmin$, \hspace{5mm} and $n=\tmax-\tmin$.\\

\noindent As
$
{\scriptsize
\left ( \begin{tabular}{c} \small$x$ \\ \small$y$ \end{tabular} \right
)
}
=
\frac {\mbox{\normalsize$x$}}{\mbox{\normalsize$y$}} \cdot
{\scriptsize \left ( \begin{tabular}{c}\small$x-1$ \\ \small$y-1$ \end{tabular} \right )
}$
we have that:

\[
\hspace*{-0.9cm}
\sum_{h_{\i..\j}=0}^{m} h_{\i..\j} \cdot
{\scriptsize
    \left ( \begin{tabular}{c} \footnotesize$l_{\i..\j}$ \\ \footnotesize$h_{\i..\j}$ \end{tabular} \right )
}
\!\cdot\!
{\scriptsize
    \left ( \begin{tabular}{c} \footnotesize$n - l_{\i..\j}$ \\ \footnotesize$m - h_{\i..\j}$ \end{tabular} \right )
}
=
\sum_{h_{\i..\j}=0}^{m} l_{\i..\j} \cdot
{\scriptsize
    \left ( \begin{tabular}{c} \footnotesize$l_{\i..\j}-1$ \\ \footnotesize$h_{\i..\j}-1$ \end{tabular} \right )
}
\!\cdot\!
{\scriptsize
    \left ( \begin{tabular}{c} \footnotesize$n - l_{\i..\j}$\\ \footnotesize$m - h_{\i..\j}$ \end{tabular} \right )
}
=
\]
\[
\hspace*{-0.9cm}
= \sum_{p_{\i..\j}=0}^{m-1} l_{\i..\j} \cdot
{\scriptsize
\left ( \begin{tabular}{c} \footnotesize$l_{\i..\j} - 1$\\ \footnotesize$p_{\i..\j}$ \end{tabular} \right )
}
\cdot
{\scriptsize
\left ( \begin{tabular}{c} \footnotesize$n - l_{\i..\j}$ \\ \footnotesize$m - 1 - p_{\i..\j}$ \end{tabular} \right )
} \ \ \ \mbox{ where: $ p_{\i..\j} = h_{\i..\j} - 1 $}
\]

\noindent By applying the {\em Vandermonde formula}:
\begin{equation}
\label{vandermonde} \sum_{i=0}^{k}
{\scriptsize
\left ( \begin{tabular}{c} \footnotesize$x$ \\ \footnotesize$i$ \end{tabular} \right )
}
\cdot
{\scriptsize
\left ( \begin{tabular}{c} \footnotesize$y$ \\ \footnotesize$k-i$ \end{tabular} \right )
}
=
{\scriptsize
\left ( \begin{tabular}{c} \footnotesize$x+y$ \\ \footnotesize$k$ \end{tabular} \right )
}
\end{equation}
\noindent we obtain:

\[
\hspace*{-0.9cm}
\sum_{p_{\i..\j}=0}^{m-1} l_{\i..\j} \cdot
{\scriptsize
\left ( \begin{tabular}{c} \footnotesize$l_{\i..\j} - 1$ \\ \footnotesize$p_{\i..\j}$ \end{tabular} \right )
}
\cdot
{\scriptsize
\left ( \begin{tabular}{c} \footnotesize$n - l_{\i..\j}$ \\ \footnotesize$m - 1 - p_{\i..\j}$ \end{tabular} \right )
}=
l_{\i..\j} \cdot
{\scriptsize
\left ( \begin{tabular}{c} \footnotesize$n - 1$ \\ \footnotesize$m - 1$ \end{tabular} \right )
}=
l_{\i..\j} \cdot \frac{m}{n} \cdot
{\scriptsize
\left ( \begin{tabular}{c} \footnotesize$n$  \\ \footnotesize$m$  \end{tabular} \right )
}
\]

\noindent and:

\[
\hspace*{-0.9cm}
\sum_{h_{\i..\j}=0}^{m} \tmin_{\i..\j} \cdot
{\scriptsize\left ( \begin{tabular}{c} \footnotesize$l_{\i..\j}$ \\ \footnotesize$h_{\i..\j}$ \end{tabular} \right )
}
\cdot
{\scriptsize\left ( \begin{tabular}{c} \footnotesize$n - l_{\i..\j}$\\ \footnotesize$m - h_{\i..\j}$ \end{tabular} \right )
}
=
\tmin_{\i..\j} \cdot
{\scriptsize \left ( \begin{tabular}{c} \footnotesize$n$ \\ \footnotesize$m$  \end{tabular} \right )
}
\]

\noindent After replacing these terms in (\ref{fildue}), we obtain:

\[\hspace*{-1cm}
E\left(S_3([\i..\j])\right)=
2\!\cdot\!
\frac {
\left( l_{\i..\j} \!\cdot \!\frac{\mbox{\small$m$}}{\mbox{\small$n$}}+\tmin_{\i..\j} \right)
\!\cdot\!\!
{\scriptsize
    \left ( \!\!\begin{tabular}{c} \footnotesize $n$ \\ \footnotesize $m$  \end{tabular}\!\! \right )
}
\cdot \frac{\mbox{\small$s$}}{\mbox{\small$t$}}
}
{
{\scriptsize \left ( \!\!\begin{tabular}{c} \footnotesize $\tmax_{\i..\j} -\tmin_{\i..\j}$ \\ \footnotesize $t_{\i..\j} - \tmin_{\i..\j}$ \end{tabular} \!\!\right )}
}
=
\left( (\tmax_{\i..\j}-\tmin_{\i..\j}) \cdot
\frac{t- \tmin}{\tmax-\tmin}+\tmin_{\i..\j}\right) \cdot
\frac{s}{t}
\]

As regards the proof of the formula expressing the variance, for the sake of presentation,
this proof is postponed in Appendix as Claim 3.
\eproof

{\bf Proof of Corollary 3}.
By applying the same arguments used in the proof of Corolaries 1 and 2, it is easy
to see that the statement of Theorem \ref{prop-sumCase2} reduces to the statement of
Corollary 3.
\eproof

The maximum estimation error $err_{S_3}^{MAX}$ which may occur
while returning $E(S_3([\i..\j]))$ as the answer of the range
query $sum([\i..\j])$ is evaluated next:

\begin{proposition}
\label{prop:maxerrsum3} \
\begin{center}
$
err_{S_3}^{MAX} = \small
max \left\{
    E(S_3([\i..\j]) - max\{\tmin_{\i..\j}, t- (\tmax-\tmax_{\i..\j}) \},\right.$\\
$   \left.
    \hspace*{3.6cm}s-min\{\tmin-\tmin_{\i..\j}, t-\tmax_{\i..\j} \} - E(S_3([\i..\j])
    \right \}
$
\end{center}
\end{proposition}
{\bf Proof.}
The maximum error can be obtained when the actual sum inside the range
of the query is either minimum or maximum.
This sum is minimum if the number of non null elements inside $[\i..\j]$
is minimum, and each of these non null elements has the minimum value
(i.e. $1$).
Thus, the minimum sum inside $[\i..\j]$ coincides with the minimum value of
$count([\i..\j])$ and is given by $max\{\tmin_{\i..\j}, t- LB_{>0}([\compij]) \}$,
that is: $t-\tmin+\tmin_{\i..\j}$.
On the other hand, the value of $sum([\i..\j])$ is maximum if the number of
non null elements outside $[\i..\j]$ is minimum, and if all of non null
elements in $[\compij]$ have value $1$.
Therefore, the maximum value of the sum inside $[\i..\j]$ is given by:
$s-max\{\tmin_{\compij}, t-\tmax_{\i..\j} \}=$
$s-max\{\tmin-\tmin_{\i..\j}, t-\tmax_{\i..\j} \}$.
\eproof

\vspace*{-.5cm}
\noindent
\textbf{Remark.}
Note that, unlike the mean values of $S_1(b_{\i..\j})$ and $S_2(b_{\i..\j})$,
the value of $E(S_3([\i..\j]))$ generally depends on $t$.
That is, when the integrity constraints provided by $LB_{>0}$ are exploited,
the estimated answer of a sum query depends on the number of non null
elements occurring in the whole block $B_\k$ .
This difference between Case 3 and previous cases can be explained as follows.
In Case 1 and 2 (when the function $LB_{>0}$ is not available or not exploited),
no information about the exact position of non null elements inside $B_\k$
is provided.
Now, the estimation of the sum query is made by considering all possible
ways of distributing $t$ non null elements in the block.
Thus, if we partition $B_\k$ into two equal halves (by splitting $B_\k$ along
one of its dimensions), for each configuration of $B_\k$ consisting of $t'$
non null elements located inside the first half of $B_\k$ and $t-t'$ non null
elements in the other half, there exists a ``symmetric" configuration where $t-t'$
non null elements are in the first half of $B_\k$ and $t'$ non null elements are
in the second half.
This implies that the only knowledge of $t$ does not make the
distribution of the sum $s$ inside the block ``unbalanced".
In contrast, the information encoded in the function $LB_{>0}$
invalidates the symmetry condition described above.
That is, given a consistent configuration of $B_\k$ containing $t'$ non null
elements inside the first half of $B_\k$ and $t-t'$ non null elements in the
other half, the ``symmetric" configuration exists only if it is consistent
according to the integrity constraint expressed by the function $LB_{>0}$.

It should be pointed out that if $LB_{>0}$ is not available or not used, the
estimate provided using Case 3 does not depend on $t$.
In fact, when only $LB_{=0}$ is exploited, the estimation process described in
Case 3 works in the same way as Cases 1 and 2, after removing from $B_\k$ all
elements which are certainly null according to $LB_{=0}$.
We can reach the same conclusion by extracting a formula for $E(S_3([\i..\j]))$
from the one provided in Theorem \ref{prop-sumCase2}, by substituting
$LB_{>0}([\i..\j])=0$ and $LB_{>0}([\compij])=0$, thus obtaining:
$E(S_3([\i..\j])) = (\tmax_{\i..\j}- \tmin_{\i..\j})\cdot \frac{s}{\tmax}$,
which is independent on $t$.

\section{Influence of integrity constraints on accuracy: some experimental results}
\label{sec:experiments}

In the analysis of the accuracy of the estimated answers for the
Cases 1 and 2, we focused our attention on discussing the
dependence of variance on the ratios $b_{\i..\j}/b$ and $t/b$. The
introduction of integrity constraints makes both the estimated
answer and variance to depend on the position of $[\i..\j]$ inside
the block (since the values of $\tmax_{\i..\j}$ and
$\tmin_{\i..\j}$ change as the boundaries of the range move), and
on the maximum number $\tmax$ and minimum number $\tmin$ of non
nulls inside the block. Therefore, it is relevant to check how
much the variance change when we use the knowledge of
$LB_{=0}([\i..\j])$, $LB_{=0}([\compij])$, $LB_{>0}([\i..\j])$ and
$LB_{>0}([\compij])$, whose values determine $\tmax_{\i..\j}$,
$\tmin_{\i..\j}$, $\tmax$ and $\tmin$. Next we perform this
analysis but, for the sake of brevity, we shall only consider the
presence of the constraints $LB_{=0}([\i..\j])$ and
$LB_{=0}([\compij])$, thus assuming $LB_{>0}([\i..\j])=0$ and
$LB_{>0}([\compij])=0$ --- indeed the dependency of the estimates
on the latter classes of constraints are quantitatively the same.

\begin{figure}[h]
\centerline{\psfig{figure=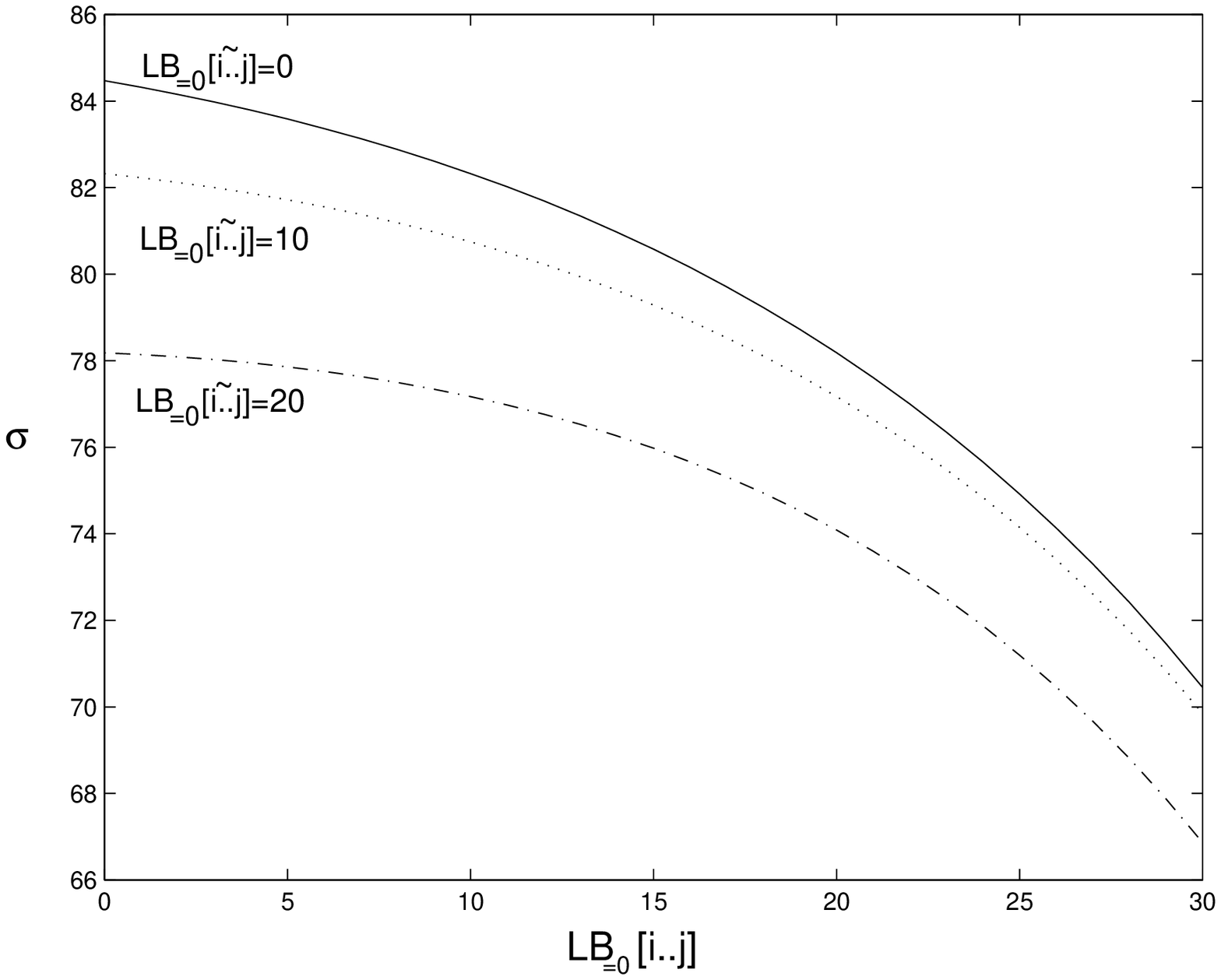,width=6cm,height=4.2cm}}
\caption{$\sigma(S_3([\i..\j]))$ versus $LB_{=0}([\i..\j])$ for
different values of $LB_{=0}([\compij])$}
\label{varconstraintsnull}
\end{figure}

Consider a sum query of size $b_{\i..\j}=500$ over a block with
$b=1000$ and $t=500$. Fig. \ref{varconstraintsnull} shows the
standard deviation of the random variable $S_3([\i..\j])$ versus
the value of $LB_{=0}([\i..\j])$, for different values of
$LB_{=0}([\compij])$: the solid line corresponds to the value $0$
of $LB_{=0}([\compij])$, the dotted line to the value $10$, and
the dash-dot line to the value $20$. The diagram shows that, when
$LB_{=0}([\compij])=0$ is fixed, $\sigma$ decreases from $84.31$
to $70.44$, as $LB_{=0}([\i..\j])$ changes from $0$ (which is
equivalent to consider no integrity constraint) to $30$. This
change corresponds to a variation of 16\% of the standard
deviation.

The decrease of the standard deviation depicted in Fig.
\ref{varconstraintsnull} corresponds to a ``restriction" of the
datacube population on which the random variable associated to the
query is applied. In fact, evaluating a query of size $b_{\i..\j}$
over a block of size $b$ whose elements have sum $s$ is equivalent
to evaluating a query of size $b_{\i..\j}-LB_{=0}([\i..\j])$ over
a block containing $b-LB_{=0}([\i..\j])-LB_{=0}([\compij])$
elements with the same value of $s$. Thus, when
$LB_{=0}([\i..\j])>0$ or $LB_{=0}([\compij])>0$, the population of
datacubes which are compatible with the given aggregate data is
restricted w.r.t. both the cases $LB_{=0}([\i..\j])=0$ and
$LB_{=0}([\compij])=0$. This restricted population of datacubes
corresponds to a lower ``degree of uncertainty" in distributing
the value of $s$ among the elements inside the blocks.

\begin{figure}[h]
\centerline{\psfig{figure=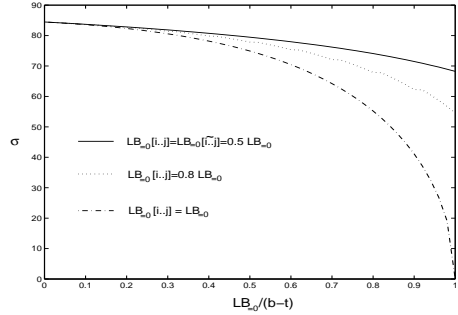,width=6cm,height=4.2cm}}
\caption{$\sigma(S_3([\i..\j]))$ versus $LB_{=0}/(b-t)$ for
$b=100$,$t=50$, $b_{\i..\j}=50$ and $s=1000$.} \label{rappLBsub-t}
\end{figure}

The  diagram in Fig. \ref{rappLBsub-t} reports
$\sigma(S_3([\i..\j]))$ versus
$(LB_{=0}([\i..\j])+LB_{=0}([\compij]))/(b-t)$, that is the ratio
between the number of the null elements localized by integrity
constraints and the total number of null elements of the block
(according to the aggregate data $t$). Fig. \ref{rappLBsub-t}
shows that the larger the number
$LB_{=0}([\i..\j])+LB_{=0}([\compij])$ of null elements localized
by integrity constraints (compared to the total number of nulls
inside the block), the lower is the value of the estimated error.
In Fig. \ref{rappLBsub-t} the sum
$LB_{=0}([\i..\j])+LB_{=0}([\compij])$ is denoted as $LB_{=0}$.

The same diagram shows that the estimated error is smaller when
$LB_{=0}([\i..\j])$ and $LB_{=0}([\compij])$ are ``unbalanced",
i.e. either $LB_{=0}([\i..\j])>LB_{=0}([\compij])$ or
$LB_{=0}([\i..\j])<$ $LB_{=0}([\compij])$. As it can be easily
intuited, knowing that most of null elements are distributed
either inside or outside the range of the query reduces the
approximation in evaluating the distribution of $s$ inside the
block.

In sum, as expected, introducing integrity constraints on the
number of null elements in each block influences ``positively" the
estimation process. We stress that our results are valid for
random data samples so  errors may be larger in real-world
applications whose data distributions can be rather \"biased", so
that the accuracy of the estimates evaluated using the framework
can be far from being accurate. Next, we present the results of
testing our estimation models to a sample consisting of ten
real-life two-dimensional datacubes which confirm the positive
influence of integrity constraints on the accuracy of
estimations.

The datacubes for our experiments contain the daily incomes
corresponding to the products sold in a chain store during periods
of two months belonging to ten different years. Each datacube
consists of a matrix made of $7580$ rows (corresponding to all
store products) and $60$ columns (corresponding to the working
days). Both count and sum queries over all the ranges of size
$100\times 20$ have been evaluated for each datacube, comparing
the exact answers to the approximate ones. In particular,
different compressed representations of every datacube have been
examined, corresponding to different sizes of the summary blocks;
for each compressed structure, both the actual and the estimated
errors obtained with and without the use of integrity constraints
have been evaluated. For sum queries, the influence of using the
parameter $t$ on the query estimation result has been studied too.

In our experiments, the integrity constraints consist of
``macro-blocks" which delimit portions of the cube consisting of
all null elements or of all non-null elements. These macro-blocks
do not identify all null [resp., non null] elements inside the
cube, but only those null [resp., non null] elements which are
inside a portion of the cube containing at least $20$ null [resp.,
non null] elements. Macro-blocks do not overlap, and can be
efficiently stored and retrieved using traditional indexing
methods for spatial access. On the average, the adopted
constraints located $40\%$ of null values and $10\%$ of non null
elements inside the examined samples.

In the tables of Figures \ref{tabellaerroricount} and
\ref{tabellaerrorisum}, results obtained for count and sum queries
are reported. The tables represent the intervals where the actual
error for queries of size $100\times20$ are contained, considering
all datacubes. That is, each entry of the table shows, in
percentage terms, the number of estimates whose actual error is
less than $3\times \sigma$, $4\times \sigma$, and $5\times
\sigma$, for each of the estimation techniques proposed in Case 1,
Case 2 and Case 3.

\begin{figure}[h]
\begin{center}
\small
\begin{tabular}{||c||c|c|c||c|c|c||}
\hline\hline
Block  \vspace*{-2mm}& \multicolumn{3}{c||}{Without constraints} & \multicolumn{3}{c||}{Using constraints}  \\
size &\multicolumn{3}{c||}{(Case 1)} & \multicolumn{3}{c||}{(Case 3)}  \\
  & $3\times\sigma$ & $4\times\sigma$ & $5\times\sigma$ & $3\times\sigma$ & $4\times\sigma$ & $5\times\sigma$\\
\hline
$10\!\!\times\!\!10$ & 63.9\% & 74.1\% & 81.8\% & 87.4\% & 95.2\% & 98.6\% \\
\hline
$12\!\!\times\!\!12$ & 72.4\% & 84.4\% & 92.1\% & 91.4\% & 97.9\% & 99.6\% \\
\hline
$14\!\!\times\!\!14$ & 77.4\% & 87.8\% & 93.5\% & 91.9\% & 98.1\% & 99.6\% \\
\hline
$16\!\!\times\!\!16$ & 59.2\% & 72.8\% & 81.9\% & 87.8\% & 95.7\% & 98.8\% \\
\hline
$18\!\!\times\!\!18$ & 51.3\% & 62.7\% & 71.8\% & 84.1\% & 93.1\% & 97.5\% \\
\hline
$20\!\!\times\!\!20$ & 58.1\% & 70.1\% & 78.8\% & 86.2\% & 94.6\% & 98.1\% \\
\hline\hline
\end{tabular}
\end{center}
\caption{Number of count queries whose actual error is less than
$3 \sigma$, $4 \sigma$, and $5  \sigma$}
\label{tabellaerroricount}
\end{figure}
\ \\

\begin{figure}[h]
\begin{center}
\small
\begin{tabular}{||c||c|c|c||c|c|c||c|c|c||}
\hline\hline
Block \vspace*{-3mm}& \multicolumn{3}{c||}{Without $t$} &\multicolumn{3}{c||}{Without constraints} & \multicolumn{3}{c||}{Using constraints}  \\
size& \multicolumn{3}{c||}{(Case 1)} &\multicolumn{3}{c||}{(Case 2)} & \multicolumn{3}{c||}{(Case 3)}  \\
  & $3\times\sigma$ & $4\times\sigma$ & $5\times\sigma$ &$3\times\sigma$ & $4\times\sigma$ & $5\times\sigma$ & $3\times\sigma$ & $4\times\sigma$ & $5\times\sigma$\\
\hline
$10\!\!\times\!\!10$ & 34.4\% & 43.7\% & 51.8\% & 70.4\% & 79.9\% & 86.9\% & 81.7\% & 90.5\% & 95.7\%\\
\hline
$12\!\!\times\!\!12$ & 33.1\% & 44.1\% & 54.2\% & 78.8\% & 90.1\% & 95.7\% & 88.6\% & 96.1\% & 98.4\%\\
\hline
$14\!\!\times\!\!14$ & 28.1\% & 37.1\% & 45.7\% & 69.1\% & 81.9\% & 89.4\% & 73.3\% & 86.1\%  & 92.8\%\\
\hline
$16\!\!\times\!\!16$ & 22.1\% & 29.2\% & 60.9\% & 73.6\% & 82.1\% & 89.4\% & 84.5\% & 91.5\% & 95.4\%\\
\hline
$18\!\!\times\!\!18$ & 23.6\% & 30.7\% & 36.9\% & 60.6\% & 72.1\% & 80.2\% & 76.9\% & 86.9\% & 92.8\%\\
\hline
$20\!\!\times\!\!20$ & 30.2\% & 38.9\% & 46.8\% & 69.3\% & 79.1\% & 85.9\% & 79.6\% & 88.9\% & 94.1\%\\
\hline\hline
\end{tabular}
\end{center}
\caption{Number of sum queries whose actual error is less than $3
\sigma$, $4 \sigma$, and $5 \sigma$} \label{tabellaerrorisum}
\end{figure}

\vspace*{-.5cm}
Results reported in the tables show that:
\begin{enumerate}
\item the use of $t$ makes the estimation of the error for sum
queries more accurate; \item for both count and sum queries, the
accuracy of estimates benefits from the use of integrity
constraints.
In particular, a smaller coefficient to \"correct" effectively
the estimate provided by $\sigma$ is needed, and the value of
this coefficient is almost independent from the particular
compressed representation of the datacube.
For instance, without using integrity constraints, the number of
estimated count queries whose actual error is less than
$5\times\sigma$ is between $71.8 \%$ and $93.5 \%$, depending on
the block size.
On the other hand, when integrity constraints are used, the number
of estimated count queries whose actual error is less than
$5 \times \sigma$ is greater than $90 \%$ for every block size.
\end{enumerate}

\section{Estimation of Range Queries on Histograms}
\label{sec:histo}
In this section we apply our framework to derive some results about
mono-dimensional histograms.
Mono-dimensional histograms are constructed to summarize the frequency
distribution of the values of a single attribute in a database relation,
and can be exploited to estimate query result sizes \cite{IoPo95,Poo97,PooIoa96}.
The estimation is accomplished on the basis of the knowledge of both
the number $t$ of non-null frequencies and the total frequency
sum $s$ in each block $B_k$ (called \textit{bucket} in the histogram
terminology).
As mentioned in the Introduction, a crucial point for providing good
estimations is the way the frequency distributions for original values
are partitioned into buckets.
Here we assume that the buckets have been already arranged using any
of the known techniques, and we therefore focus on the problem of
estimating the frequency distribution inside a bucket.

\subsection{A theory for the Continuous Value Assumption}

The most common approach to estimate frequency distribution inside
a bucket is the {\em continuous value assumption} \cite{Sac79}:
The sum of frequencies in a range of a bucket is estimated by
linear interpolation.
It corresponds to equally distributing the overall sum of
frequencies of the bucket to all attribute values occurring in it.

Corollary \ref{prop-sumj} (where both $t$ and $s$ are used to estimate
sum range queries) provides a theoretical foundation of the continuous
value assumption, as it states that the mean value of the random variable
$S_2(b_{\i..\j})$ is {\large $\frac{b_{\i..\j}}{b}$}$\cdot s$.
Thus our approach gives a model to explain the linear interpolation and,
besides, allows to evaluate the error of the estimation, thus exploiting
the knowledge about the number $t$ of non-nulls in a block --- instead
$t$ is not mentioned in the computation of  the mean.

We point out that, in order to provide a more elaborated interpolation scheme,
in \cite{PooIoa96,Poo97} another method for estimating sum of frequencies
inside a  block is proposed, based on the {\em uniform spread assumption}:
The $t$ non-null attribute values in each bucket are assumed to be located at equal distance
from each other, and the overall frequency sum is therefore
equally distributed among them.
This method does not give a correct estimation unless we assume
that nun-nulls are scattered on the block in some particular, biased way.
Next, using our theoretical framework,  we propose an unbiased
estimation inside a block which takes into account the number $t$
of non-null values.

\subsection{The 1/2-Biased Assumption}

We first recall that the classical definition of histogram requires that
both lowest and highest elements (or at least one of them) of any block
are not null \cite{Poo97} (i.e. they are attribute values occurring in the
relation).
We call {\em 2-biased} a block for which the extreme elements are not null;
if only the lowest (or the highest) element is not null then the block is
called  {\em 1-biased}.

So far linear interpolation is also used for biased blocks, thus
producing a wrong estimation --- it is the case to say a
``biased" estimation.
We next show the correct formulas, that are derived from Theorem
\ref{prop-sumCase2}.

\begin{corollary}
Let $B_\k$ be a block of a histogram, and let $S_4([\i..\j]) =
\sumr(\MF[\i..\j])$ be an integer random variable ranging from $0$
to $s$, defined by taking $\MF$ in the population
$\Pi_{LB_{>0}}(M^{-1}_{cs,F})$. Then
\begin{enumerate}
\item if the block $B_\k$ is 1-biased and $\i$ is the lowest
element of the block then mean and variance of $S_4([\i..\j])$
are, respectively:
\[
E(S_4([\i..\j])) = \frac{s}{t} + (b_{\i..\j}- 1) \cdot \frac{s}{t}
\cdot \frac{t- 1}{b - 1},
\]

\[
\begin{array}{ll}
\sigma^2(S_4([\i..\j]))=& \alpha \cdot (b_{\i..\j}-1) \cdot \frac
{t- 1}{b- 1} \cdot \left[ 1+(b_{\i..\j}-2) \cdot \frac {t- 2}{b -
2} \right ] + \\
&(\beta +2 \cdot \alpha ) \cdot (b_{\i..\j}-1) \cdot
 \frac {t- 1}{b - 1} +
 (\alpha + \beta ) - {E(S_4([\i..\j]))}^2
 \end{array}
\]
\item if the block $B_\k$ is 1-biased and $\i$ is not the lowest
element of the block then mean and variance of $S_4([\i..\j])$
are, respectively:

\[
E(S_4([\i..\j])) = b_{\i..\j} \cdot \frac{s}{t} \cdot
\frac{t-1}{b- 1},
\]
\[
\sigma^2(S_4([\i..\j]))= \alpha \cdot b_{\i..\j} \cdot \frac {t-
1}{b- 1} \cdot \left[ 1+(b_{\i..\j} -1) \cdot \frac {t- 2}{b - 2}
\right ] +
\]
\[
\hspace*{2.7cm} + \beta \cdot b_{\i..\j} \cdot
\frac {t- 1}{b - 1} - {E(S_4([\i..\j]))}^2
\]

\item if the block $B_\k$ is 2-biased and either $\i$ or $\j$ is
an extreme element of the block then mean and variance of
$S_4([\i..\j])$ are, respectively:

\[
E(S_4([\i..\j])) = \frac{s}{t} + (b_{\i..\j}- 1) \cdot \frac{s}{t}
\cdot \frac{t- 2}{b - 2},
\]

\[
\begin{array}{ll}
\sigma^2(S_4([\i..\j]))=& \alpha \cdot (b_{\i..\j}-1) \cdot \frac
{t- 2}{b- 2} \cdot \left[ 1+(b_{\i..\j}-2) \cdot \frac {t- 3}{b -
3} \right ] +\\

&+(\beta\! +\!2 \!\cdot \!\alpha )\! \cdot \!(b_{\i..\j}-\!1) \!\cdot\!
 \frac {t- 2}{b - 2} +
 (\alpha + \beta) - {E(S_4([\i..\j]))}^2
 \end{array}
\]

\item if the block $B_\k$ is 2-biased, and neither $\i$ nor $\j$
is an extreme element of the block, then mean and variance of
$S_4([\i..\j])$ are, respectively:
\[
E(S_4([\i..\j])) = b_{\i..\j} \cdot \frac{s}{t} \cdot \frac{t-
2}{b - 2},
\]

\[
\sigma^2(S_4([\i..\j]))= \alpha \cdot b_{\i..\j} \cdot \frac {t-
2}{b- 2} \cdot \left[ 1+(b_{\i..\j}-1) \cdot \frac {t- 3}{b - 3}
\right ] +
\]
\[
 \hspace*{2.8cm}+ \beta \cdot b_{\i..\j} \cdot
 \frac {t- 2}{b - 2} - {E(S_4([\i..\j]))}^2
\]
\end{enumerate}
\noindent where:

\[
\alpha=\frac{s \cdot (s+1)}{t \cdot (t+1)}, \ \mbox{and:} \
\beta=\frac{s \cdot (s-t)}{t \cdot (t+1)}.
\]

\end{corollary}
{\bf Proof.}
\begin{enumerate}
\item ({\em $B_\k$ is 1-biased and $\i$ is the lowest element of
the block}).
In this case, $E(S_4([\i..\j]))$ and $\sigma^2(S_4([\i..\j]))$ coincide
to $E(S_3([\i..\j]))$ and $\sigma^2(S_3([\i..\j]))$, respectively, computed
in Theorem \ref{prop-sumCase2}, by considering $LB_{=0}([\i..\j])=0$,
$LB_{=0}(\compij) = 0$, $LB_{>0}([\i..\j]) = 1$, and
$LB_{>0}(\compij) = 0$.
The statement of the corollary is thus obtained by considering that
$\tmax = b$, $\tmax_{\i..\j} = b_{\i..\j}$, $\tmin_{\i..\j} = 1$ and $\tmin = 1$.
\item
({\em $B_\k$ is 1-biased and $\i$ is not the lowest element of the block}).
In this case, $E(S_4([\i..\j]))$ and $\sigma^2(S_4([\i..\j]))$ coincide with
$E(S_3([\i..\j]))$ and $\sigma^2(S_3([\i..\j]))$, respectively, computed in
Theorem \ref{prop-sumCase2}, by considering $LB_{=0}([\i..\j])=0$,
$LB_{=0}(\compij) = 0$, $LB_{>0}([\i..\j]) = 0$, and $LB_{>0}(\compij) = 1$.
The statement of the corollary is thus obtained by considering that
$\tmax = b$, $\tmax_{\i..\j} = b_{\i..\j}$, $\tmin_{\i..\j} = 0$ and $\tmin = 1$.
\item
({\em $B_\k$ is 2-biased and either $\i$ or $\j$ is an extreme element
of the block}).
Suppose that $\i$ is an extreme element of the block (the other case can be
obtained by symmetry).
In this case, $E(S_4([\i..\j]))$ and $\sigma^2(S_4([\i..\j]))$ coincide with
$E(S_3([\i..\j]))$ and $\sigma^2(S_3([\i..\j]))$, respectively, computed in
Theorem \ref{prop-sumCase2}, in case $LB_{=0}([\i..\j])= LB_{=0}(\compij) = 0$, $LB_{>0}([\i..\j]) =
1$, and $LB_{>0}(\compij) = 1$.
The statement of the corollary is thus obtained by considering that $\tmax = b$,
$\tmax_{\i..\j} = b_{\i..\j}$, $\tmin_{\i..\j} = 1$ and $\tmin = 2$.
\item
({\em $B_\k$ is 2-biased and neither $\i$ nor $\j$ is an extreme element of the block}).
In this case, $E(S_4([\i..\j]))$ and $\sigma^2(S_4([\i..\j]))$ coincide to $E(S_3([\i..\j]))$
and $\sigma^2(S_3([\i..\j]))$, respectively, computed in Theorem \ref{prop-sumCase2}, in
case $LB_{=0}([\i..\j])= LB_{=0}(\compij) = 0$, $LB_{>0}([\i..\j]) = 0$, and $LB_{>0}(\compij) = 2$.
The statement of the corollary is thus obtained by considering that
$\tmax = b$, $\tmax_{\i..\j} = b_{\i..\j}$, $\tmin_{\i..\j} = 0$
and $\tmin = 2$.
\end{enumerate}
\eproof

The above formulas have been used in \cite{BuPoRoSa02} to replace
the continuous value assumption inside one of the most efficient
methods for histogram representation (the maxdiff method
\cite{PooIoa96}), and have produced some meaningful improvements
in the performance of the method.

\section{Conclusion and Future Work}
\label{sec:conclusioni}
\vspace{-4mm}
In this paper we have defined a probabilistic framework for estimating
range queries on a compressed datacube obtained by partitioning the original
datacube into a number of non-overlapping blocks and then storing, for each block,
some aggregate information on its data distribution.
The proposed estimation paradigm allows us to provide an
approximate answer of range queries (more specifically,
\emph{sum} and \emph{count} queries) together with an estimate
of the error of the returned answer, by accessing only the
compressed representation of the datacube.
The estimates of both the answer and the error depend on the
aggregate data and integrity constraints which are exploited,
without any a priori assumption on the particular data distribution
inside the original datacube.
We have investigated how the values of the answer and the estimated
error depend on the available aggregate data and integrity
constraints, by performing both an analytical and experimental study.

We remark that the idea of introducing integrity constraints is
crucial  to improve the accuracy of estimations in real
applications.
In fact, the need of integrity constraints is due to the fact that
the real-life datacubes are rather \"biased" with respect to the
\"virtual" population of datacubes on which the theoretical estimation
process is performed.
The effectiveness of the estimates improves as integrity constraints are
introduced because the estimation process is accomplished on a restricted
population of datacubes, and the examined samples are more representative
of this population than the more general one.

Therefore, further types of constraints are needed in order to catch the
actual distribution of data inside a datacube and improve the accuracy of
the estimates.
Thus, extensions of this work will follow the directions below:
\begin{itemize}
\item
extending the framework by considering further aggregate data on the
blocks of the datacube, other than the sum and the number of non null values
inside each block (for instance, the maximum and the minimum value inside
the blocks);
\item
taking into account data skew: this issue can be accomplished by storing
some information regarding the number of distinct values inside
each block, or the values with the maximum number of occurrences
in the blocks.
\end{itemize}


\newpage

\section*{APPENDIX}


\begin{claim} \[
\sigma^2(S_1(b_{\i..\j}))=b_{\i..\j}  \cdot s \cdot \frac{(b-b_{\i..\j}) \cdot (b+s) }{b^2 \cdot  (b+1)} \\
\]
\end{claim}

\bproof We start from the definition of variance:

\[
\hspace*{-.9cm}
\sigma^2(S_1(b_{\i..\j}))=\sum_{s_{\i..\j}=0}^{s}
\left ( s_{\i..\j}- \frac {b_{\i..\j}}{b} \cdot s \right )^2
\cdot
P(S_1(b_{\i..\j})=s_{\i..\j}) =\]

\[
\hspace*{-.9cm}
= \sum_{s_{\i..\j}=0}^{s}s_{\i..\j}^2 \cdot
P(S_1(b_{\i..\j})=s_{\i..\j}) - \left ( \frac {b_{\i..\j}}{b}
\cdot s \right ) ^2 =\]

\begin{equation}
\label{s1uno}
\hspace*{-.9cm}
= \sum_{s_{\i..\j}=0}^{s} s_{\i..\j}^2 \cdot
\frac {
    {\scriptsize \left ( \begin{tabular}{c} \footnotesize$b_{\i..\j}+s_{\i..\j}-1$ \\ \footnotesize$s_{\i..\j}$ \end{tabular} \right )}
    \cdot
    {\scriptsize \left ( \begin{tabular}{c} \footnotesize$b - b_{\i..\j} + s - s_{\i..\j}- 1$ \\ \footnotesize$s-s_{\i..\j}$ \end{tabular} \right )}
}
{
    {\scriptsize \left ( \begin{tabular}{c}\footnotesize$b+s-1$ \\ \footnotesize$s$ \end{tabular} \right )}
}
- \left ( \frac {b_{\i..\j}}{b} \cdot s \right )^2
\end{equation}

\noindent
As
${\scriptsize \left ( \begin{tabular}{c} \footnotesize$x$ \\ \footnotesize$y$ \end{tabular} \right )}
=
\frac {\mbox{\small$x\!-\!y\!+\!1$}}{\mbox{\small$y$}}
\cdot\!
{\scriptsize \left ( \begin{tabular}{c} \footnotesize$x$ \\ \footnotesize$y\!-\!1$ \end{tabular} \right )}$,
the term:

$\sum_{s_{\i..\j}=0}^{s}
s_{\i..\j}^2 \cdot
{\scriptsize \left ( \begin{tabular}{c} \footnotesize$b_{\i..\j}\!+\!s_{\i..\j}\!-\!1$ \\ \footnotesize$s_{\i..\j}$ \end{tabular} \right )}
\!\!\cdot\!\!
{\scriptsize \left ( \begin{tabular}{c} \footnotesize$b\!-\!b_{\i..\j}\!+\!s\!-\!s_{\i..\j}\!-\!1$ \\ \footnotesize$s-s_{\i..\j}$ \end{tabular} \right )}$

\noindent can be re-written as:

\[
\hspace*{-0.9cm}
\sum_{s_{\i..\j}=0}^{s}
b_{\i..\j}\cdot s_{\i..\j} \cdot
{\scriptsize \left ( \begin{tabular}{c} \footnotesize$b_{\i..\j}\!+\!s_{\i..\j}\!-\!1$ \\ \footnotesize$s_{\i..\j}-1$ \end{tabular} \right )}
\!\!\cdot\!\!
{\scriptsize \left ( \begin{tabular}{c} \footnotesize$b\!-\!b_{\i..\j}\!+\!s\!-\!s_{\i..\j}\!-\!1$ \\ \footnotesize$s-s_{\i..\j}$ \end{tabular} \right )}
=
\]

\[
\hspace*{-0.9cm}
= \sum_{s_{\i..\j}=0}^{s}
b_{\i..\j}\cdot (s_{\i..\j}-1) \cdot
{\scriptsize \left ( \begin{tabular}{c} \footnotesize$b_{\i..\j}\!+\!s_{\i..\j}\!-\!1$ \\ \footnotesize$s_{\i..\j}-1$ \end{tabular} \right )}
\!\!\cdot\!\!
{\scriptsize \left ( \begin{tabular}{c} \footnotesize$b\!-\!b_{\i..\j}\!+\!s\!-\!s_{\i..\j}\!-\!1$ \\ \footnotesize$s-s_{\i..\j}$ \end{tabular} \right )}
+
\]
\[
+
\sum_{s_{\i..\j}=0}^{s}
b_{\i..\j} \cdot
{\scriptsize \left ( \begin{tabular}{c} \footnotesize$b_{\i..\j}\!+\!s_{\i..\j}\!-\!1$ \\ \footnotesize$s_{\i..\j}-1$ \end{tabular} \right )}
\!\!\cdot\!\!
{\scriptsize \left ( \begin{tabular}{c} \footnotesize$b\!-\!b_{\i..\j}\!+\!s\!-\!s_{\i..\j}\!-\!1$ \\ \footnotesize$s-s_{\i..\j}$ \end{tabular} \right )}
=
\]

\[
\hspace*{-0.9cm}
= \sum_{S_{\i..\j}=1}^{s-1} b_{\i..\j} \cdot S_{\i..\j} \cdot
{\scriptsize
    \left ( \begin{tabular}{c} \footnotesize$b_{\i..\j}+S_{\i..\j}$ \\ \footnotesize$S_{\i..\j}$ \end{tabular} \right )
}
\!\!\cdot\!\!
{\scriptsize
    \left ( \begin{tabular}{c} \footnotesize$b - b_{\i..\j} + (s\!-\!1) - S_{\i..\j}- 1$ \\ \footnotesize$(s-1)-S_{\i..\j}$ \end{tabular} \right )
}
+
\]

\begin{equation}
\label{s1due}
\hspace*{-0.4cm}
+ \sum_{S_{\i..\j}=0}^{s-1}
b_{\i..\j} \cdot \!
{\scriptsize
    \left ( \begin{tabular}{c} \footnotesize$(b_{\i..\j}+1)\!\!+\!S_{\i..\j}\!-\!\!1$ \\ \footnotesize$S_{\i..\j}$ \end{tabular} \right )
}
\!\!\cdot\!\!
{\scriptsize
    \left ( \begin{tabular}{c} \footnotesize$(b\!+\!1)\!\!-\!\!(b_{\i..\j}+1)\!+\!(s\!-\!1)\!\!-\!\!S_{\i..\j}\!-\!\!1$ \\ \footnotesize$(s\!-\!1)\!-\!S_{\i..\j}$ \end{tabular} \right )
}
\end{equation}
where: $S_{\i..\j}=s_{\i..\j}-1$.\\


Then, by applying formula(\ref{s1tre}), the latter becomes:
\[
\hspace*{-.9cm}
\sum_{S_{\i..\j}=1}^{s-1}
\!b_{\i..\j} \cdot\! S_{\i..\j} \cdot
\frac{b_{\i..\j}+\!1}{S_{\i..\j}} \cdot
{\scriptsize
    \left ( \begin{tabular}{c} \footnotesize$b_{\i..\j}+S_{\i..\j}$ \\ \footnotesize$S_{\i..\j}-1$ \end{tabular} \right )
}
\!\!\cdot\!\!
{\scriptsize
    \left ( \begin{tabular}{c} \footnotesize$b\!-\!b_{\i..\j} +\!(s\!-\!1)\!-\!S_{\i..\j}-\!1$ \\ \footnotesize$(s\!-\!1)\!-\!S_{\i..\j}$ \end{tabular} \right )
}
+
\]

\[
+\ b_{\i..\j}
\cdot\!
{\scriptsize
    \left ( \begin{tabular}{c} \footnotesize$b\!+\!s\!-\!1$ \\ \footnotesize$s-1$ \end{tabular} \right )
}
=
\]

\[\hspace*{-.9cm}
= \sum_{S_{\i..\j}=1}^{s-1} b_{\i..\j} \cdot (b_{\i..\j}+\!1) \cdot\!
{\scriptsize
    \left ( \begin{tabular}{c} \footnotesize$b_{\i..\j}+S_{\i..\j}$ \\ \footnotesize$S_{\i..\j}-1$ \end{tabular} \right )
}
\!\!\cdot\!\!
{\scriptsize
    \left ( \begin{tabular}{c} \footnotesize$b\!-\!b_{\i..\j} +\!(s\!-\!1)\!-\!S_{\i..\j}-\!1$ \\ \footnotesize$(s\!-\!1)\!-\!S_{\i..\j}$ \end{tabular} \right )
}
+
\]

\[
+\ b_{\i..\j} \cdot
\frac{s}{b} \cdot
{\scriptsize
    \left ( \begin{tabular}{c} \footnotesize$b\!+\!s\!-\!1$ \\ \footnotesize$s-1$ \end{tabular} \right )
}
= \]

\[\hspace*{-.9cm}
= \sum_{\alpha=0}^{s-2} b_{\i..\j}\!\cdot\!\!(b_{\i..\j}\!+\!1)\!\cdot\!\!
{\scriptsize
    \left ( \begin{tabular}{c} \footnotesize$(b_{\i..\j}+\!2)\!+\!\alpha\!-\!1$ \\ \footnotesize$\alpha$ \end{tabular} \right )
}
\!\!\cdot\!\!
{\scriptsize
    \left ( \begin{tabular}{c} \footnotesize$(b\!+\!2)\!-\!(b_{\i..\j}+\!2)\!+\!(s\!-\!2)-\!\alpha\!-\!1$ \\ \footnotesize$(s\!-\!2)\!-\!\alpha$ \end{tabular} \right )
}
+
\]

\[
+\ b_{\i..\j} \cdot\!\frac{s}{b} \cdot
{\scriptsize
    \left ( \begin{tabular}{c} \footnotesize$b+s-1$ \\ \footnotesize$s$ \end{tabular} \right )
}
=
\hspace{10mm}[where \hspace{3mm} \alpha=S_{\i..\j}-1]
 \]

\[\hspace*{-.9cm}
= b_{\i..\j} \cdot (b_{\i..\j}+1) \cdot
{\scriptsize
    \left ( \begin{tabular}{c} \footnotesize$b + s -1$ \\ \footnotesize$s-2$ \end{tabular} \right )
}
+\  b_{\i..\j} \cdot \frac{s}{b} \cdot
{\scriptsize
    \left ( \begin{tabular}{c} \footnotesize$b +s-1$ \\ \footnotesize$s$ \end{tabular} \right )
}
=
\]

\[
\hspace*{-.9cm}
= b_{\i..\j} \cdot (b_{\i..\j}+1) \cdot \frac{s \cdot (s-1)}{b \cdot (b+1)} \cdot
{\scriptsize
    \left ( \begin{tabular}{c} \footnotesize$b + s -1$ \\ \footnotesize$s$ \end{tabular} \right )
}
+\ b_{\i..\j} \cdot \frac{s}{b} \cdot
{\scriptsize
    \left ( \begin{tabular}{c} \footnotesize$b +s-1$ \\ \footnotesize$s$ \end{tabular} \right )
}
=
\]

\begin{equation}\label{s1quattro}
\hspace*{-.9cm}
= b_{\i..\j} \cdot \frac{s}{b} \cdot
{\scriptsize
    \left ( \begin{tabular}{c} \footnotesize$b + s -1$ \\ \footnotesize$s$ \end{tabular} \right )
}
\cdot
\left[ (b_{\i..\j}+1) \cdot \frac{s-1}{b+1}+1 \right]
\end{equation}

By substituting (\ref{s1quattro}) in (\ref{s1uno}) we obtain:
\[
\hspace*{-.9cm}
\sigma^2(S_1(b_{\i..\j}))=
b_{\i..\j} \!\cdot\! \frac{s}{b} \cdot
\left[ (b_{\i..\j}+\!1) \!\cdot\! \frac{s\!-\!1}{b\!+\!1}\!+\!1 \right] -
\left( \frac{b_{\i..\j}}{b} \cdot \! s \right)^2 =
\]

\[
\hspace*{-.9cm}
= b_{\i..\j} \cdot \frac{s}{b^2} \cdot \frac{b \cdot
(b_{\i..\j}+1) \cdot (s-1)+ b \cdot (b+1)-b_{\i..\j} \cdot s \cdot
(b+1) }{b+1} =
\]

\[
\hspace*{-.9cm}
= b_{\i..\j} \cdot s \cdot \frac{(b-b_{\i..\j}) \cdot (b+s)}{b^2
\cdot (b+1)}
\]

 \eproof

\newpage
\begin{claim}\ \\
\[
\begin{array}{ll}
\hspace*{-1cm} \sigma^2(C_3([\i..\j])) =
                    \left \{
                    \begin{array}{ll}
                    \frac{\tmax_{\i..\j} - \tmin_{\i..\j}}{\tmax - \tmin}
                    \cdot
                    (t - \tmin) \cdot
                    \frac { [(\tmax - \tmin) - (\tmax_{\i..\j} - \tmin_{\i..\j})] \cdot (\tmax - t) }
                    {(\tmax - \tmin) \cdot (\tmax - \tmin -1)} &
                    \mbox{\ \ \ \small if \ $\tmax\!>\!\tmin\!+\!1$}\\
                    0 & \mbox{\ \ \ \small if \ $\tmin\!\leq\!\tmax\!\leq\!\tmin\!+\!1$}
                    \end{array}
                    \right .

\end{array}
\]
\end{claim}

\bproof
We consider the case that $\tmax\!>\!\tmin\!+\!1$.
We start from the definition of variance:

\[
\hspace*{-.9cm}
\sigma^2(C_3([\i..\j]))=\!\!\sum_{t_{\i..\j}=0}^{t}
\!\!\left ( t_{\i..\j}-\!E(C_3([\i..\j])) \right )^2
\!\cdot
\!\!\!\!\sum_{s_{\i..\j}=0}^{s}
\!\!P(C_3([\i..\j])\!=\!t_{\i..\j},S_3([\i..\j])\!=\!s_{\i..\j})
=
\]

\[
\hspace*{-.9cm}
= \!\!\sum_{t_{\i..\j}=0}^{t} t_{\i..\j}^2 \cdot
\sum_{s_{\i..\j}=0}^{s}P(C_3([\i..\j])=t_{\i..\j},S_3([\i..\j])=s_{\i..\j})
- \left ( E(C_3([\i..\j])) \right ) ^2
=
\]

\[
\hspace*{-.9cm}
= \!\!\sum_{t_{\i..\j}=\tmin_{\i..\j}}^{t-(\tmin-\tmin_{\i..\j})}
\!t_{\i..\j}^2 \cdot\!
\left [
\frac {
    {\scriptsize
        \left ( \begin{tabular}{c} \footnotesize$\tmax_{\i..\j} - \tmin_{\i..\j}$\\ \footnotesize$t_{\i..\j} - \tmin_{\i..\j}$ \end{tabular} \right )
    }
    \!\!\cdot\!\!
    {\scriptsize
        \left ( \begin{tabular}{c} \footnotesize$\tmax - \tmax_{\i..\j} - (\tmin - \tmin_{\i..\j})$ \\ \footnotesize$t - t_{\i..\j} - (\tmin - \tmin_{\i..\j})$ \end{tabular} \right )
    }
}
{
    {\scriptsize
        \left ( \begin{tabular}{c} \footnotesize$\tmax - \tmin$\\ \footnotesize$t -\tmin$ \end{tabular} \right )
    }
}
\cdot
\right .
\]

\[
\hspace*{1.8cm}
\left .
\cdot
\sum_{s_{\i..\j}=t_{\i..\j}}^{s-t+t_{\i..\j}}
\frac {
    {\scriptsize
        \left ( \begin{tabular}{c} \footnotesize$s_{\i..\j}-1$ \\ \footnotesize$s_{\i..\j}-t_{\i..\j}$ \end{tabular} \right )
    }
    \!\!\cdot\!\!
    {\scriptsize
        \left ( \begin{tabular}{c} \footnotesize$s - s_{\i..\j}- 1$ \\ \footnotesize$s-s_{\i..\j} - t + t_{\i..\j}$ \end{tabular} \right )
    }
}
{
    {\scriptsize
        \left ( \begin{tabular}{c} \footnotesize$s-1$ \\ \footnotesize$s-t$ \end{tabular} \right )
    }
}
\right ]
\!-\! \left ( E(C_3([\i..\j])) \right ) ^2
\]

\noindent By applying the substitutions:
$S_{\i..\j}=s_{\i..\j}-t_{\i..\j}$, and: $S=s-t$, the previous
expression can be rewritten as:

\[
\hspace*{-.9cm}
\sum_{t_{\i..\j}=\tmin_{\i..\j}}^{t-(\tmin-\tmin_{\i..\j})}
\!t_{\i..\j}^2 \!\cdot\!\!
\left [
\frac {
    {\scriptsize
        \left ( \begin{tabular}{c} \footnotesize$\tmax_{\i..\j} - \tmin_{\i..\j}$\\ \footnotesize$t_{\i..\j} - \tmin_{\i..\j}$ \end{tabular} \right )
    }
    \!\!\cdot\!\!
    {\scriptsize
        \left ( \begin{tabular}{c} \footnotesize$\tmax - \tmax_{\i..\j} - (\tmin - \tmin_{\i..\j})$ \\ \footnotesize$t - t_{\i..\j} - (\tmin - \tmin_{\i..\j})$ \end{tabular} \right )
    }
}
{
    {\scriptsize
        \left ( \begin{tabular}{c} \footnotesize$\tmax - \tmin$ \\ \footnotesize$t -\tmin$ \end{tabular} \right )
    }
}
\cdot
\right .
\]

\[
\hspace*{1.7cm}
\left .
\cdot\!\!
\sum_{S_{\i..\j}=0}^{S}
\frac {
    {\scriptsize
        \left ( \!\!\begin{tabular}{c}  \footnotesize$S_{\i..\j}\!+\!t_{\i..\j}\!-\!1$ \\  \footnotesize$S_{\i..\j}$ \end{tabular} \!\!\right )
    }
    \!\!\cdot\!\!
    {\scriptsize
        \left ( \!\!\begin{tabular}{c}  \footnotesize$S\!+\!t\!-\!(S_{\i..\j} +\!t_{\i..\j})\!-\!1$ \\   \footnotesize$S-S_{\i..\j}$  \end{tabular} \!\!\right )
    }
}
{
    {\scriptsize
        \left ( \begin{tabular}{c}  \footnotesize$S\!+\!t\!-\!1$ \\  \footnotesize$S$ \end{tabular} \right )
    }
}
\right ]
\!\!-\!\! \left ( E(C_3([\i..\j])) \right ) ^2 \!\!=
\]

\begin{equation}
\label{varc4uno}
\hspace*{-.9cm}
=\!\! \sum_{t_{\i..\j}=\tmin_{\i..\j}}^{t-(\tmin-\tmin_{\i..\j})}
\!\!t_{\i..\j}^2\!\cdot\!\!
\frac {
    {\scriptsize
        \left ( \begin{tabular}{c} \footnotesize$\tmax_{\i..\j} - \tmin_{\i..\j}$\\ \footnotesize$t_{\i..\j} - \tmin_{\i..\j}$ \end{tabular} \right )
    }
    \!\!\cdot\!\!
    {\scriptsize
        \left ( \begin{tabular}{c} \footnotesize$\tmax - \tmax_{\i..\j} - (\tmin - \tmin_{\i..\j})$ \\ \footnotesize$t - t_{\i..\j} - (\tmin - \tmin_{\i..\j})$ \end{tabular} \right )
    }
}
{
    {\scriptsize
        \left ( \begin{tabular}{c} \footnotesize$\tmax - \tmin$ \\ \footnotesize$t -\tmin$ \end{tabular} \right )
    }
}
\!-\!\left ( E(C_3([\i..\j])) \right ) ^2
\end{equation}

\noindent which holds since, from (\ref{s1tre}), we have that:

\[
\hspace*{-.9cm}
\sum_{S_{\i..\j}=0}^{S}
\frac {
    {\scriptsize
        \left ( \begin{tabular}{c} \footnotesize$S_{\i..\j}+t_{\i..\j}-1$ \\ \footnotesize$S_{\i..\j}$ \end{tabular} \right )
    }
    \cdot
    {\scriptsize
        \left ( \begin{tabular}{c} \footnotesize$S + t - ( S_{\i..\j} + t_{\i..\j} ) - 1$ \\ \footnotesize$S-S_{\i..\j}$ \end{tabular} \right )
    }
}
{
    {\scriptsize
        \left ( \begin{tabular}{c} \footnotesize$S+t-1$ \\ \footnotesize$S$ \end{tabular} \right )
    }
}
= 1 .
\]
\vspace*{2mm}

\noindent Let: $h_{\i..\j}= t_{\i..\j}-\tmin_{\i..\j}$,
\hspace{4mm} $l_{\i..\j}= \tmax_{\i..\j}-\tmin_{\i..\j}$,
\hspace{4mm} $m= t- \tmin$, \hspace{4mm} and: $n=\tmax-\tmin$. By
applying these substitutions in (\ref{varc4uno}) we obtain:

\[
\hspace*{-.9cm}
\sigma^2(C_3([\i..\j]))= \sum_{h_{\i..\j}=0}^{m}
(h_{\i..\j}+\tmin_{\i..\j})^2 \cdot
\frac{
    {\scriptsize
        \left ( \begin{tabular}{c} \footnotesize$l_{\i..\j}$ \\ \footnotesize$h_{\i..\j}$ \end{tabular} \right )
    }
    \cdot
    {\scriptsize
        \left ( \begin{tabular}{c} \footnotesize$n - l_{\i..\j}$\\ \footnotesize$m - h_{\i..\j}$ \end{tabular} \right )
    }
}
{
    {\scriptsize
        \left ( \begin{tabular}{c} \footnotesize$n$ \\ \footnotesize$m$ \end{tabular} \right )
    }
}
- \left ( E(C_3([\i..\j])) \right ) ^2
\]

\noindent Finally, by substituting (\ref{applvandermonde}),
(\ref{s4quattro}) and (\ref{s4cinque}) in the above expression, we
obtain:

\[
\sigma^2(C_3([\i..\j])) = \frac{\tmax_{\i..\j} -
\tmin_{\i..\j}}{\tmax - \tmin} \cdot (t - \tmin) \cdot \frac {
[(\tmax - \tmin) - (\tmax_{\i..\j} - \tmin_{\i..\j})] \cdot (\tmax
- t) } {(\tmax - \tmin) \cdot (\tmax - \tmin -1)}
\]
\eproof

\begin{claim}\ \\

$\sigma^2(S_3([\i..\j]))=
                    \left \{
                    \begin{array}{ll}
                        \begin{array}{l}
                        \alpha\!\cdot\!(\tmax_{\i..\j}\!-\!\tmin_{\i..\j})\!\cdot\!
                        \frac {t-\tmin}{\tmax- \tmin}\!\cdot\!
                        \left[ 1\!+\!(\tmax_{\i..\j}\!-\!\tmin_{\i..\j}\!-\!1)\!\cdot\!
                        \frac{t- \tmin -1}{\tmax - \tmin -1} \right ]\!+ \\
                        (\!\beta\!+\!2\!\cdot\!\alpha\!\cdot\!\tmin_{\i..\j})\!\cdot\!
                        (\!\tmax_{\i..\j}\!-\!\tmin_{\i..\j}\!)\!\cdot\!
                        \frac{t- \tmin}{\tmax - \tmin} +\!
                        (\alpha\!\cdot\!{\tmin_{\i..\j}}^2\!+\!\beta\!\cdot\!\tmin_{\i..\j})\!-\!\gamma^2
                        \end{array}
                        &
                        \mbox{\ \ \small if \ $\tmax\!\!>\!\!\tmin\!+\!1$}\\
                        & \\
                        \frac{s \cdot \tmin_{\i..\j} \cdot (t-\tmin_{\i..\j})\cdot(s-t)}{t^2 \cdot (t+1)}
                        &
                        \hspace*{-5cm}\mbox{\ \ \small if \ \ $\tmax\!=\!\tmin \vee (\tmax\!=\!\tmin\!+\!1 \ \wedge \ t\!=\!\tmin)$}\\
                        \ & \\
                        \frac{s \cdot \tmax_{\i..\j} \cdot (t-\tmax_{\i..\j})\cdot(s-t)}{t^2 \cdot (t+1)}
                        &
                        \hspace*{-5cm}\mbox{\ \ \small if \ $(\tmax\!=\!\tmin\!+\!1 \ \wedge \ t\!=\!\tmax)$}
                    \end{array}
                    \right .$

where:

\[
\alpha=\frac{s \cdot (s+1)}{t \cdot (t+1)}, \beta=\frac{s \cdot
(s-t)}{t \cdot (t+1)} \mbox{, and } \ \gamma=E \left(
S_3([\i..\j]) \right).
\]

\end{claim}

\bproof
We consider the case that $\tmax\!\!>\!\!\tmin\!+\!1$.
We start from the definition of variance:

\[
\hspace*{-.9cm}
\sigma^2\!(S_3([\i..\j]))\!\!=\!\! \sum_{s_{\i..\j}=0}^{s}
\!\!\left ( s_{\i..\j}\!-\!\!E(S_3([\i..\j])) \right )^2
\cdot
\!\!\sum_{t_{\i..\j}=0}^{t}P(C_4([\i..\j])\!=\!t_{\i..\j},S_3([\i..\j])\!=\!s_{\i..\j})
=
\]

\[
\hspace*{-.9cm}
= \sum_{s_{\i..\j}=0}^{s} s_{\i..\j}^2 \cdot
\sum_{t_{\i..\j}=0}^{t}P(C_4([\i..\j])=t_{\i..\j},S_3([\i..\j])=s_{\i..\j})
- \left ( E(S_3([\i..\j])) \right ) ^2=
\]

\[
\hspace*{-.9cm}
=\!\!\! \sum_{t_{\i..\j}=\tmin_{\i..\j}}^{t-(\tmin-\tmin_{\i..\j})}
\sum_{s_{\i..\j}=t_{\i..\j}}^{s-t+t_{\i..\j}}
\left [ s_{\i..\j}^2 \!\cdot\!\!
\frac {
    {\scriptsize
        \left ( \begin{tabular}{c} \footnotesize$\tmax_{\i..\j} - \tmin_{\i..\j}$ \\ \footnotesize$t_{\i..\j} - \tmin_{\i..\j}$ \end{tabular} \right )
    }
    \cdot
    {\scriptsize
        \left ( \begin{tabular}{c} \footnotesize$s_{\i..\j}-1$ \\ \footnotesize$s_{\i..\j}-t_{\i..\j}$ \end{tabular} \right )
    }
}
{
    {\scriptsize
        \left ( \begin{tabular}{c} \footnotesize$\tmax - \tmin$  \\ \footnotesize$t - \tmin$\end{tabular} \right )
    }
    \cdot
    {\scriptsize
        \left ( \begin{tabular}{c} \footnotesize$s-1$ \\ \footnotesize$s-t$ \end{tabular} \right )
    }
}
\cdot \right .
\]

\[
\left . \hspace{2.3cm}
\cdot
    {\scriptsize
        \left ( \!\!\!\begin{tabular}{c} \footnotesize$\tmax\!\!-\!\tmax_{\i..\j}\!-\!(\tmin\!\!-\!\tmin_{\i..\j})$ \\
                                         \footnotesize$t\!\!-\!t_{\i..\j}\!-\!(\tmin\!-\!\tmin_{\i..\j})$
                      \end{tabular} \!\!\!\right )
    }
    \!\!\cdot\!\!
    {\scriptsize
        \left ( \!\!\!\begin{tabular}{c} \footnotesize$s\!-\!s_{\i..\j}\!-\!1$ \\
                                         \footnotesize$s\!-\!s_{\i..\j}\!-\!t\!+\!t_{\i..\j}$
                      \end{tabular} \!\!\!\right )
    }
\right ]
\!-\!\left ( E(S_3([\i..\j])) \right ) ^2 =
\]

\[
\hspace*{-.9cm}
= \sum_{t_{\i..\j}=\tmin_{\i..\j}}^{t-(\tmin-\tmin_{\i..\j})}
\left [
\frac {
    {\scriptsize
        \left ( \begin{tabular}{c} \footnotesize$\tmax_{\i..\j} -\tmin_{\i..\j}$ \\ \footnotesize$t_{\i..\j} - \tmin_{\i..\j}$ \end{tabular} \right )
    }
    \cdot
    {\scriptsize
        \left ( \begin{tabular}{c} \footnotesize$\tmax - \tmax_{\i..\j} - \tmin + \tmin_{\i..\j}$ \\ \footnotesize$t - t_{\i..\j} - \tmin + \tmin_{\i..\j}$ \end{tabular} \right )
    }
}
{
    {\scriptsize
        \left ( \begin{tabular}{c} \footnotesize$\tmax -\tmin$ \\ \footnotesize$t -\tmin$ \end{tabular} \right )
    }
    \cdot
    {\scriptsize
        \left ( \begin{tabular}{c} \footnotesize$s-1$ \\ \footnotesize$s-t$ \end{tabular} \right )
    }
}
\cdot
\right .
\]

\begin{equation}\label{s4uno}
\hspace{1.6cm}
\left .
\cdot\!\!\
\sum_{s_{\i..\j}=t_{\i..\j}}^{s-t+t_{\i..\j}} \!\!s_{\i..\j}^2 \!\cdot\!
{\scriptsize
    \left ( \!\!\begin{tabular}{c} \footnotesize$s_{\i..\j}\!-\!1$ \\ \footnotesize$s_{\i..\j\!}\!-\!t_{\i..\j}$ \end{tabular} \!\!\right )
}
\!\!\cdot\!\!
{\scriptsize
    \left ( \!\!\begin{tabular}{c} \footnotesize$s\!-\!s_{\i..\j}\!-\!1$ \\ \footnotesize$s\!-\!s_{\i..\j}\! -\! t\! +\! t_{\i..\j}$ \end{tabular} \!\!\right )
}
\right ]
\!-\!\left( E(S_3([\i..\j])) \right) ^2
\end{equation}

\[
\hspace*{-.9cm}
\mbox{The term: }
\sum_{s_{\i..\j}=t_{\i..\j}}^{s\!-\!t\!+\!t_{\i..\j}}
s_{\i..\j}^2 \cdot
{\scriptsize
    \left ( \!\!\begin{tabular}{c} \footnotesize$s_{\i..\j}-1$ \\ \footnotesize$s_{\i..\j}-t_{\i..\j}$ \end{tabular} \!\!\right )
}
\!\cdot\!\!
{\scriptsize
    \left ( \!\!\begin{tabular}{c} \footnotesize$s - s_{\i..\j} - 1$ \\ \footnotesize$s-s_{\i..\j} - t + t_{\i..\j}$ \end{tabular} \!\!\right )
}
\]
\noindent can be re-written, by replacing $s_{\i..\j}$ with $S_{\i..\j}$ + $t_{\i..\j}$, obtaining:

\[
\hspace*{-.9cm}
\sum_{s_{\i..\j}=t_{\i..\j}}^{s\!-\!t\!+\!t_{\i..\j}}
s_{\i..\j}^2 \cdot
{\scriptsize
    \left ( \!\!\begin{tabular}{c} \footnotesize$s_{\i..\j}-1$ \\ \footnotesize$s_{\i..\j}-t_{\i..\j}$ \end{tabular} \!\!\right )
}
\!\cdot\!\!
{\scriptsize
    \left ( \!\!\begin{tabular}{c} \footnotesize$s - s_{\i..\j} - 1$ \\ \footnotesize$s-s_{\i..\j} - t + t_{\i..\j}$ \end{tabular} \!\!\right )
}
=
\]

\[
\hspace*{-.9cm}
= \sum_{S_{\i..\j}=0}^{s-t} (S_{\i..\j}+t_{\i..\j})^2 \cdot
{\scriptsize
    \left ( \begin{tabular}{c} \footnotesize$t_{\i..\j} + S_{\i..\j}-1$ \\ \footnotesize$S_{\i..\j}$ \end{tabular} \right )
}
\!\cdot\!\!
{\scriptsize
    \left ( \begin{tabular}{c} \footnotesize$- t_{\i..\j} + s - S_{\i..\j} - 1$ \\ \footnotesize$s-S_{\i..\j} - t$ \end{tabular} \right )
}
=
\]

\[
\hspace*{-.9cm}
= \sum_{S_{\i..\j}=0}^{S} (S_{\i..\j}+t_{\i..\j})^2 \!\!\cdot\!\!
{\scriptsize
    \left ( \!\!\begin{tabular}{c} \footnotesize$t_{\i..\j} + S_{\i..\j}-1$ \\ \footnotesize$S_{\i..\j}$ \end{tabular} \!\!\right )
}
\!\!\cdot\!\!
{\scriptsize
    \left ( \!\!\begin{tabular}{c} \footnotesize$t \!-\! t_{\i..\j} \!+\! S\! -\! S_{\i..\j}\! -\! 1$ \\ \footnotesize$S-S_{\i..\j}$ \end{tabular} \!\!\right )
}
\ \ \mbox{  where: $S\! =\! s\!-\!t$}
\]

\[
\hspace*{-.9cm}
\mbox{Since: }
{\scriptsize
    \left ( \begin{tabular}{c} \footnotesize$x$ \\ \footnotesize$y$ \end{tabular} \right )
}
=
\frac{x-y+1}{y} \cdot
{\scriptsize
    \left ( \begin{tabular}{c} \footnotesize$x$ \\ \footnotesize$y-1$ \end{tabular} \right )
}
\mbox{, it results that:}
\]

\[
\hspace*{-.9cm}
\sum_{S_{\i..\j}=0}^{S} (S_{\i..\j})^2 \cdot
{\scriptsize
    \left ( \begin{tabular}{c} \footnotesize$t_{\i..\j} + S_{\i..\j}-1$ \\ \footnotesize$S_{\i..\j}$ \end{tabular} \right )
}
\cdot
{\scriptsize
    \left ( \begin{tabular}{c} \footnotesize$t - t_{\i..\j}+ S - S_{\i..\j} - 1$ \\ \footnotesize$S-S_{\i..\j}$ \end{tabular} \right )
}
=
\]

\[
\hspace*{-.9cm}
= \sum_{S_{\i..\j}=1}^{S} t_{\i..\j} \cdot S_{\i..\j} \cdot
{\scriptsize
\left ( \begin{tabular}{c} \footnotesize$t_{\i..\j} + S_{\i..\j}-1$ \\ \footnotesize$S_{\i..\j} - 1$ \end{tabular} \right )
}
\cdot
{\scriptsize
\left ( \begin{tabular}{c} \footnotesize$t - t_{\i..\j} + S - S_{\i..\j} - 1$ \\ \footnotesize$S-S_{\i..\j}$ \end{tabular} \right )
}=
\]

\[
\hspace*{-.9cm}
= \sum_{S_{\i..\j}=2}^{S} t_{\i..\j} \cdot ( S_{\i..\j} - 1 )
\cdot
{\scriptsize
\left ( \begin{tabular}{c} \footnotesize$t_{\i..\j} + S_{\i..\j}-1$ \\ \footnotesize$S_{\i..\j} - 1$ \end{tabular} \right )
}
\cdot
{\scriptsize
\left ( \begin{tabular}{c} \footnotesize$t - t_{\i..\j} + S - S_{\i..\j} - 1$ \\ \footnotesize$S-S_{\i..\j}$ \end{tabular} \right )
}+
\]

\[
+ \sum_{S_{\i..\j}=1}^{S} t_{\i..\j} \cdot
{\scriptsize
    \left ( \begin{tabular}{c} \footnotesize$t_{\i..\j} + S_{\i..\j}-1$ \\ \footnotesize$S_{\i..\j} - 1$ \end{tabular} \right )
}
\cdot
{\scriptsize
    \left ( \begin{tabular}{c} \footnotesize$t - t_{\i..\j} + S - S_{\i..\j} - 1$ \\ \footnotesize$S-S_{\i..\j}$ \end{tabular} \right )
}
=
\]

\[
\hspace*{-.9cm}
= \sum_{Q_{\i..\j}=1}^{S-1} t_{\i..\j} \cdot Q_{\i..\j} \cdot
{\scriptsize
    \left ( \begin{tabular}{c} \footnotesize$t_{\i..\j} + Q_{\i..\j}$ \\ \footnotesize$Q_{\i..\j}$ \end{tabular} \right )
}
\cdot
{\scriptsize
    \left ( \begin{tabular}{c} \footnotesize$t - t_{\i..\j} + (S -1) - Q_{\i..\j} - 1$ \\ \footnotesize$S - 1 - Q_{\i..\j}$ \end{tabular} \right )
}
+
\]
\[
+ \!\!\sum_{Q_{\i..\j}=0}^{S-1} t_{\i..\j} \!\cdot\!\!
{\scriptsize
    \left ( \!\!\begin{tabular}{c} \footnotesize$(t_{\i..\j}\!+\!1)\!+\!Q_{\i..\j}\!-\!1$ \\ \footnotesize$Q_{\i..\j}$ \end{tabular} \!\!\right )
}
\!\!\cdot\!\!
{\scriptsize
    \left (\!\! \begin{tabular}{c} \footnotesize$(t\!+\!1)\!-\!(t_{\i..\j}\!+\!1)\!+\!(S\!-\!1)\!-\!Q_{\i..\j}\!-\!1$ \\ \footnotesize$(S-1) - Q_{\i..\j}$ \end{tabular} \!\!\right )
}
\]
\noindent
 where: $Q_{\i..\j} = S_{\i..\j} -1$.

\noindent By applying formula (\ref{s1tre}), we obtain that:
\[
\hspace*{-.9cm}
\sum_{Q_{\i..\j}=0}^{S-1} t_{\i..\j} \!\cdot\!\!
{\scriptsize
    \left ( \!\!\begin{tabular}{c} \footnotesize$(t_{\i..\j}\!+\!1)\!+\!Q_{\i..\j}\!-\!1$ \\ \footnotesize$Q_{\i..\j}$ \end{tabular} \!\!\right )
}
\!\!\cdot\!\!
{\scriptsize
    \left (\!\! \begin{tabular}{c} \footnotesize$(t\!+\!1)\!-\!(t_{\i..\j}\!+\!1)\!+\!(S\!-\!1)\!-\!Q_{\i..\j}\!-\!1$ \\ \footnotesize$(S-1) - Q_{\i..\j}$ \end{tabular} \!\!\right )
}
=
\]
\[
\hspace*{-.9cm}=
t_{\i..\j}\!\cdot\!\!
{\scriptsize
    \left ( \!\!\begin{tabular}{c} \footnotesize$t\!+\!S\!-\!1$ \\ \footnotesize$S\!-\!1$ \end{tabular} \!\!\right )
}
\]

\noindent On the other hand,

\[
\hspace*{-.9cm}
\sum_{Q_{\i..\j}=1}^{S-1} t_{\i..\j} \cdot Q_{\i..\j} \cdot
{\scriptsize
    \left ( \begin{tabular}{c} \footnotesize$t_{\i..\j} + Q_{\i..\j}$ \\ \footnotesize$Q_{\i..\j}$ \end{tabular} \right )
}
\cdot
{\scriptsize
    \left ( \begin{tabular}{c} \footnotesize$t - t_{\i..\j} + (S -1) - Q_{\i..\j} - 1$ \\ \footnotesize$S - 1 - Q_{\i..\j}$ \end{tabular} \right )
}=
\]

\[
\hspace*{-.9cm}
= \sum_{Q_{\i..\j}=1}^{S-1} t_{\i..\j} \cdot Q_{\i..\j} \cdot
\frac {t_{\i..\j}+1}{Q_{\i..\j}} \cdot
{\scriptsize
    \left ( \begin{tabular}{c} \footnotesize$t_{\i..\j} + Q_{\i..\j}$ \\ \footnotesize$Q_{\i..\j} -1$ \end{tabular} \right )
}
\cdot
{\scriptsize
    \left ( \begin{tabular}{c} \footnotesize$t - t_{\i..\j} + (S -1) - Q_{\i..\j} - 1$ \\ \footnotesize$S - 1 - Q_{\i..\j}$ \end{tabular} \right )
}
\]

\noindent Substituting $R_{\i..\j}= Q_{\i..\j} -1 $ the latter
becomes:

\[
\hspace*{-.9cm}
\sum_{R_{\i..\j}=0}^{S-2} t_{\i..\j} \!\cdot\!\! (t_{\i..\j}\!+\!1) \!\!\cdot\!\!
{\scriptsize
    \left ( \!\!\begin{tabular}{c} \footnotesize$(t_{\i..\j}\!+\!2)\! +\! R_{\i..\j}\! -\!1$ \\ \footnotesize$R_{\i..\j}$ \end{tabular} \!\!\right )
}
\!\!\cdot\!\!
{\scriptsize
    \left ( \!\!\begin{tabular}{c} \footnotesize$(t\!+\!2)\! -\! (t_{\i..\j}\!+\!2)\! +\! (S\! -\!2)\! - \!R_{\i..\j}\! -\! 1$ \\ \footnotesize$S - 2 - R_{\i..\j}$ \end{tabular} \!\!\right )
}=
\]

\[
\hspace*{-.9cm}
=t_{\i..\j} \cdot (t_{\i..\j}+1) \cdot
{\scriptsize
    \left ( \begin{tabular}{c} \footnotesize$t+S-1$ \\ \footnotesize$S - 2$ \end{tabular} \right )
}
\]

\noindent Thus we obtain that:

\[
\hspace*{-.9cm}
\sum_{S_{\i..\j}=0}^{S} (S_{\i..\j})^2 \cdot
{\scriptsize
    \left ( \begin{tabular}{c} \footnotesize$t_{\i..\j} + S_{\i..\j}-1$\\ \footnotesize$S_{\i..\j}$ \end{tabular} \right )
}
\cdot
{\scriptsize
    \left ( \begin{tabular}{c} \footnotesize$t - t_{\i..\j} + S - S_{\i..\j} - 1$ \\ \footnotesize$S-S_{\i..\j}$ \end{tabular} \right )
}
=
\]

\[
\hspace*{-.9cm}
=t_{\i..\j} \cdot (t_{\i..\j}+1) \cdot
{\scriptsize
    \left ( \begin{tabular}{c} \footnotesize$t+S-1$ \\ \footnotesize$ S - 2$ \end{tabular} \right )
}
+ t_{\i..\j} \!\cdot\!\!
{\scriptsize
    \left ( \begin{tabular}{c} \footnotesize$t+S-1$ \\ \footnotesize$S-1$ \end{tabular} \right )
}
=
\]

\[
\hspace*{-.9cm}
= t_{\i..\j} \cdot \frac {S} { t }
{\scriptsize
    \left ( \begin{tabular}{c} \footnotesize$t+S-1$ \\ \footnotesize$S-1$ \end{tabular} \right )
}
\cdot \left[ ( t_{\i..\j} +1 ) \cdot \frac {S-1}{t+1} +1 \right]
\]

\noindent It also holds that:

\[
\hspace*{-.9cm}
\sum_{S_{\i..\j}=1}^{S} 2 \cdot t_{\i..\j} \cdot S_{\i..\j} \cdot
{\scriptsize
    \left ( \begin{tabular}{c} \footnotesize$t_{\i..\j} + S_{\i..\j}-1$ \\ \footnotesize$S_{\i..\j}$ \end{tabular} \right )
}
\cdot
{\scriptsize
    \left ( \begin{tabular}{c} \footnotesize$t - t_{\i..\j} + S - S_{\i..\j} - 1$ \\ \footnotesize$S-S_{\i..\j}$ \end{tabular} \right )
}
=
\]

\[
\hspace*{-.9cm}
= 2 \cdot t_{\i..\j} \cdot \sum_{S_{\i..\j}=1}^{S} S_{\i..\j}
\cdot \frac {1}{S_{\i..\j}} \cdot
{\scriptsize
    \left ( \begin{tabular}{c} \footnotesize$t_{\i..\j} + S_{\i..\j}-1$ \\ \footnotesize$S_{\i..\j} -1$ \end{tabular} \right )
}
\cdot
{\scriptsize
    \left ( \begin{tabular}{c} \footnotesize$t - t_{\i..\j} + S - S_{\i..\j} - 1$ \\ \footnotesize$S-S_{\i..\j}$ \end{tabular} \right )
}=
\]

\[
\hspace*{-.9cm}
= 2 \cdot t_{\i..\j} \cdot \sum_{Q_{\i..\j}=0}^{S-1}
{\scriptsize
    \left ( \begin{tabular}{c} \footnotesize$t_{\i..\j} + Q_{\i..\j}$  \\ \footnotesize$Q_{\i..\j}$ \end{tabular} \right )
}
\cdot
{\scriptsize
    \left ( \begin{tabular}{c} \footnotesize$t - t_{\i..\j} + (S-1) - Q_{\i..\j} - 1$ \\ \footnotesize$(S-1)-Q_{\i..\j}$ \end{tabular} \right )
}
=
\]

\[
\hspace*{-.9cm}
=
2 \cdot t_{\i..\j} \cdot
{\scriptsize
    \left ( \begin{tabular}{c} \footnotesize$t+S-1$ \\ \footnotesize$S$ \end{tabular} \right )
}
\cdot \frac {t_{\i..\j}}{t} \cdot S
\]
\noindent and, from (\ref{s1tre}):

\[
\hspace*{-.9cm}
\sum_{S_{\i..\j}=1}^{S} t_{\i..\j}^2 \!\cdot\!\!
{\scriptsize
    \left ( \!\!\begin{tabular}{c} \footnotesize$t_{\i..\j}\! +\! S_{\i..\j}\!-\!1$ \\ \footnotesize$S_{\i..\j}$ \end{tabular} \!\!\right )
}
\!\!\cdot\!\!
{\scriptsize
    \left ( \!\!\begin{tabular}{c} \footnotesize$t\! -\! t_{\i..\j}\! + \!S\! -\! S_{\i..\j}\! -\! 1$ \\ \footnotesize$S-S_{\i..\j}$ \end{tabular} \!\!\right )
}
=
t_{\i..\j}^2 \!\cdot\!\!
{\scriptsize
    \left ( \!\!\begin{tabular}{c} \footnotesize$t_{\i..\j}\! +\! S_{\i..\j}\!-\!1$ \\ \footnotesize$S_{\i..\j}$ \end{tabular}\!\! \right )
}
\]

\noindent  Thus, we have that:

\[
\hspace*{-.9cm}
\sum_{s_{\i..\j}=t_{\i..\j}}^{s-t+t_{\i..\j}} s_{\i..\j}^2 \cdot
{\scriptsize
    \left ( \begin{tabular}{c} \footnotesize$s_{\i..\j}-1$ \\ \footnotesize$s_{\i..\j}-t_{\i..\j}$ \end{tabular} \right )
}
\cdot
{\scriptsize
    \left ( \begin{tabular}{c} \footnotesize$s - s_{\i..\j} - 1$ \\ \footnotesize$s-s_{\i..\j} - t + t_{\i..\j}$ \end{tabular} \right )
}
=
\]

\[
\hspace*{-.9cm}
=
\sum_{S_{\i..\j}=0}^{S} (S_{\i..\j}+ t_{\i..\j})^2 \cdot
{\scriptsize
    \left ( \begin{tabular}{c} \footnotesize$t_{\i..\j} + S_{\i..\j}-1$ \\ \footnotesize$S_{\i..\j}$ \end{tabular} \right )
}
\cdot
{\scriptsize
    \left ( \begin{tabular}{c} \footnotesize$t - t_{\i..\j} + S - S_{\i..\j} - 1$ \\ \footnotesize$S-S_{\i..\j}$ \end{tabular} \right )
}
=
\]

 \[
\hspace*{-.9cm}
=
{\scriptsize
    \left ( \begin{tabular}{c} \footnotesize$t + S -1$ \\ \footnotesize$S$ \end{tabular} \right )
}
\cdot
\left [ \left ( \frac {S \cdot (S-1)}{t \cdot (t+1)} + 1 + 2 \cdot \frac {S}{t} \right )
        \cdot t_{\i..\j}^2 + \frac {S \cdot (S+t)}{t \cdot (t+1)} \cdot t_{\i..\j}
\right ]
=
\]

\[
\hspace*{-.9cm}
=
{\scriptsize
    \left ( \begin{tabular}{c} \footnotesize$t + S -1$ \\ \footnotesize$S$ \end{tabular} \right )
}
\cdot
\left [ \frac{s \cdot (s+1)}{t \cdot (t+1)} \cdot t_{\i..\j}^2 +
        \frac {s \cdot (s-t)}{t \cdot (t+1)} \cdot t_{\i..\j}
\right ]
=
\]

\[
\hspace*{-.9cm}
=
{\scriptsize
    \left ( \!\begin{tabular}{c} \footnotesize$t\! +\! S\! -\!1$ \\ \footnotesize$S$ \end{tabular} \!\right )
}
\!\!\cdot\!\!
\left(  \alpha \cdot t_{\i..\j}^2 + \beta \cdot t_{\i..\j} \right ),
\mbox{  where: }
\alpha= \frac{s \cdot (s+1)}{t \cdot (t+1)}, \hspace{5mm}
\beta=\frac {s \cdot (s-t)}{t \cdot (t+1)}
\]

\noindent Substituting this term in (\ref{s4uno}) we obtain:
\[
\hspace*{-.9cm}
\sigma^2(S_3[\i..\j])=
\frac {1}
{
    {\scriptsize
        \left ( \!\! \begin{tabular}{c} \footnotesize$\tmax\!\! -\!\!\tmin$ \\ \footnotesize$t\!\! -\!\! \tmin$ \end{tabular} \!\!\right )
    }
}
\!\cdot\!\!
\left [ \alpha \!\cdot\!\! \sum_{t_{\i..\j}=\tmin_{\i..\j}}^{t-(\tmin-\tmin_{\i..\j})}
t_{\i..\j}^2 \cdot\!\!
{\scriptsize
    \left ( \!\!\begin{tabular}{c} \footnotesize$\tmax_{\i..\j}\!  -\!\! \tmin_{\i..\j}$ \\ \footnotesize$t_{\i..\j}\! -\!\! \tmin_{\i..\j}$ \end{tabular} \right )
}
\!\!\cdot\!\!
{\scriptsize
    \left ( \!\!\begin{tabular}{c} \footnotesize$\tmax\!\! -\! \tmin\!\! -\! \tmax_{\i..\j} \!+\! \tmin_{\i..\j}$ \\ \footnotesize$t\!\! - \!\tmin\!\! -\! t_{\i..\j} + \tmin_{\i..\j}$ \end{tabular} \right )
}
+
\right .
\]

\[
\hspace*{3.6cm}
\left .
+ \beta \cdot\!\!\!
\sum_{t_{\i..\j}=\tmin_{\i..\j}}^{t-(\tmin-\tmin_{\i..\j})}
\!\!t_{\i..\j} \cdot\!\!
{\scriptsize
    \left ( \!\!\begin{tabular}{c} \footnotesize$\tmax_{\i..\j}\!  -\!\! \tmin_{\i..\j}$ \\ \footnotesize$t_{\i..\j}\! -\!\! \tmin_{\i..\j}$ \end{tabular} \right )
}
\!\!\cdot\!\!
{\scriptsize
    \left ( \!\!\begin{tabular}{c} \footnotesize$\tmax\!\! -\! \tmin\!\! -\! \tmax_{\i..\j} \!+\! \tmin_{\i..\j}$ \\ \footnotesize$t\!\! - \!\tmin\!\! -\! t_{\i..\j} + \tmin_{\i..\j}$ \end{tabular} \right )
}
\right ]
-
\]
\[
\hspace*{1.7cm}
-\left( E ( S_3([\i..\j])) \right ) ^2 =
\]

\[
\hspace*{-.9cm}
= \frac {1}
{
    {\scriptsize
        \left ( \!\! \begin{tabular}{c} \footnotesize$\tmax\!\! -\!\!\tmin$ \\ \footnotesize$t\!\! -\!\! \tmin$ \end{tabular} \!\!\right )
    }
}
\!\cdot\!\!
\left [
\alpha \cdot \!\!\!
\sum_{h_{\i..\j}=0}^{m} (h_{\i..\j}+\tmin_{\i..\j})^2 \!\cdot\!\!
{\scriptsize
    \left ( \!\!\begin{tabular}{c} \footnotesize$l_{\i..\j}$ \\ \footnotesize$h_{\i..\j}$ \end{tabular} \!\!\right )
}
\!\!\cdot\!\!
{\scriptsize
    \left ( \!\!\begin{tabular}{c} \footnotesize$\tmax - \tmin - l_{\i..\j}$ \\ \footnotesize$m - h_{\i..\j}$ \end{tabular} \!\!\right )
}
+
\right .
\]

\[
\hspace*{1.6cm}
\left .
+
\beta \cdot \!\!\!\sum_{h_{\i..\j}=0}^{m}
(h_{\i..\j}\!+\!\tmin_{\i..\j})
\!\cdot\!\!
{\scriptsize
    \left ( \!\!\begin{tabular}{c} \footnotesize$l_{\i..\j}$ \\ \footnotesize$h_{\i..\j}$ \end{tabular} \!\!\right )
}
\!\!\cdot\!\!
{\scriptsize
    \left ( \!\!\begin{tabular}{c} \footnotesize$\tmax\!\! -\! \tmin\!\! -\! l_{\i..\j}$\\ \footnotesize$m - h_{\i..\j}$ \end{tabular}\!\! \right )
}
\right ]
- \left( E ( S_3([\i..\j])) \right ) ^2
\]

\noindent that is:

\begin{equation}\label{s4tre}
\hspace*{-.9cm}
\sigma^2(S_3([\i..\j]))=
\frac {1}
{
    {\scriptsize
        \left ( \!\! \begin{tabular}{c} \footnotesize$\tmax\!\! -\!\!\tmin$ \\ \footnotesize$t\!\! -\!\! \tmin$ \end{tabular} \!\!\right )
    }
}
\!\cdot\!\!
\left [ \alpha \cdot \!\!\!\sum_{h_{\i..\j}=0}^{m} h_{\i..\j}^2 \!\cdot\!\!
{\scriptsize
    \left ( \!\!\begin{tabular}{c} \footnotesize$l_{\i..\j}$ \\ \footnotesize$h_{\i..\j}$ \end{tabular} \!\!\right )
}
\!\!\cdot\!\!
{\scriptsize
    \left ( \!\!\begin{tabular}{c} \footnotesize$n - l_{\i..\j}$ \\ \footnotesize$m - h_{\i..\j}$ \end{tabular} \!\!\right )
}
+ \right .
\end{equation}

\[
\hspace*{1.9cm}
+
\left ( \beta \!+\! 2\! \cdot\! \tmin_{\i..\j} \!\cdot\! \alpha \right )
\!\cdot\!\!\!
\sum_{h_{\i..\j}=0}^{m} \!h_{\i..\j} \!\cdot\!\!
{\scriptsize
    \left ( \begin{tabular}{c} \footnotesize$l_{\i..\j}$ \\ \footnotesize$h_{\i..\j}$ \end{tabular} \right )
}
\!\!\cdot\!\!
{\scriptsize
    \left ( \begin{tabular}{c} \footnotesize$n - l_{\i..\j}$ \\ \footnotesize$m - h_{\i..\j}$ \end{tabular} \right )
}
+
\]

\[
\hspace*{1.9cm}
\left .
+
\left ( \alpha \cdot {\tmin_{\i..\j}}^2 + \beta \cdot \tmin_{\i..\j} \right )
\!\cdot\!\!\!
\sum_{h_{\i..\j}=0}^{m}\!\!
{\scriptsize
    \left ( \!\!\begin{tabular}{c} \footnotesize$l_{\i..\j}$ \\ \footnotesize$h_{\i..\j}$ \end{tabular} \!\!\right )
}
\!\!\cdot\!\!
{\scriptsize
    \left ( \!\!\begin{tabular}{c} \footnotesize$n\! -\! l_{\i..\j}$\\ \footnotesize$m\! -\! h_{\i..\j}$ \end{tabular} \!\!\right )
}
\right ]
\!-\!\! \left( E ( S_3([\i..\j])) \right ) ^2
\]

\noindent where: $h_{\i..\j}= t_{\i..\j}-\tmin_{\i..\j}$,
\hspace{5mm} $l_{\i..\j}= \tmax_{\i..\j}-\tmin_{\i..\j}$,
\hspace{5mm}
$m= t- \tmin$, \hspace{5mm} and: $n=\tmax-\tmin$.\\

\[
\hspace*{-.9cm}
\mbox{Since: }
{\scriptsize
    \left ( \begin{tabular}{c} \footnotesize$x$ \\ \footnotesize$y$ \end{tabular} \right )
}
=
\frac {x}{y} \cdot
{\scriptsize
    \left ( \begin{tabular}{c} \footnotesize$x-1$ \\ \footnotesize$y-1$ \end{tabular} \right )
},
\mbox{ we have that:}
\]

 \[
\hspace*{-.9cm}
\sum_{h_{\i..\j}=0}^{m} h_{\i..\j}^2 \!\cdot\!\!
{\scriptsize
    \left ( \!\begin{tabular}{c} \footnotesize$l_{\i..\j}$ \\ \footnotesize$h_{\i..\j}$ \end{tabular} \!\right )
}
\!\!\cdot\!\!
{\scriptsize
    \left ( \!\begin{tabular}{c} \footnotesize$n - l_{\i..\j}$ \\ \footnotesize$m - h_{\i..\j}$ \end{tabular} \!\right )
}
=
\sum_{h_{\i..\j}=1}^{m} l_{\i..\j} \cdot
h_{\i..\j} \cdot\!\!
{\scriptsize
    \left (\! \begin{tabular}{c} \footnotesize$l_{\i..\j} - 1$ \\ \footnotesize$h_{\i..\j} - 1$ \end{tabular} \!\right )
}
\!\!\cdot\!\!
{\scriptsize
    \left (\! \begin{tabular}{c} \footnotesize$n - l_{\i..\j}$ \\ \footnotesize$m - h_{\i..\j}$ \end{tabular}\! \right )
}=
\]

\[
\hspace*{-.9cm}
=\!\!
\sum_{h_{\i..\j}=1}^{m} l_{\i..\j} \!\cdot\!\! (h_{\i..\j}\! -\!1) \!\cdot\!\!
{\scriptsize
    \left ( \!\!\begin{tabular}{c} \footnotesize$l_{\i..\j} - 1$ \\ \footnotesize$h_{\i..\j} - 1$ \end{tabular} \!\!\right )
}
\!\!\cdot\!\!
{\scriptsize
    \left ( \!\!\begin{tabular}{c} \footnotesize$n - l_{\i..\j}$ \\ \footnotesize$m - h_{\i..\j}$ \end{tabular} \!\!\right )
}
\!\!+\!\!
\sum_{h_{\i..\j}=1}^{m} l_{\i..\j} \!\cdot\!\!
{\scriptsize
    \left (\!\! \begin{tabular}{c} \footnotesize$l_{\i..\j} - 1$ \\ \footnotesize$h_{\i..\j} - 1$ \end{tabular} \!\!\right )
}
\!\!\cdot\!\!
{\scriptsize
    \left ( \!\!\begin{tabular}{c} \footnotesize$n - l_{\i..\j}$ \\ \footnotesize$m - h_{\i..\j}$ \end{tabular} \!\!\right )
}
\!=
\]

\[
\hspace*{-.9cm}
= \!\!
\sum_{p_{\i..\j}=0}^{m-1} l_{\i..\j} \cdot\! p_{\i..\j} \cdot\!
{\scriptsize
    \left ( \!\!\begin{tabular}{c} \footnotesize$l_{\i..\j} -\! 1$ \\ \footnotesize$p_{\i..\j}$ \end{tabular} \!\!\right )
}
\!\!\cdot\!\!
{\scriptsize
    \left ( \!\!\begin{tabular}{c} \footnotesize$n\! -\! l_{\i..\j}$ \\ \footnotesize$m\! -\! 1\! -\! p_{\i..\j}$ \end{tabular} \!\!\right )
}
\!\!+\!\!
\sum_{p_{\i..\j}=0}^{m-1} l_{\i..\j} \cdot\!
{ \scriptsize
    \left (\!\! \begin{tabular}{c} \footnotesize$l_{\i..\j} -\! 1$ \\ \footnotesize$p_{\i..\j}$ \end{tabular} \!\!\right )
}
\!\!\cdot\!\!
{\scriptsize
    \left ( \!\!\begin{tabular}{c} \footnotesize$n\! -\! l_{\i..\j}$ \\ \footnotesize$m\! -\! 1\! - \!p_{\i..\j}$ \end{tabular} \!\!\right )
}
\]
\noindent where: $ p_{\i..\j} = h_{\i..\j} - 1 $.

\noindent By applying the Vandermonde formula (\ref{vandermonde})
we obtain:

\begin{equation}
\label{applvandermonde}
\hspace*{-.9cm}
\sum_{h_{\i..\j}=0}^{m}
{ \scriptsize
    \left ( \begin{tabular}{c} \footnotesize$l_{\i..\j}$ \\ \footnotesize$h_{\i..\j}$ \end{tabular} \right )
}
\cdot
{ \scriptsize
    \left ( \begin{tabular}{c} \footnotesize$n - l_{\i..\j}$ \\ \footnotesize$m - h_{\i..\j}$ \end{tabular} \right )
}
=
{ \scriptsize
    \left ( \begin{tabular}{c} \footnotesize$n$ \\ \footnotesize$m$ \end{tabular} \right )
}
\end{equation}

\noindent and:

\[
\hspace*{-.9cm}
\sum_{p_{\i..\j}=0}^{m-1} l_{\i..\j} \cdot
{ \scriptsize
    \left ( \begin{tabular}{c} \footnotesize$l_{\i..\j} - 1$ \\ \footnotesize$p_{\i..\j}$ \end{tabular} \right )
}
\cdot
{ \scriptsize
    \left ( \begin{tabular}{c} \footnotesize$n - l_{\i..\j}$ \\ \footnotesize$m - 1 - p_{\i..\j}$ \end{tabular} \right )
}
=
l_{\i..\j} \cdot
{ \scriptsize
    \left ( \begin{tabular}{c} \footnotesize$n - 1$ \\ \footnotesize$m - 1$ \end{tabular} \right )
}
\]

\noindent On the other hand:

\[
\hspace*{-.9cm}
\sum_{p_{\i..\j}=0}^{m-1} l_{\i..\j} \cdot\! p_{\i..\j} \cdot\!\!
{ \scriptsize
    \left ( \!\!\begin{tabular}{c} \footnotesize$l_{\i..\j}\! -\! 1$ \\ \footnotesize$p_{\i..\j}$ \end{tabular} \!\!\right )
}
\!\!\cdot\!\!
{ \scriptsize
    \left ( \!\!\begin{tabular}{c} \footnotesize$n\! -\! l_{\i..\j}$ \\ \footnotesize$m\! -\! 1\! -\! p_{\i..\j}$ \end{tabular} \!\!\right )
}
=\]
\[
\hspace*{-.9cm}
=
\sum_{p_{\i..\j}=1}^{m-1} l_{\i..\j} \cdot\! (l_{\i..\j}-\!1)\! \cdot\!\!
{ \scriptsize
    \left ( \!\!\begin{tabular}{c} \footnotesize$l_{\i..\j} - \!2$ \\ \footnotesize$p_{\i..\j} -\!1$ \end{tabular} \!\!\right )
}
\!\!\cdot\!\!
{ \scriptsize
    \left ( \!\!\begin{tabular}{c} \footnotesize$n \!-\! l_{\i..\j}$ \\ \footnotesize$m\! -\! 1\! -\! p_{\i..\j}$ \end{tabular} \!\!\right )
}
=
 \]

\[
\hspace*{-.9cm}
= \sum_{U_{\i..\j}=0}^{m-2} l_{\i..\j}\! \cdot \!(l_{\i..\j}-\!1)\! \cdot\!\!
{ \scriptsize
    \left ( \!\! \begin{tabular}{c} \footnotesize$l_{\i..\j} \!-\! 2$ \\ \footnotesize$U_{\i..\j}$ \end{tabular} \!\!\right )
}
\!\cdot\!
{ \scriptsize
    \left ( \!\!\begin{tabular}{c} \footnotesize$n\! -\! l_{\i..\j}$ \\ \footnotesize$m\! -\! 2\! - \!U_{\i..\j}$ \end{tabular} \!\!\right )
}
\!=\ \
l_{\i..\j} \!\cdot \!(l_{\i..\j}-\!1)\!\cdot\!\!
{ \scriptsize
    \left ( \!\!\begin{tabular}{c} \footnotesize$n\! -\! 2$ \\ \footnotesize$m\! -\! 2$ \end{tabular} \!\!\right )
}
\]

\noindent  Thus:

\[
\hspace*{-.9cm}
\sum_{h_{\i..\j}=0}^{m}\! h_{\i..\j}^2 \!\cdot\!\!
{ \scriptsize
    \left ( \!\!\begin{tabular}{c} \footnotesize$l_{\i..\j}$ \\ \footnotesize$h_{\i..\j}$ \end{tabular} \!\!\right )
}
\!\!\cdot\!\!
{ \scriptsize
    \left ( \!\!\begin{tabular}{c} \footnotesize$n - l_{\i..\j}$ \\ \footnotesize$m - h_{\i..\j}$ \end{tabular} \!\!\right )
}
=
l_{\i..\j} \cdot\!\!
{ \scriptsize
    \left ( \!\!\begin{tabular}{c} \footnotesize$n\! -\! 1$ \\ \footnotesize$m\! -\! 1$ \end{tabular} \!\!\right )
}
\!+
l_{\i..\j} \cdot \!\!(l_{\i..\j}-\!1)\!\cdot\!\!
{ \scriptsize
    \left ( \!\!\begin{tabular}{c} \footnotesize$n\! -\! 2$ \\ \footnotesize$m\!-\!2$ \end{tabular} \!\!\right )
}
=
\]

\[
\hspace*{-.9cm}
= l_{\i..\j} \cdot \frac {m}{n} \cdot
{ \scriptsize
    \left ( \begin{tabular}{c} \footnotesize$n$ \\ \footnotesize$m$ \end{tabular} \right )
}
\!+ l_{\i..\j} \cdot (l_{\i..\j} - 1)
\cdot \frac {m}{n} \cdot \frac {m-1}{n - 1} \cdot
{ \scriptsize
    \left ( \begin{tabular}{c} \footnotesize$n$ \\ \footnotesize$m$ \end{tabular} \right )
}=
\]

\begin{equation}\label{s4quattro}
\hspace*{-.9cm}
= \!h_{\i..\j} \cdot\!
\frac {t\!-\!\tmin}{\tmax\!\!-\!\tmin}\! \cdot\!\!
{ \scriptsize
    \left ( \!\!\begin{tabular}{c} \footnotesize$\tmax\!\!-\!\tmin$ \\ \footnotesize$t\!-\!\tmin$ \end{tabular} \!\!\right )
}
\!+
l_{\i..\j} \cdot \!(l_{\i..\j} -\! 1) \!\cdot\!
\frac{t\!-\!\tmin}{\tmax\!\!-\!\tmin} \cdot \frac {t\!-\!\tmin\!-\!1}{\tmax\!\!-\!\tmin\! - \!1}
\!\cdot\!\!
{ \scriptsize
    \left (\!\! \begin{tabular}{c} \footnotesize$\tmax\!\!-\!\tmin$ \\ \footnotesize$t\!-\!\tmin$ \end{tabular} \!\!\right )
}
\end{equation}

\noindent It also holds that:

\[
\hspace*{-.9cm}
\sum_{h_{\i..\j}=0}^{m} h_{\i..\j} \cdot\!
{ \scriptsize
    \left ( \begin{tabular}{c} \footnotesize$l_{\i..\j}$ \\ \footnotesize$h_{\i..\j}$ \end{tabular} \right )
}
\!\!\cdot\!\!
{ \scriptsize
    \left ( \begin{tabular}{c} \footnotesize$n - l_{\i..\j}$ \\ \footnotesize$m - h_{\i..\j}$ \end{tabular} \right )
}
\!=\!\!
\sum_{T_{\i..\j}=0}^{m-1} \! \tmax_{\i..\j} \cdot\!
{ \scriptsize
    \left ( \begin{tabular}{c} \footnotesize$\tmax_{\i..\j}-1$ \\ \footnotesize$T_{\i..\j}$ \end{tabular} \right )
}
\!\!\cdot\!\!
{ \scriptsize
    \left ( \begin{tabular}{c} \footnotesize$n - l_{\i..\j}$ \\ \footnotesize$m - 1 - T_{\i..\j}$ \end{tabular} \right )
}
=
\]

\begin{equation}\label{s4cinque}
\hspace*{-.9cm}
= l_{\i..\j} \cdot \frac {m}{n} \cdot
{ \scriptsize
    \left ( \begin{tabular}{c} \footnotesize$n$ \\ \footnotesize$m$ \end{tabular} \right )
}
\!\cdot\!
(\tmax_{\i..\j}-\!\tmin_{\i..\j}) \!\cdot\!
\frac {t\!-\! \tmin}{\tmax\!-\! \tmin} \!\cdot\!
{ \scriptsize
    \left ( \begin{tabular}{c} \footnotesize$\tmax\!-\! \tmin$ \\ \footnotesize$t\!-\! \tmin$ \end{tabular} \right )
}
\end{equation}

\noindent where: $T_{\i..\j}= h_{\i..\j}-1$.

\noindent Finally, substituting (\ref{applvandermonde}),
(\ref{s4quattro}) and (\ref{s4cinque}) in (\ref{s4tre}) we obtain:

\[
\hspace*{-.9cm}
\begin{array}{ll}
\sigma^2(S_3([\i..\j]))= &
\alpha \cdot (\tmax_{\i..\j}-\tmin_{\i..\j}) \cdot
\frac {t- \tmin}{\tmax- \tmin} \cdot
\left[ 1+(\tmax_{\i..\j}-\tmin_{\i..\j} -1) \cdot
       \frac {t- \tmin -1}{\tmax - \tmin -1}
\right ] + \\
& \\
&+ \ (\beta \!+2\! \cdot\! \alpha \!\cdot\! \tmin_{\i..\j}) \cdot
(\tmax_{\i..\j}-\tmin_{\i..\j}) \cdot
 \frac {t- \tmin}{\tmax - \tmin} +
 (\alpha \cdot {\tmin_{\i..\j}}^2 + \beta \cdot \tmin_{\i..\j}) + \\
& \\
&- (E(S_3([\i..\j])))^2
 \end{array}
\]
\eproof


%
%
%
%

%
%

\end{document}